\documentclass[12pt]{article}
\usepackage{authblk}
\usepackage{hyperref}
\usepackage{url}
\usepackage{array}
\usepackage{makecell}
\usepackage{amsmath,amsfonts,amssymb,amsthm}
\usepackage{graphicx}
\usepackage[numbers]{natbib}
\usepackage{soul} 

\usepackage{booktabs} 
\usepackage{subcaption}
\usepackage{bbm}
\usepackage{bm}
\usepackage{mwe}
\usepackage{paralist} 
\usepackage{indentfirst} 
\usepackage{xcolor}

\usepackage[font=footnotesize,labelfont=bf]{caption}

\usepackage[margin=1in]{geometry}

\DeclareMathOperator*{\argmin}{arg\,min}

\title{ Precrop Payoffs: Causal machine learning reveals large but variable yield benefits of crop rotation in major breadbaskets}

\author[1,2]{Dan M. Kluger \thanks{Corresponding author: dkluger@mit.edu}}
\author[3]{Stefania Di Tommaso}
\author[3,4] {David B. Lobell}
\affil[1]{\small Department of Statistics, Stanford University, Stanford CA, 94305}
\affil[2]{\small Institute for Data, Systems, and Society, Massachusetts Institute of Technology, Cambridge MA, 02139}
\affil[3]{\small Center on Food Security and the Environment, Stanford University, Stanford CA, 94305}
\affil[4]{\small Department of Earth System Science, Stanford University, Stanford CA, 94305}
\date{\small October 2025}

\newcommand{\e}{\mathbb{E}}

\newcommand{\tran}{\mathsf{T}}
\newcommand{\indep}{\raisebox{0.05em}{\rotatebox[origin=c]{90}{$\models$}}}

\newcommand{\YSCYM}{\tilde{Y}_{\text{SCYM}}}

\newcommand{\omSCYM}{\omega_{\text{SCYM}}}

\makeatletter
\newcommand{\skipitems}[1]{\addtocounter{\@enumctr}{#1}}
\makeatother

\newcommand{\giv}{\!\mid\!} 

\theoremstyle{definition}

\begin{document}

\maketitle

\begin{abstract}

Building sustainable food systems that are resilient to climate change will require improved agricultural management and policy. One common practice that is well-known to benefit crop yields is crop rotation, yet there remains limited understanding of how the benefits of crop rotation vary for different crop sequences and for different weather conditions. To address these gaps, we leverage crop type maps, satellite data, and causal machine learning to study how precrop effects on subsequent yields vary with cropping sequence choice and weather. Complementing and going beyond what is known from randomized field trials, we find that (i) for those farmers who do rotate, the most common precrop choices tend to be among the most beneficial, (ii) the effects of switching from a simple rotation (which alternates between two crops) to a more diverse rotation were typically small and sometimes even negative, (iii) precrop effects tended to be greater under rainier conditions, (iv) precrop effects were greater under warmer conditions for soybean yields but not for other crops, and (v) legume precrops conferred smaller benefits under warmer conditions. Our results and the methods we use can enable farmers and policy makers to identify which rotations will be most effective at improving crop yields in a changing climate.

\end{abstract}


\section{Introduction}\label{sec:Intro}

Crop rotation is a well-established agronomic practice that typically raises crop yields while reducing the need for fertilizer and pesticide inputs. Yield benefits can arise from several mechanisms, including suppression of harmful weeds, pests, and diseases \cite{CropRotationReducesWeedDensity,BennettDiscussesMechanismsOfRotationBenefit}, increase of soil availability of essential nutrients such as nitrogen \cite{LegumesHelpSubsequentCropYields,LegumeMetanalysisCernay,LegumePrecropMetanalysis}, and improved soil health and microbial biomass \cite{CropRotationHelpsSoil,RotationMicrobialBiomassMetanalysis}. Although these mechanisms are well established, less is known about how the benefits of crop rotation depend on the specific cropping sequence and on the weather conditions. This knowledge gap makes it difficult to assess how sub-optimal current farming practices are (in many regions farmers often grow the same crop each year or use a simple 2-crop rotation) and whether shifts in crop rotation could help farmers adapt to ongoing climate changes. Answering these questions is crucial but will require moving beyond traditional research approaches that leverage data from multiple randomized field experiments.

Randomized field experiments are a highly reliable approach for assessing the causal effects of crop rotation. However, due to the limited number of experiments with publicly available yield data and the wide variety of management practices, study designs and crop rotations used in these experiments, it is difficult to study how the benefit of any particular type of crop rotation varies with climatic conditions using experimental data alone. Studies that make claims about how rotation benefit varies with weather either use a limited number of experiments \cite{EuropeERL_WeatherInteractions7LTE} or aggregate the analysis across multiple precrops or outcome crops \cite{EuropeERL_WeatherInteractions7LTE,ChinaExpirementalMetanalysis}. Moreover, metanalyses that study the effects of diversifying crop rotation sequences (by adding at least one additional crop type to the sequence) typically have a different set of simple control rotations and diverse treatment rotations for each experiment in the study \cite{BowlesLTEpaper,SmithEtAlFromLabMeeting,LegumePrecropMetanalysis}. These studies use metrics of rotational diversity and fit a model to estimate the benefit of diversifying crop rotations without considering the particular crop type sequence. To our knowledge, how the benefit of diversification varies with precrop and outcome crop choices has not been studied and is a limitation noted in \cite{SmithEtAlFromLabMeeting}.

Recent advances in crop type mapping, satellite imagery, and causal machine learning have enabled researchers to assess the benefit of crop rotations using much larger sample sizes which allow examination of how rotation benefits vary with weather, rotational diversity, and the specific precrop and outcome crop. A number of recent studies used crop type maps and satellite imagery to assess the benefit of crop rotations \cite{CREO_paper,ChinaRiceToBeanRotation_SatelliteBased,ChinaRiceToCottonRotation_SatelliteBased,UkraineRotation_satelliteBased,FinlandSatellite_based,FinlandSatellite_based2024,ObservationalApproachAustria,BelgiumRotation_NPP_Giannarkis,JunxiongPaper,LawesEtAl_AustraliaMLToStudyRotBenefitsNotCausal}, and the accuracy of some of these observational approaches have been validated against data from randomized field experiments \cite{CREO_paper} or expert recommendations \cite{ObservationalApproachAustria}. However, all of these studies have limitations in terms of their scope, their methodology, or both (see Table \ref{table:ObsSatLitReview} for details). In summary, none of these studies assess the impact of switching from a simple rotation to a diverse rotation, and only a few of them assess how the benefit of rotation varies with weather \cite{CREO_paper,BelgiumRotation_NPP_Giannarkis,FinlandSatellite_based2024,JunxiongPaper}. Moreover, most of these studies do not use a formal causal inference framework, and instead often rely on simple comparisons between rotated and non-rotated samples. Only \cite{CREO_paper,BelgiumRotation_NPP_Giannarkis,JunxiongPaper} use a formal causal inference framework by leveraging modern causal machine learning tools. Of these studies, only \cite{JunxiongPaper} quantifies the uncertainty in their estimates of how the benefit of rotation varies with weather. As with the current study, most of these studies focus on precrop effects (defined as the effect of the previous year's crop type on the yield of the current crop) rather than total or long-term yield effects of rotation on all crops in the sequence. The former is a major (but not the only) component of rotation effects, while the latter are difficult to detect given the limited number of years in which large-scale, high-resolution crop type maps are available. 

\begin{figure}[t]
    \centering \footnotesize
    \includegraphics[width=0.95 \hsize]{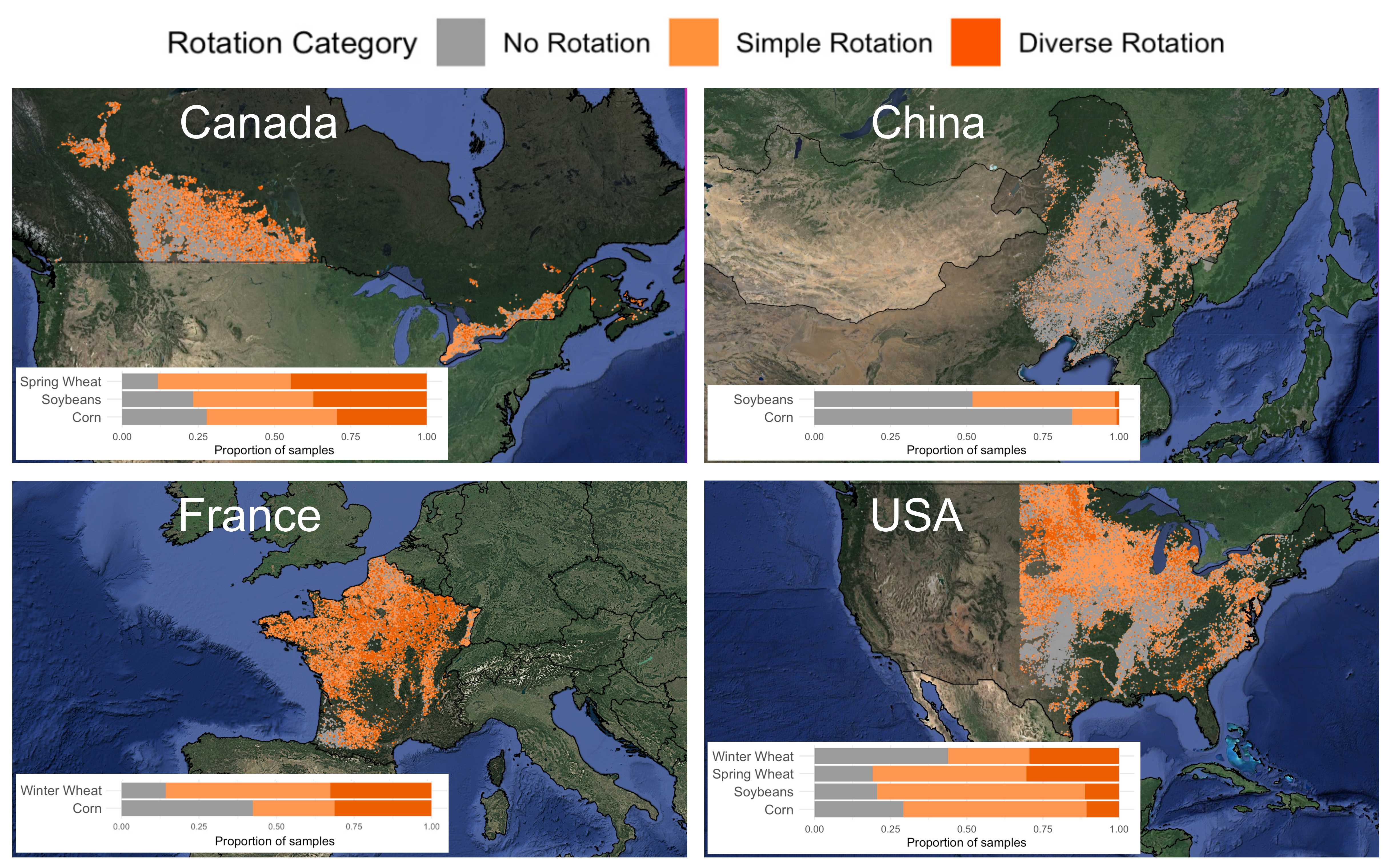}
    \caption{Rotation frequencies in study regions for a typical year. The maps for each country only show points in which one of the outcome crop types of interest was grown in 2019, and uses the crop type history from 2017-2019 to determine the rotation category. Here, the rotation categories are no rotation (i.e., with a 2-year sequence of the form B$\to$B), simple rotation (i.e., any 3-year sequence of the form B$\to$A$\to$B or A$\to$A$\to$B where crop A is different than crop B), and diverse rotation (i.e., any 3-year sequence of three distinct crops). For each country and outcome crop of interest, the barcharts give the proportion of all samples (across all years, not just 2019) in our study that had each type of rotation. Note that the proportions displayed in the barcharts are not precise estimates of the prevalence of rotation, because they are calculated from crop type maps with classification errors, and the samples were taken to over-represent rotations involving crops that we considered studying as outcome crops (Section \ref{sec:DatasetAndSample} and Appendix \ref{sec:samplingScheme}). }
    \label{fig:RotationFrequencyMap}
\end{figure}

\begin{figure}[hbt!]
    \centering \footnotesize
    \includegraphics[width=0.95 \hsize]{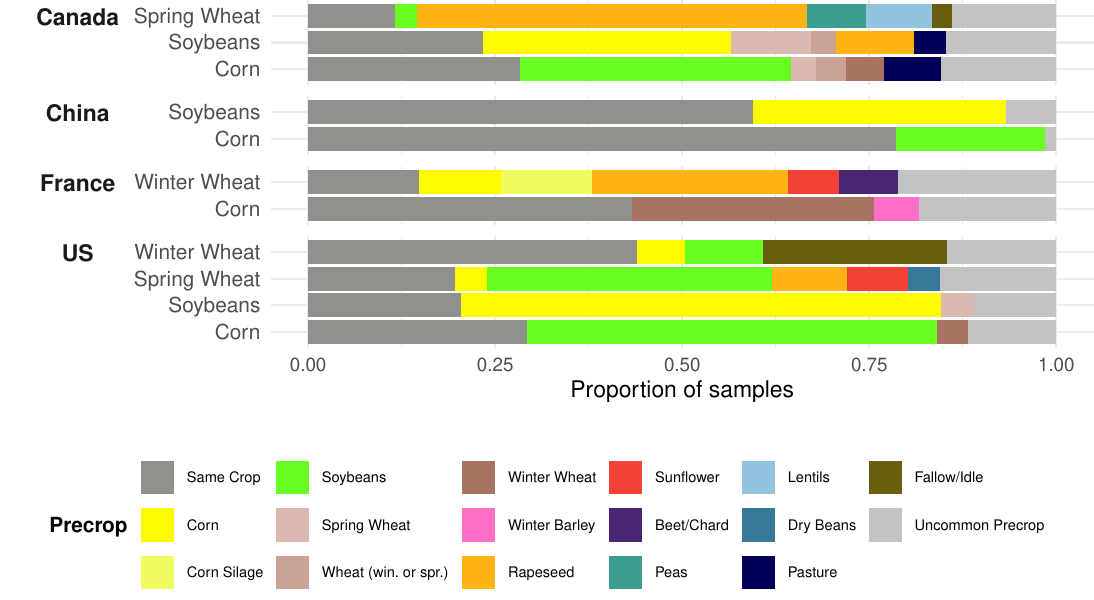}
    \caption{Distribution of precrops in our sample. Each sample corresponds to a unique pixel and year pair and was drawn according to the sampling scheme described in Section \ref{sec:DatasetAndSample} and Appendix \ref{sec:samplingScheme} (pixels were of size $\leq 30$m $\times$ 30m). For each country and outcome crop of interest, we plot the distribution of the precrop type among our samples from that country for which that outcome crop was growing in the corresponding year. The dark grey represents the proportion of samples where the precrop and the outcome crop are the same crop type. The light grey represents the proportion of samples that were excluded from our study due to an uncommon precrop (Section  \ref{sec:RotationsStudied}). We remark that the proportions displayed here are not precise estimates of the prevalence of rotation. In particular, the samples were taken in such a way that overrepresents rotations involving crops that we considered studying as outcome crops (Section \ref{sec:DatasetAndSample} and Appendix \ref{sec:samplingScheme}). In addition, in Canada, China and the US, the presented proportions are inferred using satellite-based crop type maps with classification errors (e.g., as noted in \cite{BarleyWheatDistinguishingMethod} barley and wheat are difficult to distinguish with satellite imagery which can render barley underrepresented in the figure).}
    \label{fig:PrecropDistribution}
\end{figure}

In this paper, we leverage crop type maps, satellite imagery, and a recently developed causal machine learning tool \cite{WagerAthey18_CausalForest} to study the precrop effects and benefits of diversification in four different countries where crop rotation is common but not universally adopted (Figure \ref{fig:RotationFrequencyMap}). For each outcome crop and commonly used precrop, we estimate the precrop effect and the impact of switching from a simple cropping sequence to a more diverse sequence. We also study how the precrop effects vary with weather and provide uncertainty quantifications for these estimates. Within each country, our study focuses on major outcome crops (among corn, soybean, winter wheat and spring wheat) and the most prevalent precrops used for these outcome crops (Figure \ref{fig:PrecropDistribution}). In addition, our methodology can readily be deployed in any country with annual crop type maps or to study the benefits of rotations involving other precrop and outcome crop combinations not investigated here.

\section{Overview of methods}

This section gives an overview of the data and methods. For a more detailed description of the data and methods used and for explanations of some of the concepts and choices in the analysis, see the supplementary Materials and Methods section.

\subsection{Dataset and rotations studied}

We used annual crop type maps from northeastern China \cite{YouChinaCropMap} (2017--2019), Canada \cite{CanadaACI} (2011--2020), France \cite{FrenchParcel} (2015--2021), and the United States \cite{CDL_Boryan} (2008--2021). Using these crop maps and Google Earth Engine (GEE), we extracted a random sample of geographical points among croplands of interest. At each sampled location, monthly precipitation, monthly maximum and minimum temperature averages, and monthly vapor pressure deficit (VPD) were extracted from the TerraClimate gridded weather dataset \cite{TerraClim}. For each sampled location and year, we extracted a time series of various spectral and quality assessment bands from Sentinel-2 (in China and France) and Landsat (in Canada and the US). For more details on the data sources, how they were harmonized across multiple spatial resolutions, and the sampling scheme, see Section \ref{sec:DatasetAndSample}.

For each sampled location and year, the time series of multispectral satellite data was converted to a normalized measure of peak greenness, which serves as a proxy for crop yield (see Section \ref{sec:SPH_GCVI} for details). In summary, we first computed a time series of the Green Chlorophyll Vegetation 
Index (GCVI) \cite{GCVIPaper}. We then estimated the peak GCVI in each location and year by removing observations that were corrupted by cloud cover (according to the satellites' quality assessment bands), fitting a harmonic regression, and calculating the peak of the best fit curve. GCVI was originally designed to capture chlorophyll content in crops \cite{GCVIPaper} and peak GCVI has been shown to be a useful predictor of crop yields \cite{originalSCYM, SCYM_CornJillVersion,PeakVIMalawi,JinPeakGCVIKenya}. 
For increased interpretability, estimated peak GCVI values were normalized to have mean $1$ for each crop type in each country.

For each location and year in our sample, we defined the precrop to be the crop grown in the previous year and the outcome crop to be the crop grown in the current year. Samples where the precrop and outcome crop were the same were classified as having no rotation and were used as control units. Samples where the outcome crop B and precrop A were different were considered treated units and were categorized as having an A$\to$B cropping sequence. In each country, we restricted our attention to at most four outcome crops of interest and for each outcome crop and country we studied the impact of the most prevalent precrops in the sample (Figure \ref{fig:PrecropDistribution}). See Section \ref{sec:RotationsStudied} for more details about the precrop and outcome crop selection criteria used.

\subsection{Estimating precrop and diversification effects}

We fit causal forests \cite{WagerAthey18_CausalForest,AtheyTibWager_GRF} to estimate the precrop effects. A causal forest is a recently developed method for conducting causal inference in observational studies, and it can be used to estimate the average treatment effect as well as the treatment effect as a function of the covariates. Causal forests have been used to study the effects of various agronomic practices such as tillage \cite{Tillage_CausalForest}, cover cropping \cite{CoverCrop_CausalForest}, and crop rotation \cite{CREO_paper,JunxiongPaper} on crop yields as well as the impact of crop rotation on net primary productivity \cite{BelgiumRotation_NPP_Giannarkis}. In the context of assessing the effect of the Soybean$\to$Corn crop rotation on corn yield, rotation benefits estimated by fitting a causal forest to a satellite-derived dataset were found to have a statistically significant positive correlation with estimates from actual randomized field experiments \cite{CREO_paper}.

For each of four countries, and for each common combination of precrop A and outcome crop B, we conducted the following procedure to estimate the impact of the A$\to$B sequence compared to the B$\to$B sequence on a proxy for the yield of B in the second year. First, we subsetted our data to only include samples with A$\to$B or B$\to$B sequences and designated a binary treatment variable $Z$ to be $1$ for samples with A$\to$B sequences and $0$ for samples with B$\to$B sequences. Next, we set the outcome variable $V$ to be the normalized peak GCVI variable described in Section \ref{sec:SPH_GCVI}, which serves as a proxy for crop yield. We then let $X$ denote a vector of year, latitude, longitude, 7 weather covariates, and (in the US only) irrigation status (from the Landsat-based Irrigation Dataset \cite{IrrigationDat_ESSD}). The weather covariates captured early precipitation, growing season precipitation, average maximum and minimum daily temperatures in the growing season, and VPD for the three peak months of the growing season. We then fit a causal forest, using the \texttt{grf} package in R \cite{GrfPackage}, to estimate the function $$\omega(x)= \e [V| X=x, Z=1] -\e [V| X=x, Z= 0].$$ In words, $\omega(\cdot)$ is the difference in the mean normalized peak GCVI in the treated group minus that in the control group written as a function of the covariates $x$. Under the assumption that there are no unmeasured confounders, $\omega(x)$ is equal to the conditional average treatment effect (which is the mean causal effect of the treatment $Z$ on the outcome $V$ as a function of $x$). Finally the average precrop effects were estimated by taking an overlap-weighted average of the estimated $\omega(X)$ values, assigning little weight to locations in space and time where only $Z=1$ or only $Z=0$ was observed and $\omega(X)$ could not have been properly estimated.

We refer readers to Section \ref{sec:CausalForestAnalysis} for a more explicit description of the causal forest approach, the parameters and software used, the estimand, how standard errors were calculated, and the specific weather covariates used. That section describes various justifications and robustness checks regarding the choice of outcome variable, the choice of an overlap-weighted estimand, and our choice to not account for spatial correlations when computing standard errors (see also Appendix \ref{sec:AccountForSpatialCorrelations}). We also compared some of the causal forest results to those from a simpler approach of calculating within each 1°$\times$1° grid cell the difference between the mean of $V$ in the treated (rotated) versus control (non-rotated) group, finding similar qualitative results (see Appendix \ref{sec:DiffInMeanClustered}). For model fit accuracy metrics, see Appendix \ref{sec:CausalForestGoodnessOfFits} and Table \ref{table:ModelGoodnessOfFitMetrics}.

To estimate the impact of diverse crop rotations compared to simple rotations, we used the same causal forest methodology with a different choice of treatment variable $Z$ that was based on the three year rotation history. Within each country (except China), and for each combination of precrop A and outcome crop B, samples where a crop other than A or B was grown before A were considered to have a diversified rotation and were deemed as treated units $(Z=1)$ whereas samples where B$\to$A$\to$B was the three year rotation history were considered as the simple rotation control units $(Z=0)$. See Section \ref{sec:RotDiversityMethodDescription} for more details.

\subsection{Estimating heterogeneity of precrop effects with weather}

To assess how the benefit of the A$\to$B sequence versus the control B$\to$B varied with weather, we ran the following linear regression of $\omega(X)$ on the year $t$, the growing season precipitation $P$, and the growing season temperature $T$ 
$$ \omega(X) = \beta_0 +\beta_{\text{Year}} t  + \beta_{\text{Precip}} P+\beta_{\text{Temp}} T + \varepsilon.$$ The above regression was implemented using the \texttt{best\_linear\_projection} function in the \texttt{grf} package \cite{GrfPackage}, which provided standard errors and $p$-values that appropriately accounted for the fact that the function $\omega(\cdot)$ was unknown and had to be estimated from the data. The regression was fit on the subsample of points where the two year sequence was either A$\to$B or B$\to$B and excluded regions where the estimated propensity score was between $0.05$ and $0.95$ (allowing us to ignore regions where only A$\to$B or only B$\to$B was observed, but not both). For more details about the subsample used, the growing season weather variables, how confidence intervals were constructed, and for a justification of the linear modeling choices see Section \ref{sec:HeterogeneityAnalysisMethod}. As a supplementary sensitivity analysis, variants of the above regression model that included geographical fixed effects were also considered (Appendix \ref{sec:HeterogeneityWithWeatherGeographyControls}).

To rescale the estimated regression coefficients to be on a more interpretable scale, on the same subsample, we computed the 25th percentile and 75th percentile for growing season precipitation and for growing season temperatures denoted by $P_{0.25}$, $P_{0.75}$, $T_{0.25}$, and $T_{0.75}$, respectively. Letting $\hat{\beta}_{\text{Precip}}$ and $\hat{\beta}_{\text{Temp}}$ denote the estimated regression coefficients from the above regression, our rescaled heterogeneity coefficients for temperature and weather were calculated with $$ \hat{\beta}^{\text{(resc.)}}_{\text{Precip}} = \hat{\beta}_{\text{Precip}} \times (P_{0.75}-P_{0.25})\quad \text{and} \quad \hat{\beta}^{\text{(resc.)}}_{\text{Temp}} = \hat{\beta}_{\text{Temp}} \times (T_{0.75}-T_{0.25}).$$ The rescaled heterogeneity coefficients can be roughly interpreted as the amount by which the estimated benefit of rotation on our normalized yield proxy are expected to change as the growing season precipitation (or temperature) increases from the 25th percentile to the 75th percentile in the regions where both the rotation A$\to$B and the no rotation control B$\to$B are commonly observed. The above process was repeated for each precrop A, outcome crop B, and country listed in Table \ref{table:ResultsTable}.

\subsection{Calibration with crop yield data}

To improve the interpretability of our results, we converted all estimated precrop effects, diversification effects, and heterogeneity coefficients from a unitless GCVI scale to the crop yield scale by using a post-hoc, linear calibration with subnational-level crop yield data. In particular, we fit linear regressions of annual county-level crop yield data from \cite{NASSQuickStats} on county-level averages of normalized peak GCVI. Best fit linear regression coefficients were then used to convert all results to be in units of \% of the average yield for the outcome crop and country of interest. For more details about the calibration and sensitivity analyses with alternative calibration schemes see Section \ref{sec:ConvertingToYieldScale} and Appendix \ref{sec:TransferabilityOfUSYieldCalib}.

\section{Results}\label{sec:Results}

\subsection{Estimates of impacts of 2-year crop sequences}

\begin{figure}[hbt!]
    \centering
    \includegraphics[width=0.95 \hsize]{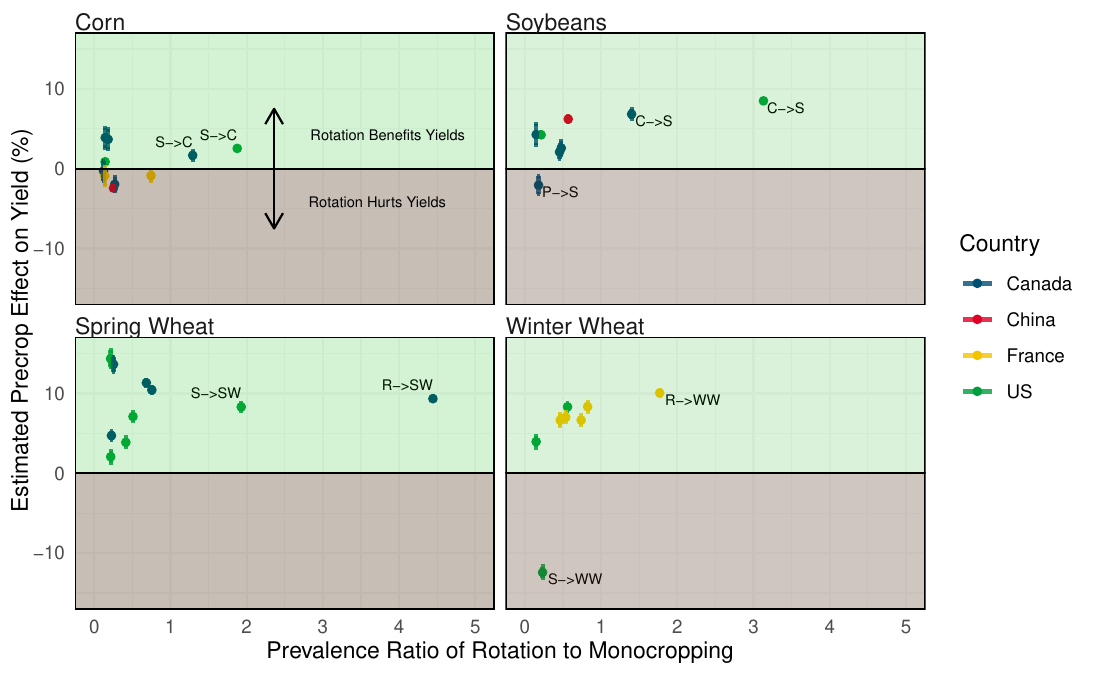}
    \caption{Estimated precrop effects. Each point represents a different country, precrop, and outcome crop combination, with the panel titles giving the outcome crop and the colors giving the country. For each point, the y-axis value gives the estimated average effect of a 2-year cropping sequence on crop yield (expressed as a percentage of the mean crop yield of the corresponding outcome crop and study region) and the error bars give 95\% confidence intervals. The estimates and standard errors were computed using the approach described in Section \ref{sec:CausalForestAnalysis} and were converted to the yield scale using the approach described in Section \ref{sec:ConvertingToYieldScale}. The confidence intervals are based on large samples and are much narrower than one might expect because they cannot reflect uncertainty in the potential bias due to unmeasured confounders. The x-axis gives the prevalence ratio, which we define as the number of samples that were used as treated units and had a given 2-year cropping sequence divided by the number of samples that were used as control units where the precrop and outcome crop matched. The specific type of rotation is labeled in cases where the prevalence ratio was greater than 1 and in cases where the rotation had a uniquely negative estimated effect. For point labels, S=Soybean, C=Corn, P=Pasture and Forages, R=Rapeseed, SW=Spring Wheat, WW=Winter Wheat.}
    \label{fig:TwoYearRotBenefit}
\end{figure}

The results (Table \ref{table:ResultsTable} (Column 6); Figure \ref{fig:TwoYearRotBenefit}) indicate that rotations generally benefit crop yields, but there are considerable differences in these benefits for different precrop and outcome crop combinations. Of the four outcome crops considered, the rotation benefit estimates were generally largest for spring wheat and smallest for corn. For winter wheat outcome crops, soybean precrops had a negative estimated effect whereas rapeseed or fallow precrops had a positive estimated effect. For spring wheat outcome crops, legume precrops and rapeseed precrops were generally the most effective, whereas for soybean outcome crops, cereals were the most effective types of precrops (Figure \ref{fig:TwoYearRotBenefitPrecropCategory}).

The results also suggest that beneficial precrops are being exploited by farmers. In Figure \ref{fig:TwoYearRotBenefit}, we see that the most common precrops have either the highest or nearly the highest benefits. There are some exceptions in which less common precrop choices are estimated to be more effective than the most common precrop choices when spring wheat or corn is the outcome crop. In addition, only relatively common precrops are included in the analysis (Section \ref{sec:RotationsStudied}), so our analysis does not rule out the possibility that some very rare precrops are more effective than commonly used precrops.

\subsection{Variability of rotation benefits with weather}

\begin{figure}[hbt!]
    \centering
    \includegraphics[width=0.95 \hsize]{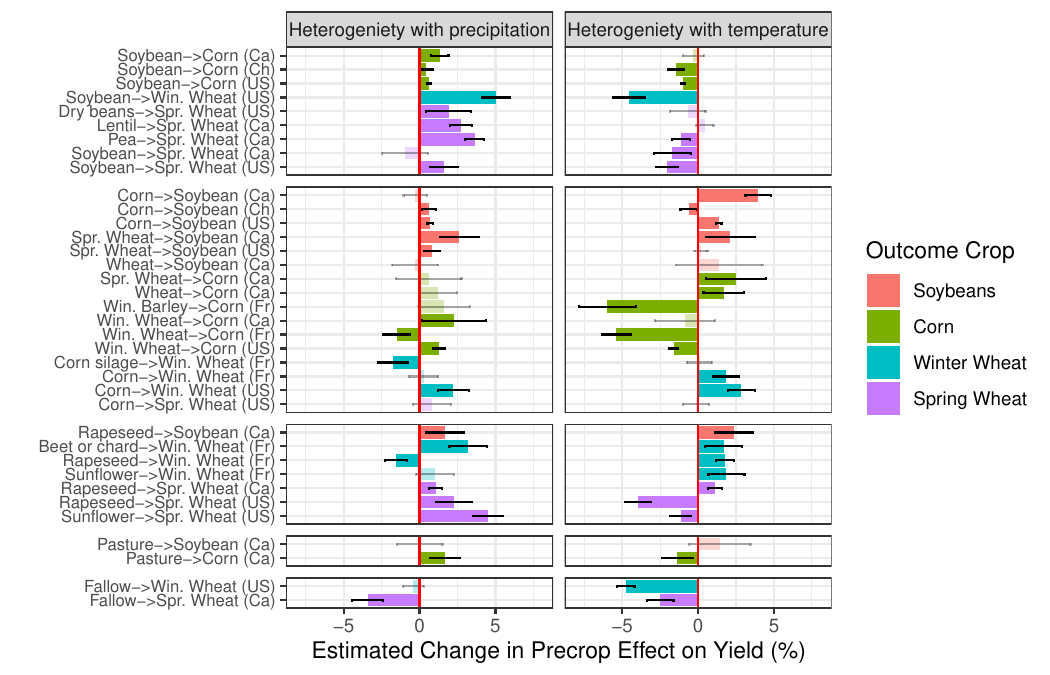}
    \caption{Estimates of the heterogeneity of rotation benefits with weather. Each row gives the two year crop sequence, the country from which the study sample was taken (Ca=Canada, Ch=China, Fr=France, US=United States), and the rescaled regression coefficients for growing season precipitation (left column) and temperature (right column) which estimate how much the precrop effect changes when the precipitation or temperature increases from the 25th percentile to the 75th percentile of observed values. The x-axis values are in units of percent of the mean crop yield of the corresponding outcome crop in the corresponding study region and can be converted to units of tons per hectare using the yield estimates in Table \ref{Table:YieldToTonsPerHectare}. The error bars give 95\% confidence intervals (based on heteroskedasticity-robust (HC3) estimation), and coefficient estimates that are not statistically significant (level $\alpha=0.05$ based on two-sided testing) are given a lighter, translucent color. The rows are grouped into blocks based on 5 precrop categories of interest: legumes, cereals, broadleaf crops, pasture/forages, and fallow.}
    \label{fig:Heterogeneity}
\end{figure}

The results (Table \ref{table:ResultsTable} (Columns 8--9); Figure \ref{fig:Heterogeneity}) indicate that at higher growing season precipitation there is generally a greater benefit due to rotation. In particular, with a few exceptions in France and for fallow precrops, cases where the rotation benefit is greater at lower precipitation levels did not have statistically significant heterogeneity coefficients. In addition, based on a sensitivity analysis where we use the log of estimated crop yield as the outcome variable (Appendix \ref{sec:HeterogeneityWithWeatherLogScale}; Figure \ref{fig:logEstimatedYieldOutcome}, bottom left panel), we find that this result is not merely driven by the fact that crop productivity is higher in rainier conditions. Instead, we see increases in both the relative effectiveness and absolute effectiveness of crop rotation in rainier conditions. For the rotations with a legume or broadleaf precrop, the heterogeneity of the rotation benefit with precipitation was larger and more consistently statistically significant than it was for the rotations where the precrop was a cereal. These findings are generally robust to sensitivity analyses where we include fixed effects to control for geographical subregions when estimating the heterogeneity of the rotation benefits with weather (Appendix \ref{sec:HeterogeneityWithWeatherGeographyControls}; Figure \ref{fig:SensitivityCheckHeterogeneityWithPrecip}), although for cereal precrops our sensitivity analyses did not clearly suggest that the rotation benefit was higher at higher precipitation.

The direction of how the rotation benefit varies with temperature was not consistently positive or negative; however, some notable patterns exist. When soybean is the outcome crop (red bars in Figure \ref{fig:Heterogeneity}), the benefit of rotation is generally greater at higher growing season temperatures. In addition, when there is a legume precrop, the benefit of rotation is generally smaller at higher growing season temperatures (top right panel in Figure \ref{fig:Heterogeneity}). The only exceptions are cases where the heterogeneity coefficient is not statistically significant. Moreover, sensitivity analyses (Appendix \ref{sec:HeterogeneityWithWeatherGeographyControls}, Figure \ref{fig:SensitivityCheckHeterogeneityWithTemp}) demonstrate that these findings are robust to the addition of various geographical fixed effects in our model used to assess the heterogeneity of the rotation benefits.

\subsection{Estimated impacts of diversified crop rotations}\label{sec:DiversificationResults}

\begin{figure}[hbt!]
    \centering
    \includegraphics[width=0.95 \hsize]{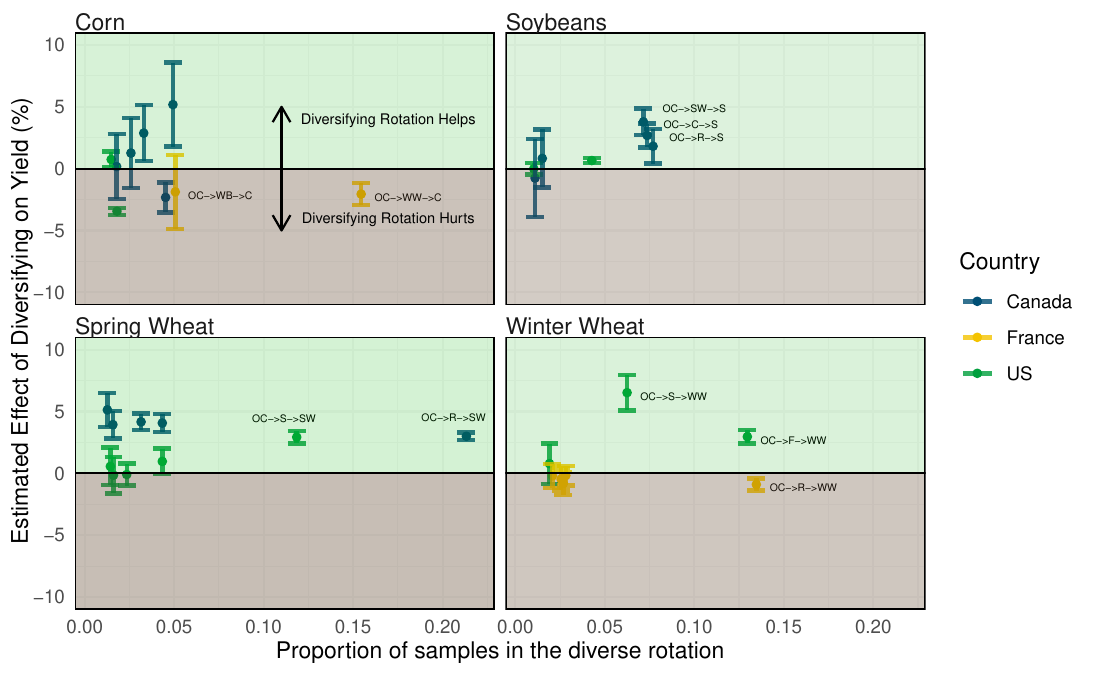}
    \caption{Estimated benefits of diversifying crop rotations. Each point represents a different country, precrop, and outcome crop combination, with the panel titles giving the outcome crop and the colors giving the country. For each point, the y-axis value gives the estimated effect of diversifying the rotation (by growing a crop other than the precrop or outcome crop in the year prior to the precrop) on crop yield, and the error bars give 95\% confidence intervals. The estimates and standard errors were computed using the approach described in Sections \ref{sec:CausalForestAnalysis} and \ref{sec:RotDiversityMethodDescription} and were rescaled to be on the crop yield scale (expressed as a percentage of the mean crop yield of the corresponding outcome crop and study region) using the approach described in Section \ref{sec:ConvertingToYieldScale}. The x-axis gives the proportion of samples that had a diversified rotation and a particular precrop, among all samples with the given outcome crop in the given country. The specific type of diversified rotation is labeled in cases where this proportion exceeded $0.05$. For point labels, S=Soybean, C=Corn, SW=Spring Wheat, WW=Winter Wheat, WB=Winter Barley, F=Fallow, R=Rapeseed, and OC=Other Crop which groups together \textit{all} other crops besides the specified precrop and outcome crop.}
    \label{fig:DiversificationBenefit}
\end{figure}

 The results suggest that diversification beyond a simple rotation confers large additional benefits in some cases, but more often the benefits are small, not statistically significant, or even negative (Table \ref{table:ResultsTable} (Column 7); Figure \ref{fig:DiversificationBenefit}). Spring wheat appears to benefit most consistently from more diverse rotations, whereas when corn is the outcome crop, the estimated impact of diversification is sometimes negative, particularly for common precrop choices (soybean in Canada and the US, and winter wheat in France). Supplementary results also showed that the impact of diversification beyond a simple rotation tended to be larger in rainier conditions (Figure \ref{fig:HeterogeneityDiversificationFig}), although many estimates of the heterogeneity of the diversification effects were not statistically significant due to limited sample sizes in the diversification analyses.

\section{Discussion} \label{sec:Discussion}

\subsection{Discussion of rotation benefit estimate findings}\label{sec:DiscussionOfPrecropEffects}

Our results are consistent with the well established principle that switching from no rotation to crop rotation generally benefits the subsequent crop's yields, providing some validation to our approach. For corn yields, our results suggest that winter wheat and soybean were beneficial precrops in the United States and Canada, although this conclusion could not be drawn in China and France or for spring wheat precrops. For soybean yields, corn was a particularly beneficial precrop (in the US, Canada, and China), while wheat (in the US and Canada) and rapeseed (in Canada) precrops were found to have smaller, yet still positive benefits on yields. For spring wheat yields in the US and Canada, rapeseed and legumes such as soybeans, peas, lentils and drybeans were found to be particularly beneficial precrops, whereas in the US sunflower and corn  precrops had more modest benefits. For winter wheat yields in France, rapeseed was again found to be a particularly effective precrop, and sunflower, and chard/beet, corn silage, and corn, were also quite effective as precrops, with the latter finding for corn precrops also holding in the US. Overall, the precrop effects that we estimated were generally positive and statistically significant, with two exceptions when corn was the outcome crop and soybean or wheat was the precrop, two exceptions when pasture was the precrop, and a notable exception of the Soybean$\to$Winter Wheat sequence in the US. 

The negative effects for corn can be partly explained by reduced fertilizer use in rotations. For example, in the US, farmers growing Soybean$\to$Corn are advised to use less fertilizer than those growing Corn$\to$Corn \cite{sawyerNitrogenGuidelines}. Prior work \cite{CREO_paper} found that estimates of the impact of the Soybean$\to$Corn rotation using a similar observational approach that omitted fertilizer use underestimated the rotation benefit estimates based on randomized field experiments (in the experiments, fertilizer use was typically the same in rotated and non-rotated subplots). Omission of fertilizer also explains why the estimated precrop effects of soybean on corn were much smaller than the estimated precrop effects of corn on soybean, contrary to findings from randomized field experiments. Negative effects in Canada for cases where pasture was a precrop could be explained by the fact that Pasture$\to$Soybean and Pasture$\to$Corn samples often reflected a conversion of pastoral land to cropland rather than a true rotation, and thus may correspond to less productive land. Indeed, among our samples from Canada that had either a Pasture$\to$Corn or a Pasture$\to$Soybean sequence, roughly 64\% observed a switch from a pastoral classification to a nonpastoral classification only once between 2011-2020, and 29\% observed such a switch twice. Further, when using a different satellite-based proxy for yield in our sensitivity analysis in Appendix \ref{sec:UsingNIRvInsteadOfGCVI}, the negative estimated effects of the Pasture$\to$Corn and Pasture$\to$Soybean rotations were no longer statistically significant.

We estimated a large negative effect of the Soybean$\to$Winter Wheat rotation on subsequent winter wheat yields in the United States. One possible explanation for this is that winter wheat has higher yields at earlier planting dates \cite{WinterWheatLatePlantingDateHurts}, but the Soybean$\to$Winter Wheat rotation can prevent early planting of winter wheat. 
According to a USDA report \cite{NASS_TypicalPlantingAndHarvestDateUS}, the typical winter wheat planting window started and ended earlier than the typical soybean harvest window in most of the states that were highly represented in our study's Soybean$\to$Winter Wheat and Winter Wheat$\to$Winter Wheat samples (Figure \ref{fig:HarvestPlantingDatesSoyWW}).
In addition, some experiments have suggested that soybean precrops lead to reduced soil water content in the subsequent spring because their residues are ineffective at trapping snow compared to the residues of cereals such as corn and sorghum \cite{NorwoodResiduesSnow}. Meanwhile, residues from winter wheat (typically harvested in the summer) are effective at trapping snow (in the subsequent winter), particularly when the residues are left standing \cite{WinterWheatResiduesTrapSnowWell}. Therefore, in semi-arid regions with snow and in fields that use favorable residue management practices, a winter wheat precrop leads to more soil water availability in the subsequent spring than a soybean precrop does. Low yields in the Soybean$\to$Winter Wheat rotation have also been reported in the experimental literature, although the results are mixed. A field experiment in the north China plain found that the Soybean$\to$Winter Wheat rotation led to lower Winter Wheat yields than in a Corn$\to$Winter Wheat rotation, particularly in dry years \cite{ChinaStoWWversusCtoWW} which the authors attributed to soybean reducing the soil available water for subsequent wheat growth. 
Contrary to our findings, an experiment in Woodslee, Ontario found higher winter wheat yields in a Soybean-Winter Wheat rotation than in monocropped winter wheat \cite{WoodsleeOntarioContWWversusOthers}; however, the study was conducted in a location with high mean annual precipitation, and to our knowledge, the treated and control plots had identical planting dates. The fact that some Soybean$\to$Winter Wheat versus Winter Wheat$\to$Winter Wheat experiments use the same planting dates for the treatment and control subplots and are conducted in highly favorable weather conditions for the rotation serves as an example of how our observational approach can be used to uncover external validity issues in the experimental crop rotation literature.

We also find that some rotations are particularly beneficial for the outcome crop. For example, legumes appear particularly effective at improving subsequent spring wheat yields in the US and Canada (Figure \ref{fig:TwoYearRotBenefitPrecropCategory}, Table \ref{table:ResultsTable}). This is consistent with field experiments in the region, which have shown that a variety of legumes, including soybeans and dry beans, increase nitrogen availability and yield of the subsequent spring wheat crop \cite{LegumeEffectsOnNitrogenAndSubsequentSpringWheat,SpringWheat10PrecropsInclDryBeans}. A global metanalysis of randomized field experiments found legumes to be the most effective precrops for wheat \cite{AngusPrecropForWheatMetanalysis}, although the study did not distinguish between spring and winter wheat.

\subsection{Discussion of heterogeneity with weather findings}

Our results suggest that in rainier conditions, common crop rotations involving corn, soybean, and wheat generally have a greater impact on yields than in drier conditions. One likely reason is that higher precipitation increases pest and weed pressures, which crop rotations can help to address. In addition, the residues of certain precrops, such as sunflower and soybean, are not effective at trapping precipitation from snow for subsequent soil absorption \cite{NorwoodResiduesSnow}.  Therefore, assuming no residue removal, rotations with such precrops would be less effective when there is too little precipitation during the growing season to compensate for the reduction in soil moisture. Meanwhile, wheat stalks are particularly effective at trapping snow \cite{WinterWheatResiduesTrapSnowWell}, so non-rotated wheat controls would perform relatively better under low levels of rain. 

Increased benefits of rotation for soybean yields at higher temperatures likely relate to pest pressures \cite{SoyExpsRotationBenefitLargelyDrivenByCystNematodeReduction,RotationHelpsWithCystNematode,HighTemperatureImpliesHigherPestPressure}, as these harmful pressures (that are mitigated by crop rotation) generally increase with warming. In contrast to soybean, we observe mixed effects of warming on rotation benefits for wheat and corn, and we find that rotations involving legume precrops are less effective at higher temperatures. The latter finding likely relates to the effect of warming on nitrogen limitations. At higher temperatures, residues of the precrop mineralizes at a faster rate, thereby reducing nitrogen limitations \cite{NitrogenMineralizationInSoilArrheniusEqBased,CornResidueDecomp}. The benefit from a legume precrop is therefore reduced with warming. Another potential explanation is that high temperatures can render soybean precrops as less effective in fixing nitrogen \cite{RootTemperatureNitrogenFixation}, and samples with high growing season temperatures are more likely to have had high temperatures while the soybean precrop was growing.

While these findings of how rotation effects vary with weather are consistent with the existing literature  (Appendix \ref{sec:HeterogeneityWithWeatherUSversusLiterature}), we note that our findings go far beyond the scope of most prior work. In particular, existing claims in the experimental literature looked at weather interactions after aggregating over many different precrop and outcome crop combinations, due to limitations in the number and types of experiments \cite{BowlesLTEpaper,LegumePrecropMetanalysis,ChinaExpirementalMetanalysis,EuropeERL_WeatherInteractions7LTE}. Such claims can therefore reflect how precrops and outcome crops in experiments vary across regions with differing climates rather than how the benefit of rotation varies with weather. In addition, claims in the observational literature about how rotation benefits varied with weather all focused on a single country and only one of them provided uncertainty quantifications for these claims \cite{CREO_paper,FinlandSatellite_based2024,BelgiumRotation_NPP_Giannarkis,JunxiongPaper} (Table \ref{table:ObsSatLitReview}).

\subsection{Discussion of diversification benefit findings}

Our results suggest that diversification beyond a simple rotation has diminishing returns (see Table \ref{table:ResultsTable} or Figures \ref{fig:TwoYearRotBenefit} and \ref{fig:DiversificationBenefit}). In particular, we find that the average benefit of a 2-crop rotation across our 36 rotation-country pairs is about 3.7 times larger than the average additional benefit of switching from 2-crop rotations to a more diverse sequence. This finding agrees with a recent metanalysis of 45 experiments in China \cite{ChinaExpirementalMetanalysis} that found crop rotation increased yields by 20$\%$ when compared to monoculture on average but extended (diversified) rotations provided only an additional 4$\%$ yield boost. Similarly, a recent global metanalysis of 462 crop rotation experiments involving legumes found that the inclusion of an additional legume precrop in a rotation had lower benefits when the control rotation was already diverse \cite{LegumePrecropMetanalysis}. A recent analysis of field experiments in the European Union and Africa \cite{MacLarenConservationAg} found no additional benefits from increased diversification beyond a switch from monocropping to simple rotations. A recent analysis of data from 32 long-term experiments from North America and Europe \cite{SmithEtAlFromLabMeeting} found that crop yields for maize and small grain cereals increased at higher values of an index used to measure rotational diversity. However, when fitting quadratic regression models of crop yield on their diversity measure and other controls, the coefficient corresponding to the squared diversity measure was negative for all 3 outcome crop classes (maize, small grain winter cereal and small grain spring cereal) suggesting diminishing returns with each additional unit of diversity.

 Due to experimental data limitations, recent metanalyses studying crop rotation diversification did not explore how the diversification benefit varied with each precrop and outcome crop combination \cite{SmithEtAlFromLabMeeting,LegumePrecropMetanalysis,ChinaExpirementalMetanalysis,MacLarenConservationAg}. Our results suggest that the benefit of diversification varies considerably, and can even be negative (Table \ref{table:ResultsTable}; Figure \ref{fig:DiversificationBenefit}). For example, we estimate a negative effect of diversification in settings where soybean is the precrop and corn is the outcome crop whereas we find considerable diversification benefits when there is a legume precrop and spring wheat is the outcome crop (Table \ref{table:ResultsTable}).

\subsection{Limitations}

This study has several limitations. First, our causal inference approach relies on the assumption that there are no unmeasured confounders. Due to data availability constraints we did not control for potential confounders such as soil quality (although, including soil variables in the US barley changed our results (Appendix \ref{sec:AddingSoilCovariatesSensitivityCheck})). We also did not account for management practices such as tillage, cover cropping, residue management, fertilizer use, and pesticide use; however, these unmeasured management practices can be viewed as unmeasured mediators as opposed to unmeasured confounders, and we estimate the net rather than direct effects of crop sequence choice (see Section \ref{sec:NetVSDirect}). Moreover, farms that rotate tend to use less fertilizer \cite{sawyerNitrogenGuidelines} and pesticide \cite{FranceLongerRotationsLessPesticide}, so our analysis does not capture the cost savings reductions in chemical inputs that are associated with rotations. 

Second, we do not have exact measurements of the treatment and outcome variables which are instead estimated using satellite data. Misclassifications in the treatment variable likely lead to attenuation bias \cite{CREO_paper,lewbel2007estimation}, which as with the issue of unmeasured fertilizer and pesticide use, results in underestimation of rotation benefits. Appendix \ref{sec:VI_versus_SCYM} suggests that our choice to use a vegetation index as the outcome variable rather than crop yield estimates from \cite{originalSCYM,QDANNPaper} does not substantially influence the results and we also find robustness to the choice of vegetation index (Appendix \ref{sec:UsingNIRvInsteadOfGCVI}). 

Third, the study has some limitations in its scope. We do not study the well-documented longer-term benefits of crop rotation \cite{EuropeERL_WeatherInteractions7LTE,SmithEtAlFromLabMeeting}. In addition, our analysis only considers the impact of a cropping sequence on the final crop in a cropping sequence
rather than all crops in the sequence. Ultimately, farmers make decisions about which
rotations to use based on many factors, including the expected yield and price of each
of the crops in the rotation sequence; however, future work can couple our results and methods with crop rotation decision support systems \cite{DecisionSupportROTAT,DecisionSupportROTOR,DecisionSupportCropRota,DecisionSupportFinland,DecisionSupportFruchtfolge}.

A more comprehensive and detailed discussion of the limitations can be found in Appendix \ref{sec:LimitationDiscussion}. That appendix also summarizes sensitivity analyses and mathematical results, which assess whether or not the limitations discussed are severe.

\section{Conclusion}

Using satellite data and causal machine learning, we estimated the impacts of various cropping sequences and how they varied with weather. While randomized field experiments can reliably estimate the effects of cropping sequences on crop yield, given the limited geographical span of experimental data, drawing conclusions of how rotation benefits vary with weather remains a substantial challenge. Our results using observational approaches suggest that precrop effects are larger under rainier conditions. Moreover, the results suggest that if the temperature increases, precrop effects will decrease in cases where the precrop is a legume but increase in cases where the outcome crop is soybean. Estimates of how rotation benefits vary with weather can be coupled with climate models to forecast how the effectiveness of crop rotation could change in a changing climate (as is done for the Soybean-Corn rotation in \cite{JunxiongPaper}) and can inform agricultural adaptation to climate change.

In addition, our observational approach can uncover or fill in gaps in the experimental crop rotation literature. For example, as discussed in Section \ref{sec:DiscussionOfPrecropEffects} our results for the Soybean$\to$Winter Wheat sequence point to possible external validity issues regarding planting dates and weather conditions in the experimental literature. In addition, our approach can be used in any region with multiple years of crop type maps in order to identify cropping sequences that may be highly beneficial but have been understudied in randomized field experiments.

\section*{Acknowledgments}
This work was supported by a Stanford Interdisciplinary Graduate Fellowship and the NASA Harvest and ACRES Consortia (NASA Applied Sciences Grant No. 80NSSC23M0032 (sub-award 125062-Z6521205) and 80NSSC23M0034 (sub-award 124245-Z6512205)). The authors wish to thank Kevin Guo and Stefan Wager for helpful discussions and anonymous reviewers for helpful comments.

\subsection*{Data and code availability}

Data and code that can be used to reproduce the analysis in this paper can be downloaded from Zenodo: \url{https://zenodo.org/records/15579276}.

\bibliographystyle{unsrt}
\bibliography{RotationPaper}

\begin{thebibliography}{10}

\bibitem{CropRotationReducesWeedDensity}
David Weisberger, Virginia Nichols, and Matt Liebman.
\newblock Does diversifying crop rotations suppress weeds? {A} meta-analysis.
\newblock {\em PLOS ONE}, 14(7):1--12, 2019.

\bibitem{BennettDiscussesMechanismsOfRotationBenefit}
Amanda~J. Bennett, Gary~D. Bending, David Chandler, Sally Hilton, and Peter Mills.
\newblock Meeting the demand for crop production: The challenge of yield decline in crops grown in short rotations.
\newblock {\em Biological Reviews}, 87(1):52 – 71, 2012.

\bibitem{LegumesHelpSubsequentCropYields}
M.~B. Peoples, J.~Brockwell, D.~F. Herridge, I.~J. Rochester, B.~J.~R. Alves, S.~Urquiaga, R.~M. Boddey, F.~D. Dakora, S.~Bhattarai, S.~L. Maskey, C.~Sampet, B.~Rerkasem, D.~F. Khan, H.~Hauggaard-Nielsen, and E.~S. Jensen.
\newblock The contributions of nitrogen-fixing crop legumes to the productivity of agricultural systems.
\newblock {\em Symbiosis}, 48(1):1--17, 2009.

\bibitem{LegumeMetanalysisCernay}
Charles Cernay, David Makowski, and Elise Pelzer.
\newblock Preceding cultivation of grain legumes increases cereal yields under low nitrogen input conditions.
\newblock {\em Environmental Chemistry Letters}, 16(2):631--636, 2018.

\bibitem{LegumePrecropMetanalysis}
Jie Zhao, Ji~Chen, Damien Beillouin, Hans Lambers, Yadong Yang, Pete Smith, Zhaohai Zeng, J{\o}rgen~E. Olesen, and Huadong Zang.
\newblock Global systematic review with meta-analysis reveals yield advantage of legume-based rotations and its drivers.
\newblock {\em Nature Communications}, 13(1):4926, 2022.

\bibitem{CropRotationHelpsSoil}
L.~K. Tiemann, A.~S. Grandy, E.~E. Atkinson, E.~Marin-Spiotta, and M.~D. McDaniel.
\newblock Crop rotational diversity enhances belowground communities and functions in an agroecosystem.
\newblock {\em Ecology Letters}, 18(8):761--771, 2015.

\bibitem{RotationMicrobialBiomassMetanalysis}
M.~D. McDaniel, L.~K. Tiemann, and A.~S. Grandy.
\newblock Does agricultural crop diversity enhance soil microbial biomass and organic matter dynamics? {A} meta-analysis.
\newblock {\em Ecological Applications}, 24(3):560--570, 2014.

\bibitem{EuropeERL_WeatherInteractions7LTE}
Lorenzo Marini, Audrey St-Martin, Giulia Vico, Guido Baldoni, Antonio Berti, Andrzej Blecharczyk, Irena Malecka-Jankowiak, Francesco Morari, Zuzanna Sawinska, and Riccardo Bommarco.
\newblock Crop rotations sustain cereal yields under a changing climate.
\newblock {\em Environmental Research Letters}, 15(12):124011, 2020.

\bibitem{ChinaExpirementalMetanalysis}
Jie Zhao, Yadong Yang, Kai Zhang, Jaehak Jeong, Zhaohai Zeng, and Huadong Zang.
\newblock Does crop rotation yield more in {C}hina? {A} meta-analysis.
\newblock {\em Field Crops Research}, 245:107659, 2020.

\bibitem{BowlesLTEpaper}
Timothy~M. Bowles, Maria Mooshammer, Yvonne Socolar, Francisco Calderón, Michel~A. Cavigelli, Steve~W. Culman, William Deen, Craig~F. Drury, Axel {Garcia y Garcia}, Amélie~C.M. Gaudin, W.~Scott Harkcom, R.~Michael Lehman, Shannon~L. Osborne, G.~Philip Robertson, Jonathan Salerno, Marty~R. Schmer, Jeffrey Strock, and A.~Stuart Grandy.
\newblock Long-term evidence shows that crop-rotation diversification increases agricultural resilience to adverse growing conditions in {N}orth {A}merica.
\newblock {\em One Earth}, 2(3):284--293, 2020.

\bibitem{SmithEtAlFromLabMeeting}
Monique~E. Smith, Giulia Vico, Alessio Costa, Timothy Bowles, Am{\'e}lie C.~M. Gaudin, Sara Hallin, Christine~A. Watson, Remedios Alarc{\`o}n, Antonio Berti, Andrzej Blecharczyk, Francisco~J. Calderon, Steve Culman, William Deen, Craig~F. Drury, Axel Garcia~y. Garcia, Andr{\'e}s Garc{\'\i}a-D{\'\i}az, Eva~Hern{\'a}ndez Plaza, Krzysztof Jonczyk, Ortrud J{\"a}ck, R.~Michael Lehman, Francesco Montemurro, Francesco Morari, Andrea Onofri, Shannon~L. Osborne, Jos{\'e} Luis~Tenorio Pasam{\'o}n, Bo{\"e}l Sandstr{\"o}m, In{\'e}s Sant{\'\i}n-Montany{\'a}, Zuzanna Sawinska, Marty~R. Schmer, Jaroslaw Stalenga, Jeffrey Strock, Francesco Tei, Cairistiona F.~E. Topp, Domenico Ventrella, Robin~L. Walker, and Riccardo Bommarco.
\newblock Increasing crop rotational diversity can enhance cereal yields.
\newblock {\em Communications Earth \& Environment}, 4(1):89, 2023.

\bibitem{CREO_paper}
Dan~M Kluger, Art~B Owen, and David~B Lobell.
\newblock Combining randomized field experiments with observational satellite data to assess the benefits of crop rotations on yields.
\newblock {\em Environmental Research Letters}, 17(4):044066, 2022.

\bibitem{ChinaRiceToBeanRotation_SatelliteBased}
Ling Sun and Zesheng Zhu.
\newblock Evaluation of rotational effect of bean in large-scale rice-bean rotation using satellite remote sensing experiment.
\newblock {\em IOP Conference Series: Earth and Environmental Science}, 69(1):012055, 2017.

\bibitem{ChinaRiceToCottonRotation_SatelliteBased}
Ling Sun and Zesheng Zhu.
\newblock Using spectral vegetation index to estimate continuous cotton and rice-cotton rotation effects on cotton yield.
\newblock In {\em 2019 8th International Conference on Agro-Geoinformatics (Agro-Geoinformatics)}, pages 1--4, 2019.

\bibitem{UkraineRotation_satelliteBased}
Klaus Deininger, Daniel~Ayalew Ali, Nataliia Kussul, Mykola Lavreniuk, and Oleg Nivievskyi.
\newblock Using machine learning to assess yield impacts of crop rotation: Combining satellite and statistical data for {U}kraine.
\newblock Technical report, World Bank, Washington, DC, 2020.

\bibitem{FinlandSatellite_based}
Pirjo Peltonen-Sainio, Lauri Jauhiainen, Eija Honkavaara, Samantha Wittke, Mika Karjalainen, and Eetu Puttonen.
\newblock Pre-crop values from satellite images for various previous and subsequent crop combinations.
\newblock {\em Frontiers in Plant Science}, 10, 2019.

\bibitem{FinlandSatellite_based2024}
Pirjo Peltonen-Sainio, Mari Niemi, and Lauri Jauhiainen.
\newblock Legacy effects of crop sequencing on biomass and their variability on farmers' fields in {F}inland are shaped by weather, farm conditions and rationales for land use.
\newblock {\em Agricultural Systems}, 215:103850, 2024.

\bibitem{ObservationalApproachAustria}
Stefan Fenz, Thomas Neubauer, Johannes Heurix, Jürgen~Kurt Friedel, and Marie-Luise Wohlmuth.
\newblock Ai- and data-driven pre-crop values and crop rotation matrices.
\newblock {\em European Journal of Agronomy}, 150:126949, 2023.

\bibitem{BelgiumRotation_NPP_Giannarkis}
Georgios Giannarakis, Vasileios Sitokonstantinou, Roxanne~Suzette Lorilla, and Charalampos Kontoes.
\newblock Towards assessing agricultural land suitability with causal machine learning.
\newblock In {\em Proceedings of the IEEE/CVF Conference on Computer Vision and Pattern Recognition (CVPR) Workshops}, pages 1442--1452, 2022.

\bibitem{JunxiongPaper}
Junxiong Zhou, Peng Zhu, Dan~M. Kluger, David~B. Lobell, and Zhenong Jin.
\newblock Changes in the yield effect of the preceding crop in the us corn belt under a warming climate.
\newblock {\em Global Change Biology}, 30(11):e17556, 2024.

\bibitem{LawesEtAl_AustraliaMLToStudyRotBenefitsNotCausal}
Roger Lawes, Gonzalo Mata, Jonathan Richetti, Andrew Fletcher, and Chris Herrmann.
\newblock Using remote sensing, process-based crop models, and machine learning to evaluate crop rotations across 20 million hectares in western australia.
\newblock {\em Agronomy for Sustainable Development}, 42(6):120, 2022.

\bibitem{WagerAthey18_CausalForest}
Stefan Wager and Susan Athey.
\newblock Estimation and inference of heterogeneous treatment effects using random forests.
\newblock {\em Journal of the American Statistical Association}, 113(523):1228--1242, 2018.

\bibitem{BarleyWheatDistinguishingMethod}
Davoud Ashourloo, Hamed Nematollahi, Alfredo Huete, Hossein Aghighi, Mohsen Azadbakht, Hamid~Salehi Shahrabi, and Salman Goodarzdashti.
\newblock A new phenology-based method for mapping wheat and barley using time-series of {S}entinel-2 images.
\newblock {\em Remote Sensing of Environment}, 280:113206, 2022.

\bibitem{YouChinaCropMap}
Nanshan You, Jinwei Dong, Jianxi Huang, Guoming Du, Geli Zhang, Yingli He, Tong Yang, Yuanyuan Di, and Xiangming Xiao.
\newblock The 10-m crop type maps in northeast {C}hina during 2017--2019.
\newblock {\em Scientific Data}, 8(1):41, 2021.

\bibitem{CanadaACI}
{Agriculture and Agri-Food Canada}.
\newblock Annual crop inventory.
\newblock Last accessed December 2022.

\bibitem{FrenchParcel}
French~Republic Geoservices.
\newblock Geographical database used as a reference for the appraisal of aid under the common agricultural policy.
\newblock Last accessed February 2025.

\bibitem{CDL_Boryan}
Claire Boryan, Zhengwei Yang, Rick Mueller, and Mike Craig.
\newblock Monitoring {US} agriculture: the {US} {D}epartment of {A}griculture, {N}ational {A}gricultural {S}tatistics {S}ervice, {C}ropland {D}ata {L}ayer {P}rogram.
\newblock {\em Geocarto International}, 26(5):341--358, 2011.

\bibitem{TerraClim}
John~T. Abatzoglou, Solomon~Z. Dobrowski, Sean~A. Parks, and Katherine~C. Hegewisch.
\newblock Terra{C}limate, a high-resolution global dataset of monthly climate and climatic water balance from 1958--2015.
\newblock {\em Scientific Data}, 5(1):170191, 2018.

\bibitem{GCVIPaper}
Anatoly~A. Gitelson, Andrés Viña, Verónica Ciganda, Donald~C. Rundquist, and Timothy~J. Arkebauer.
\newblock Remote estimation of canopy chlorophyll content in crops.
\newblock {\em Geophysical Research Letters}, 32(8), 2005.

\bibitem{originalSCYM}
David~B. Lobell, David Thau, Christopher Seifert, Eric Engle, and Bertis Little.
\newblock A scalable satellite-based crop yield mapper.
\newblock {\em Remote Sensing of Environment}, 164:324--333, 2015.

\bibitem{SCYM_CornJillVersion}
Jillian~M. Deines, Rinkal Patel, Sang-Zi Liang, Walter Dado, and David~B. Lobell.
\newblock A million kernels of truth: Insights into scalable satellite maize yield mapping and yield gap analysis from an extensive ground dataset in the {US} {C}orn {B}elt.
\newblock {\em Remote Sensing of Environment}, 253:112174, 2021.

\bibitem{PeakVIMalawi}
Chengxiu Li, Ellasy~Gulule Chimimba, Oscar Kambombe, Luke~A. Brown, Tendai~Polite Chibarabada, Yang Lu, Daniela Anghileri, Cosmo Ngongondo, Justin Sheffield, and Jadunandan Dash.
\newblock Maize yield estimation in intercropped smallholder fields using satellite data in southern malawi.
\newblock {\em Remote Sensing}, 14(10), 2022.

\bibitem{JinPeakGCVIKenya}
Zhenong Jin, George Azzari, Calum You, Stefania {Di Tommaso}, Stephen Aston, Marshall Burke, and David~B. Lobell.
\newblock Smallholder maize area and yield mapping at national scales with google earth engine.
\newblock {\em Remote Sensing of Environment}, 228:115--128, 2019.

\bibitem{AtheyTibWager_GRF}
Susan Athey, Julie Tibshirani, and Stefan Wager.
\newblock {Generalized random forests}.
\newblock {\em The Annals of Statistics}, 47(2):1148 -- 1178, 2019.

\bibitem{Tillage_CausalForest}
Jillian~M Deines, Sherrie Wang, and David~B Lobell.
\newblock Satellites reveal a small positive yield effect from conservation tillage across the us corn belt.
\newblock {\em Environmental Research Letters}, 14(12):124038, 2019.

\bibitem{CoverCrop_CausalForest}
Jillian~M. Deines, Kaiyu Guan, Bruno Lopez, Qu~Zhou, Cambria~S. White, Sheng Wang, and David~B. Lobell.
\newblock Recent cover crop adoption is associated with small maize and soybean yield losses in the {U}nited {S}tates.
\newblock {\em Global Change Biology}, 29(3):794--807, 2023.

\bibitem{IrrigationDat_ESSD}
Y.~Xie, H.~K. Gibbs, and T.~J. Lark.
\newblock Landsat-based irrigation dataset ({LANID}): 30\,m resolution maps of irrigation distribution, frequency, and change for the {US}, 1997--2017.
\newblock {\em Earth System Science Data}, 13(12):5689--5710, 2021.

\bibitem{GrfPackage}
Julie Tibshirani, Susan Athey, Erik Sverdrup, and Stefan Wager.
\newblock {\em grf: Generalized Random Forests}, 2022.
\newblock R package version 2.2.1.

\bibitem{NASSQuickStats}
{USDA National Agricultural Statistics Service}.
\newblock Quick stats, 2024.
\newblock Accessed on 07-19-2024 from \url{https://quickstats.nass.usda.gov}.

\bibitem{sawyerNitrogenGuidelines}
J~Sawyer, E~Nafziger, G~Randall, L~Bundy, G~Rehm, and B~Joern.
\newblock Concepts and rationale for regional nitrogen guidelines for corn.
\newblock {\em Iowa State Univ. Extension Publ. PM2015}, 2006.

\bibitem{WinterWheatLatePlantingDateHurts}
B.~J. Dahlke, E.~S. Oplinger, J.~M. Gaska, and M.~J. Martinka.
\newblock Influence of planting date and seeding rate on winter wheat grain yield and yield components.
\newblock {\em Journal of Production Agriculture}, 6(3):408--414, 1993.

\bibitem{NASS_TypicalPlantingAndHarvestDateUS}
{United States Department of Agriculture, National Agricultural Statistics Service}.
\newblock {\em Field Crops: Usual Planting and Harvesting Dates (October 2010)}.
\newblock https://usda.library.cornell.edu/concern/publications/vm40xr56k.

\bibitem{NorwoodResiduesSnow}
Charles~A. Norwood.
\newblock Dryland winter wheat as affected by previous crops.
\newblock {\em Agronomy Journal}, 92(1):121--127, 2000.

\bibitem{WinterWheatResiduesTrapSnowWell}
Luana~M. Simão, Amanda~C. Easterly, Greg~R. Kruger, and Cody~F. Creech.
\newblock Winter wheat residue impact on soil water storage and subsequent corn yield.
\newblock {\em Agronomy Journal}, 113(1):276--286, 2021.

\bibitem{ChinaStoWWversusCtoWW}
Jiangwen Nie, Jie Zhou, Jie Zhao, Xiquan Wang, Ke~Liu, Peixin Wang, Shang Wang, Lei Yang, Huadong Zang, Matthew~Tom Harrison, Yadong Yang, and Zhaohai Zeng.
\newblock Soybean crops penalize subsequent wheat yield during drought in the {N}orth {C}hina {P}lain.
\newblock {\em Frontiers in Plant Science}, 13, 2022.

\bibitem{WoodsleeOntarioContWWversusOthers}
Ikechukwu~V. Agomoh, Craig~F. Drury, Lori~A. Phillips, W.~Daniel Reynolds, and Xueming Yang.
\newblock Increasing crop diversity in wheat rotations increases yields but decreases soil health.
\newblock {\em Soil Science Society of America Journal}, 84(1):170--181, 2020.

\bibitem{LegumeEffectsOnNitrogenAndSubsequentSpringWheat}
M.~Badaruddin and D.~W. Meyer.
\newblock Grain legume effects on soil nitrogen, grain yield, and nitrogen nutrition of wheat.
\newblock {\em Crop Science}, 34(5), 1994.

\bibitem{SpringWheat10PrecropsInclDryBeans}
J.M. Krupinsky, D.L. Tanaka, S.D. Merrill, M.A. Liebig, and J.D. Hanson.
\newblock Crop sequence effects of 10 crops in the northern great plains.
\newblock {\em Agricultural Systems}, 88(2):227--254, 2006.

\bibitem{AngusPrecropForWheatMetanalysis}
J.~F. Angus, J.~A. Kirkegaard, J.~R. Hunt, M.~H. Ryan, L.~Ohlander, and M.~B. Peoples.
\newblock Break crops and rotations for wheat.
\newblock {\em Crop and Pasture Science}, 66(6):523--552, 2015.

\bibitem{SoyExpsRotationBenefitLargelyDrivenByCystNematodeReduction}
S.~M. Dabney, E.~C. McGawley, D.~J. Boethel, and D.~A. Berger.
\newblock Short-term crop rotation systems for soybean production.
\newblock {\em Agronomy Journal}, 80(2):197--204, 1988.

\bibitem{RotationHelpsWithCystNematode}
J.~Earl~Creech, Andreas Westphal, Virginia~R. Ferris, Jamal Faghihi, Tony~J. Vyn, Judith~B. Santini, and William~G. Johnson.
\newblock Influence of winter annual weed management and crop rotation on soybean cyst nematode (heterodera glycines) and winter annual weeds.
\newblock {\em Weed Science}, 56(1):103–111, 2008.

\bibitem{HighTemperatureImpliesHigherPestPressure}
Lewis~H Ziska.
\newblock Increasing minimum daily temperatures are associated with enhanced pesticide use in cultivated soybean along a latitudinal gradient in the mid-western {U}nited {S}tates.
\newblock {\em PLoS One}, 9(6), 2014.

\bibitem{NitrogenMineralizationInSoilArrheniusEqBased}
George Stanford, M.~H. Frere, and D.~H. Schwaninger.
\newblock Temperature coefficient of soil nitrogen mineralization.
\newblock {\em Soil Science}, 115(4), 1973.

\bibitem{CornResidueDecomp}
Mahdi~M. Al-Kaisi, David Kwaw-Mensah, and En~Ci.
\newblock Effect of nitrogen fertilizer application on corn residue decomposition in {I}owa.
\newblock {\em Agronomy Journal}, 109(5):2415--2427, 2017.

\bibitem{RootTemperatureNitrogenFixation}
F.~Munévar and A.~G. Wollum~II.
\newblock Effect of high root temperature and rhizobium strain on nodulation, nitrogen fixation, and growth of soybeans.
\newblock {\em Soil Science Society of America Journal}, 45(6):1113--1120, 1981.

\bibitem{MacLarenConservationAg}
Chloe MacLaren, Andrew Mead, Derk van Balen, Lieven Claessens, Ararso Etana, Janjo de~Haan, Wiepie Haagsma, Ortrud J{\"a}ck, Thomas Keller, Johan Labuschagne, {\AA}sa Myrbeck, Magdalena Necpalova, Generose Nziguheba, Johan Six, Johann Strauss, Pieter~Andreas Swanepoel, Christian Thierfelder, Cairistiona Topp, Flackson Tshuma, Harry Verstegen, Robin Walker, Christine Watson, Marie Wesselink, and Jonathan Storkey.
\newblock Long-term evidence for ecological intensification as a pathway to sustainable agriculture.
\newblock {\em Nature Sustainability}, 5(9):770--779, 2022.

\bibitem{FranceLongerRotationsLessPesticide}
Benjamin Nowak, Audrey Michaud, and Gaëlle Marliac.
\newblock Assessment of the diversity of crop rotations based on network analysis indicators.
\newblock {\em Agricultural Systems}, 199:103402, 2022.

\bibitem{lewbel2007estimation}
Arthur Lewbel.
\newblock Estimation of average treatment effects with misclassification.
\newblock {\em Econometrica}, 75(2):537--551, 2007.

\bibitem{QDANNPaper}
Yuchi Ma, Sang-Zi Liang, D.~Brenton Myers, Anu Swatantran, and David~B. Lobell.
\newblock Subfield-level crop yield mapping without ground truth data: A scale transfer framework.
\newblock {\em Remote Sensing of Environment}, 315:114427, 2024.

\bibitem{DecisionSupportROTAT}
S~Dogliotti, W.A.H Rossing, and M.K {van Ittersum}.
\newblock Rotat, a tool for systematically generating crop rotations.
\newblock {\em European Journal of Agronomy}, 19(2):239--250, 2003.

\bibitem{DecisionSupportROTOR}
Johann Bachinger and Peter Zander.
\newblock Rotor, a tool for generating and evaluating crop rotations for organic farming systems.
\newblock {\em European Journal of Agronomy}, 26(2):130--143, 2007.

\bibitem{DecisionSupportCropRota}
Martin Schönhart, Erwin Schmid, and Uwe~A. Schneider.
\newblock Croprota – a crop rotation model to support integrated land use assessments.
\newblock {\em European Journal of Agronomy}, 34(4):263--277, 2011.

\bibitem{DecisionSupportFinland}
Lauri~Jauhiainen Pirjo Peltonen-Sainio and Arto Latukka.
\newblock Interactive tool for farmers to diversify high-latitude cereal-dominated crop rotations.
\newblock {\em International Journal of Agricultural Sustainability}, 18(4):319--333, 2020.

\bibitem{DecisionSupportFruchtfolge}
C.~Pahmeyer, T.~Kuhn, and W.~Britz.
\newblock ‘fruchtfolge’: A crop rotation decision support system for optimizing cropping choices with big data and spatially explicit modeling.
\newblock {\em Computers and Electronics in Agriculture}, 181:105948, 2021.

\bibitem{DoubleCroppingRareInCanada}
Eric~R. Page, Sydney Meloche, and Jamie Larsen.
\newblock Evaluating the potential for double cropping in {C}anada: effect of seeding date and relative maturity on the development and yield of maize, white bean, and soybean.
\newblock {\em Canadian Journal of Plant Science}, 99(5):751--760, 2019.

\bibitem{MultiCroppingRareInFrance}
Bernhard Schauberger, Hiromi Kato, Tomomichi Kato, Daiki Watanabe, and Philippe Ciais.
\newblock French crop yield, area and production data for ten staple crops from 1900 to 2018 at county resolution.
\newblock {\em Scientific Data}, 9(1):38, 2022.

\bibitem{NIRvPaper}
Grayson Badgley, Christopher~B. Field, and Joseph~A. Berry.
\newblock Canopy near-infrared reflectance and terrestrial photosynthesis.
\newblock {\em Science Advances}, 3(3), 2017.

\bibitem{ChinaNIRvGoodPerformance}
Linchao Li, Bin Wang, Puyu Feng, De~{Li Liu}, Qinsi He, Yajie Zhang, Yakai Wang, Siyi Li, Xiaoliang Lu, Chao Yue, Yi~Li, Jianqiang He, Hao Feng, Guijun Yang, and Qiang Yu.
\newblock Developing machine learning models with multi-source environmental data to predict wheat yield in {C}hina.
\newblock {\em Computers and Electronics in Agriculture}, 194:106790, 2022.

\bibitem{CrumpOverlap}
Richard~K. Crump, V.~Joseph Hotz, Guido~W. Imbens, and Oscar~A. Mitnik.
\newblock Dealing with limited overlap in estimation of average treatment effects.
\newblock {\em Biometrika}, 96(1):187--199, 2009.

\bibitem{OverlapWeights_ATO_proposal}
Kari Lock~Morgan Fan~Li and Alan~M. Zaslavsky.
\newblock Balancing covariates via propensity score weighting.
\newblock {\em Journal of the American Statistical Association}, 113(521):390--400, 2018.

\bibitem{SCYM_SoyDadoVersion}
Walter~T. Dado, Jillian~M. Deines, Rinkal Patel, Sang-Zi Liang, and David~B. Lobell.
\newblock High-resolution soybean yield mapping across the {US} {M}idwest using subfield harvester data.
\newblock {\em Remote Sensing}, 12(21), 2020.

\bibitem{SemenovaChernozhukovTheoryForBestLinearProjection}
Vira Semenova and Victor Chernozhukov.
\newblock {Debiased machine learning of conditional average treatment effects and other causal functions}.
\newblock {\em The Econometrics Journal}, 24(2):264--289, 2020.

\bibitem{MACKINNONandWhiteHC3}
James~G MacKinnon and Halbert White.
\newblock Some heteroskedasticity-consistent covariance matrix estimators with improved finite sample properties.
\newblock {\em Journal of Econometrics}, 29(3):305--325, 1985.

\bibitem{DoublyRobustCILinearModel}
Michele~Jonsson Funk, Daniel Westreich, Chris Wiesen, Til Stürmer, M.~Alan Brookhart, and Marie Davidian.
\newblock Doubly robust estimation of causal effects.
\newblock {\em American Journal of Epidemiology}, 173(7):761--767, 2011.

\bibitem{SSURGO}
{Soil Survey Staff, Natural Resources Conservation Service, United States Department of Agriculture}.
\newblock Web soil survey, 2021.
\newblock https://websoilsurvey.nrcs.usda.gov/. Accessed on 08-16-2021.

\bibitem{BeillouinMetaMetaAnalysis}
Damien Beillouin, Tamara Ben-Ari, Eric Malézieux, Verena Seufert, and David Makowski.
\newblock Positive but variable effects of crop diversification on biodiversity and ecosystem services.
\newblock {\em Global Change Biology}, 27(19):4697--4710, 2021.

\bibitem{BealCohenObservationalRotUS}
Allegra~A. Beal~Cohen, Christopher~A. Seifert, George Azzari, and David~B. Lobell.
\newblock Rotation effects on corn and soybean yield inferred from satellite and field-level data.
\newblock {\em Agronomy Journal}, 111(6):2940--2948, 2019.

\bibitem{FixestPackage}
Laurent Berge, Sebastian Krantz, Grant McDermott, and Russell Lenth.
\newblock {\em fixest: Fast Fixed-Effects Estimations}, 2022.
\newblock R package version 0.10.3.

\bibitem{APSIM}
Dean~P. Holzworth, Neil~I. Huth, Peter~G. deVoil, Eric~J. Zurcher, Neville~I. Herrmann, Greg McLean, Karine Chenu, Erik~J. {van Oosterom}, Val Snow, Chris Murphy, Andrew~D. Moore, Hamish Brown, Jeremy~P.M. Whish, Shaun Verrall, Justin Fainges, Lindsay~W. Bell, Allan~S. Peake, Perry~L. Poulton, Zvi Hochman, Peter~J. Thorburn, Donald~S. Gaydon, Neal~P. Dalgliesh, Daniel Rodriguez, Howard Cox, Scott Chapman, Alastair Doherty, Edmar Teixeira, Joanna Sharp, Rogerio Cichota, Iris Vogeler, Frank~Y. Li, Enli Wang, Graeme~L. Hammer, Michael~J. Robertson, John~P. Dimes, Anthony~M. Whitbread, James Hunt, Harm {van Rees}, Tim McClelland, Peter~S. Carberry, John~N.G. Hargreaves, Neil MacLeod, Cam McDonald, Justin Harsdorf, Sara Wedgwood, and Brian~A. Keating.
\newblock {APSIM} – evolution towards a new generation of agricultural systems simulation.
\newblock {\em Environmental Modelling \& Software}, 62:327--350, 2014.

\bibitem{LeafAreaIndexToGCVI}
Anthony Nguy-Robertson, Anatoly Gitelson, Yi~Peng, Andrés Viña, Timothy Arkebauer, and Donald Rundquist.
\newblock Green leaf area index estimation in maize and soybean: Combining vegetation indices to achieve maximal sensitivity.
\newblock {\em Agronomy Journal}, 104(5):1336--1347, 2012.

\bibitem{SCYM_AzzarriVersion}
George Azzari, Meha Jain, and David~B. Lobell.
\newblock Towards fine resolution global maps of crop yields: Testing multiple methods and satellites in three countries.
\newblock {\em Remote Sensing of Environment}, 202:129--141, 2017.

\bibitem{SCYM_JinVersion}
Zhenong Jin, George Azzari, and David~B. Lobell.
\newblock Improving the accuracy of satellite-based high-resolution yield estimation: A test of multiple scalable approaches.
\newblock {\em Agricultural and Forest Meteorology}, 247:207--220, 2017.

\bibitem{PeakEVI2GoodProxyCanada}
Jiangui Liu, Jiali Shang, Budong Qian, Ted Huffman, Yinsuo Zhang, Taifeng Dong, Qi~Jing, and Tim Martin.
\newblock Crop yield estimation using time-series modis data and the effects of cropland masks in {O}ntario, {C}anada.
\newblock {\em Remote Sensing}, 11(20), 2019.

\bibitem{ChinaMaizeGCVIBetter}
Liangliang Zhang, Zhao Zhang, Yuchuan Luo, Juan Cao, Ruizhi Xie, and Shaokun Li.
\newblock Integrating satellite-derived climatic and vegetation indices to predict smallholder maize yield using deep learning.
\newblock {\em Agricultural and Forest Meteorology}, 311:108666, 2021.

\bibitem{CanadaSurveyData}
{Agriculture and Agri-Food Canada}.
\newblock {Canadian Crop Yields - Historic Yields of Major Crops (Pre-packaged CSV files)}, 2024.
\newblock Accessed on 07-19-2024 from \url{ open.canada.ca/data/en/dataset/9253a01b-f1d9-4b67-ba98-857667827c7b/resource/b04e8717-50da-4f24-b4a1-d6c4bf283a9e}.

\bibitem{FranceSurveyData}
{French Ministry of Agriculture and Food, Agreste}.
\newblock {Cultivated Crops}, 2024.
\newblock Accessed on 07-19-2024 from \url{agreste.agriculture.gouv.fr/agreste-saiku/?plugin=true&query=query/open/SAANR_DEVELOPPE_2#query/open/SAANR_DEVELOPPE_2}.

\bibitem{ReviewPaperWithECEmetric}
Jakob Gawlikowski, Cedrique Rovile~Njieutcheu Tassi, Mohsin Ali, Jongseok Lee, Matthias Humt, Jianxiang Feng, Anna Kruspe, Rudolph Triebel, Peter Jung, Ribana Roscher, Muhammad Shahzad, Wen Yang, Richard Bamler, and Xiao~Xiang Zhu.
\newblock A survey of uncertainty in deep neural networks.
\newblock {\em Artificial Intelligence Review}, 56(1):1513--1589, 2023.

\bibitem{ChinaSurveyData}
{National Bureau of Statistics of {C}hina}.
\newblock {Output of Major Farm Products Per Hectare}, 2024.
\newblock Accessed on 07-24-2024 from \url{https://data.stats.gov.cn/english/easyquery.htm?cn=E0103}.

\end{thebibliography}

\appendix
\setcounter{table}{0}
\renewcommand{\thetable}{S\arabic{table}}
\setcounter{figure}{0}
\renewcommand{\thefigure}{S\arabic{figure}}

\newpage

\section*{Supplement for ``Precrop payoffs: causal machine learning reveals large but
variable yield benefits of crop rotation in major breadbaskets"}

\section{Materials and Methods} 

\subsection{Data sources and sample} \label{sec:DatasetAndSample}

We used country-specific crop type maps (Table \ref{Table:CropMapInfo}) to construct our study sample and to infer the crop type at each year and location within our study sample. These annual crop type maps had spatial resolutions of at most 30m $\times$ 30m and gave the primary crop for each year and pixel, ignoring cover crops. Note that double cropping was not accounted for as a separate category in the crop type maps from China and Canada; however, this practice is considered uncommon in northeast China \cite{YouChinaCropMap}, Canada \cite{DoubleCroppingRareInCanada}, and France \cite{MultiCroppingRareInFrance}. We considered pixels that were classified as ``double crop" in the US or as a ``mixture" in France, as separate crop type categories, but these categories were not so common and thus were ultimately excluded from the analysis (Section \ref{sec:RotationsStudied}).

\begin{table}[!h]\caption{Sources for annual crop type maps used in this study. The 3rd column reports the spatial resolution of these crop type maps and the 4th column reports the years used in this study. The final column indicates whether the crop map had a separate category for instances where there were multiple primary crop types.}\label{Table:CropMapInfo}
\centering \footnotesize
\begin{tabular}{lllll}
  \hline
Country & Crop Type Map [Ref.] & Spatial Resolution & Years & Multi-crop labels? \\ 
  \hline
  Canada & Annual Crop Inventory \cite{CanadaACI} & 30m $\times$ 30m  & 2011--2020 & No \\ 
    China (Northeast) & You et al. \cite{YouChinaCropMap} &  10m $\times$ 10m & 2017--2019 & No \\ 
  France & Parcel Dataset \cite{FrenchParcel} & 10m $\times$ $\text{10m}^{\dag}$ & 2015--2021 & Yes: ``Mixture" \\ 
  US & Cropland Data Layer \cite{CDL_Boryan} &  30m $\times$ 30m & 2008--2021 & Yes: ``Double crop" \\ 
   \hline 
\end{tabular}
\vspace{-0.5cm}
\smallskip
\dag We rasterized this parcel-level dataset to have 10m $\times$ 10m resolution.

\end{table}

We used Google Earth Engine (GEE) and these country-specific crop type maps to construct a sample of geographical points corresponding to croplands. To include more samples with crop types of interest, we randomly sampled geographical points after masking for locations in which crop types of interest were observed in at least 20\% of the years. Each country had a slightly different random sampling scheme due to differences across countries in the number of years with available crop type maps and in the crop types of interest (see Appendix \ref{sec:samplingScheme} for details). For each country and sampled geographical point, the country-specific crop type maps were again used to infer the crop that was grown during every year with an available crop type classification.

For each location and year in our sample, we used a collection of satellite images in GEE to extract a time series of the Green Chlorophyll Vegetation Index (GCVI) using the formula $\text{GCVI}=\text{NIR}/\text{GREEN}-1$ where $\text{NIR}$ and $\text{GREEN}$ are bands from the multispectral satellite image. GCVI was originally designed to capture chlorophyll content in crops \cite{GCVIPaper} and has been shown to be useful for estimating crop yields \cite{originalSCYM}. For each sample in China and France, we used Sentinel-2 top of atmosphere (S2 TOA) 
to obtain the GCVI time series. 
The choice of the S2 TOA, instead of surface reflectance (SR) was due to lack of coverage of SR imagery in our regions before 2017.
The Sentinel-2A and Sentinel-2B satellites acquire images with a spatial resolution of 10m $\times$ 10m (Blue, Green, Red, and NIR bands) and 20m $\times$ 20m (for other bands that were not used), and together they provide images every 5 days. Clouds in the S2 dataset were masked out using the S2 Cloud Probability dataset provided by SentinelHub in GEE. Due to lack of availability of Sentinel-2 data prior to 2015 and in order to maintain a within-country consistency of GCVI measurements, we instead used Landsat to obtain the GCVI time series for each year and location in our Canada and the US samples. 
Landsat sensors acquire images with a spatial resolution of 30m $\times$ 30m. Each Landsat satellite has a revisit cycle of 16 days, which is reduced to eight days when two satellites are operating simultaneously.
In GEE, we accessed all available Landsat Collection 2 Tier 1 SR imagery from four Landsat satellites (Landsat 5,7,8 and 9) and filtered cloudy observations using the “pixel\_qa” band provided with Landsat SR products. 

 For the purposes of a sensitivity analysis (in Appendix \ref{sec:UsingNIRvInsteadOfGCVI}), at each location and year in our sample, we also extracted a time series of the near infrared reflectance of vegetation (NIRv) as an alternative to GCVI. NIRv, a vegetation index developed in \cite{NIRvPaper}, is the product of the Normalized Difference Vegetation Index and the NIR band, and NIRv has shown strong performance for predicting crop yields \cite{ChinaNIRvGoodPerformance}. In particular, to extract NIRv values, we used the formula NIRv=NIR$\times$(NIR$-$RED)/(NIR$+$RED) where NIR and RED are bands from the multispectral satellite images. 

For each location and year in our sample, we accessed the TerraClimate \cite{TerraClim} gridded weather dataset via GEE to extract weather variables. In particular, we extracted monthly precipitation, monthly vapor pressure deficit (VPD), the monthly average of maximum daily temperature, and the monthly average of minimum daily temperature. 

The dataset was harmonized using latitude, longitude, and year. The raw TerraClimate weather dataset had a spatial resolution of $1/24$° $\times$ $1/24$° (approximately 4km $\times$ 4km) which was a substantially lower spatial resolution than that of the crop type maps. Therefore, the latitude and longitude of each sampled pixel from the crop type map was used to determine the grid cell from which the TerraClimate weather data was extracted. The satellite-based GCVI in each country had the same spatial resolution as the corresponding crop type map, and we extracted the GCVI data from the Landsat (or Sentinel-2) pixels that intersected our sampled pixels from the crop type maps.

\subsection{Normalized peak greenness as an outcome variable}\label{sec:SPH_GCVI}

We used the peak GCVI value as a proxy for crop yield. To account for the fact that different samples had a different set of days with cloud free images, we computed the peak GCVI value at each location and year by fitting a harmonic regression to the January-December GCVI time series of cloud free measurements, and calculated the peak of the fitted time series curve. Specifically, for each location $i$ and year $t$ with $n(i,t)$ cloud free GCVI measurements $G_{1}^{(i,t)},\dots,G_{n(i,t)}^{(i,t)}$ at fractional times $u_{1}^{(i,t)},\dots, u_{n(i,t)}^{(i,t)}$ respectively (with the $u_{j}^{(i,t)}$ being the time of year on a scale from $0$ on January 1st to $1$ on December 31st), we ran ordinary least squares regression to find the coefficients minimizing $$\sum_{j=1}^{n(i,t)} \Big( G_{j}^{(i,t)} - c_{i,t} - \sum_{k=1}^3  \big[ a_{i,t,k} \cos (2k \pi u_{j}^{(i,t)}) + b_{i,t,k} \sin (2k \pi u_{j}^{(i,t)}) \big] \Big)^2.$$  Letting $\hat{c}_{i,t}, \hat{a}_{i,t,1}, \hat{a}_{i,t,2},\hat{a}_{i,t,3}, \hat{b}_{i,t,1}, \hat{b}_{i,t,2},\hat{b}_{i,t,3}$ be the coefficients minimizing the above expression, we computed the peak GCVI values by taking the maximum of the fitted harmonic curves: $$\text{peak GCVI}_{i,t} = \max_{0 \leq u \leq 1} \Bigl\{ \hat{c}_{i,t} + \sum_{k=1}^3  \big[ \hat{a}_{i,t,k} \cos (2k \pi u) + \hat{b}_{i,t,k} \sin (2k \pi u) \big] \Bigr\}.$$ 

For some locations $i$ and years $t$ in our sample, we could not compute the peak GCVI value because the number of cloud free observations $n(i,t)$ was too small to implement ordinary least squares regression. In other cases, likely also due to a dearth of cloud free observations, outlier peak GCVI values were observed, which we removed from the analysis (see Appendix \ref{sec:outlierGCVI}). Ultimately, when restricting our attention in each country to cases where the outcome crops of interest are grown and when averaging across each country, $0.1 \%$ percent of observations were dropped for the former reason and an additional $0.5 \%$ percent of samples were dropped for the latter reason. 

Finally, we constructed a normalized peak GCVI-based yield proxy that would be comparable across crop types and across countries in which different satellite sources were used. To do so, for each crop type and country pair, we divided the peak GCVI values by the mean peak GCVI value among samples in that country where that crop type was grown.

\subsection{Types of rotations studied}\label{sec:RotationsStudied}

For each location and year in our sample, we defined the precrop to be the crop grown in the previous year and the outcome crop to be the crop grown in the current year. In each country, we restricted our attention to at most four outcome crops of interest among corn, soybean, winter wheat, and spring wheat. In the US we studied all four outcome crops; however, we did not study winter wheat or spring wheat as an outcome crop in China because the crop type map used included wheat within its ``other" category, and we also did not consider spring wheat in France, soybean in France, and winter wheat in Canada as outcome crops due to insufficient sample sizes.

Samples where the precrop and outcome crop were the same were classified as having no rotation and were used as control units. For each country and outcome crop of interest, Figure \ref{fig:PrecropDistribution} depicts the distribution of precrops in the sample. Samples where the outcome crop and precrop were different were classified as having a rotation; however, because we wanted to compare the effects of rotation across different precrop and outcome crop combinations, each precrop and outcome crop combination was categorized as a different type of rotation. In particular, a location and year in our sample with precrop A and outcome crop B was categorized as having an A$\to$B rotation. In order to focus our study on the most common crop rotations, we applied two selection criteria. For each country and outcome crop of interest within the country, we first only considered precrops that were among the 5 most prevalent precrops used in rotation. Second, we only considered precrops for which the number of samples with that precrop exceeded one tenth of the number of samples with the no rotation control. Table \ref{table:ResultsTable} shows the categories of rotation considered within each country after applying these selection criteria. It also shows the sample sizes of the treatment and control groups.

\subsection{Causal forest analysis}\label{sec:CausalForestAnalysis}

We assessed the benefit of the various crop rotations on crop yield by fitting causal forests \cite{WagerAthey18_CausalForest,AtheyTibWager_GRF} using version 2.2.1 of the \texttt{grf} package in R \cite{GrfPackage}. A causal forest is a recently developed method for conducting causal inference in observational studies, and it can be used to estimate the average treatment effect as well as the treatment effect as a function of the covariates. 
Causal forests have been used to study the effects of various agronomic practices such as tillage \cite{Tillage_CausalForest}, cover cropping \cite{CoverCrop_CausalForest}, and crop rotation \cite{CREO_paper} on crop yields as well as the impact of crop rotation on net primary productivity \cite{BelgiumRotation_NPP_Giannarkis}. In the context of assessing the effect of the Soybean$\to$Corn crop rotation on crop yield, rotation benefits estimated by fitting a causal forest to a satellite-derived dataset were found to have a statistically significant positive correlation with estimates from actual randomized field experiments \cite{CREO_paper}.

For each country, precrop A, and outcome crop B (corresponding to a unique row in Table \ref{table:ResultsTable}), we conducted the following procedure to estimate the benefit of the A$\to$B rotation. First, we subsetted our data to only include samples with A$\to$B rotations or B$\to$B rotations and designated a binary treatment variable $Z$ to be $1$ for samples with A$\to$B rotations and $0$ for samples with B$\to$B. Next we set the outcome variable $V$ to be the normalized peak GCVI variable described in Section \ref{sec:SPH_GCVI}. Finally we let $X$ denote a vector of year, latitude, longitude, 7 weather covariates, and (in the US only) irrigation status. The weather covariates captured early precipitation, growing season precipitation, average maximum and minimum daily temperatures in the growing season, and VPD for the three peak months of the growing season. In particular, when winter wheat was not the outcome crop (versus when winter wheat was the outcome crop) our 7 weather covariates were: total January-April (January-March) precipitation, total May-September (April-June) precipitation, average maximum daily temperature during April-September (April-June), average minimum daily temperature during April-September (April-June), June (April) VPD, July (May) VPD, and August (June) VPD. For samples in the US we also included irrigation status as an 11th covariate in $X$, which we retrieved from the Landsat-based Irrigation Dataset \cite{IrrigationDat_ESSD} in the United States. Due to lack of estimates from this data product after the year 2017, and because irrigation status of fields do not typically change in the US, we imputed the irrigation status of each pixel after the year 2017 based on the proportion of years in which that pixel was classified as having irrigation between 2011 and 2017 (see Appendix \ref{sec:IrrigationImputationScheme}). Due to lack of irrigation data as well as greater homogeneity of irrigation status in the other study regions, we did not include irrigation as a covariate in Canada, France, or northeastern China. 

After defining our binary treatment variable $Z$, our outcome variable $V$, and our covariates $X$ we then ran the \texttt{causal\_forest} function with 200 trees (and otherwise using the default settings), returning an object containing random forest-based approximations to the following two functions of the covariates $x$: \begin{equation}
    \pi(x)=  \mathbb{P}[Z= 1| X=x], \text{ and }
\end{equation}
\begin{equation}\label{eq:omegaDefinition}
    \omega(x)= \e [V| X=x, Z=1] -\e [V| X=x, Z= 0].
\end{equation} In particular, when using the \texttt{causal\_forest} function, $\pi(x)$ was estimated by training a standard random forest for binary classification. Meanwhile $\omega(x)$ was estimated using a random forest of $B=200$ honest causal trees. That is, for each $b=1,\dots,B$, a random subsample without replacement was drawn, and then was further randomly split in half, where half the subsamples were used to determine the optimal splits and leaf assignment function $L_b(x)$ and the $V_i$ values from the remaining, withheld half of the subsamples (whose indices we denote by $\mathcal{W}_b$) were used to determine the function $$\hat{\omega}_b(x)= \frac{1}{\vert \mathcal{I}_{x,b}^{(1)}  \vert} \sum_{i \in \mathcal{I}_{x,b}^{(1)}}  V_i-\frac{1}{\vert \mathcal{I}_{x,b}^{(0)}  \vert} \sum_{i \in \mathcal{I}_{x,b}^{(0)}}  V_i \quad \text{where} \quad  \mathcal{I}_{x,b}^{(j)} \equiv \{ i \in \mathcal{W}_b \ : \ Z_i=j, X_i \in L_b(x) \}.$$ In words, if $x$ belongs to leaf $L_b(x)$, $\hat{\omega}_b(x)$ is the average of $V$ for samples with $Z=1$ minus that for samples with $Z=0$ when restricting our attention to the withheld samples belonging to leaf $L_b(x)$.
The causal forest estimated the function $\omega(x)$ via averaging honest trees according to $\hat{\omega}(x)=\frac{1}{B} \sum_{b=1}^B \hat{\omega}_b(x)$. \cite{WagerAthey18_CausalForest} provides theoretical guarantees that under certain technical conditions and if the sample size is large, the function $\hat{\omega}(\cdot)$ returned by the causal forest gives a good estimate of the function $\omega(\cdot)$ defined in Equation \eqref{eq:omegaDefinition}.
Note that when running the \texttt{causal\_forest} function, we used the default setting of tune.parameters=``none", meaning that no hyperparameters were tuned in the training of the causal forests and only prespecified or default hyperparameter values were used. Nonetheless, we found that the estimates of $\pi(\cdot)$ and $\omega(\cdot)$ produced were well calibrated and reasonably accurate (see Appendix \ref{sec:CausalForestGoodnessOfFits} and Table \ref{table:ModelGoodnessOfFitMetrics}).

The functions $\pi(\cdot)$ and $\omega(\cdot)$ that were estimated can be intuitively described as follows. $\pi(\cdot)$ gives the propensity score, which is the probability that the treatment occurred as a function of the covariates $x$. $\omega(\cdot)$ can be thought of as the difference in the mean normalized peak GCVI in the treated group minus that in the control group written as a function of the covariates $x$. Under the assumption that there are no unmeasured confounders, $\omega(x)$ is equal to the conditional average treatment effect (CATE), which is the actual mean causal effect of the treatment $Z$ on the outcome $V$ as a function of $x$.

\begin{figure}[!t]
    \centering
    \includegraphics[width=0.95 \hsize]{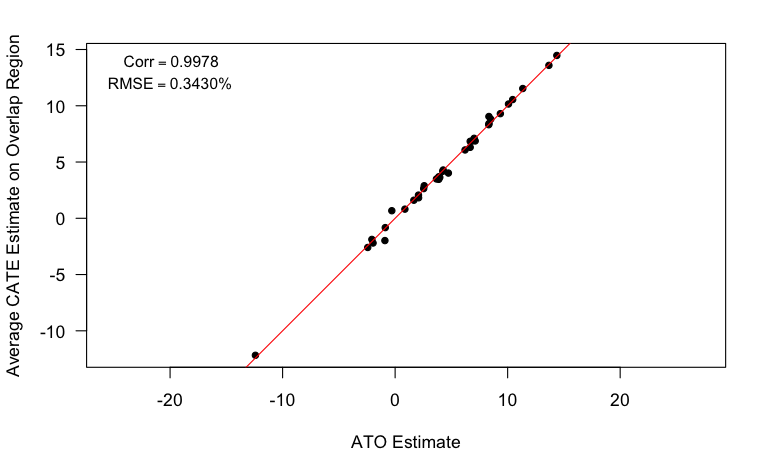}
    \caption{ATO estimates versus average CATE estimates on the overlap samples. In this scatter plot, each of the 36 points corresponds to a unique country-rotation pair in our study, the x-axis gives the ATO estimate which was used throughout the main text, and the y-axis gives an average of the estimated CATE among samples where the estimated propensity score was between $0.05$ and $0.95$. The quantities plotted were converted from the normalized peak GCVI scale to the crop yield scale (in units of percent of mean crop yield of the corresponding outcome crop in the corresponding study region) using the approach described in Section \ref{sec:ConvertingToYieldScale}. The red line gives a line through the origin with slope one and the text in the top left gives the sample Pearson correlation and the RMSE when comparing the two plotted quantities.}
    \label{fig:ATOvsMeanCATEOverlap}
\end{figure}

A standard way to summarize the treatment effects is the average treatment effect, given by $\e[\text{CATE}(X)]$; however, estimation of this quantity is unstable in settings where some of the propensity scores are near $0$ or $1$ \cite{CrumpOverlap}. Because in our setting many samples had estimated propensity scores $\hat{\pi}(x)$ near $0$ or $1$ (Table \ref{table:OverlapTable}), we instead estimated the overlap-weighted average treatment effect (ATO), given by $$\text{ATO}=\frac{\e[\text{CATE}(X) \pi(X)(1-\pi(X))]}{\e[ \pi(X)(1-\pi(X))]}.$$ The ATO is essentially a weighted average of the of the CATE that gives higher weight to points in regions of covariate space where the treatment and control are observed with similar frequency but no weight to points in regions of covariate space that have only treated units or only control units. The ATO was proposed in \cite{OverlapWeights_ATO_proposal} and was shown in that paper to have the desirable property of minimizing asymptotic variance among a class of weighted average treatment effect estimands. We estimated the ATO using the \texttt{average\_treatment\_effect} function in the \texttt{grf} package. We briefly note that empirically, our ATO estimates were very similar to the mean estimated CATE when the mean is restricted to the sample of points with $\hat{\pi}(x)$ between 0.05 and 0.95 (Figure \ref{fig:ATOvsMeanCATEOverlap}).

We used the default settings of the \texttt{average\_treatment\_effect} function to calculate the standard errors of the estimated ATO. These standard errors accounted for the overlap-weighting scheme, but notably, as with all standard errors reported in the main text, they did not account for possible spatial correlations between the samples. In Appendix \ref{sec:AccountForSpatialCorrelations}, we justify our choice to not account for spatial correlations. In that appendix, we also show that the samples used in each analysis are relatively well spread out (median distance to nearest neighbor $>1$km) and conduct a sensitivity analysis where we find that accounting for spatial correlations increases the confidence interval widths by a factor of at most 1.7, but typically does so by a smaller factor (Figure \ref{fig:ClusterSE_CIchanges}).

We computed the estimated ATO and its confidence interval using the above procedure for each rotation A$\to$B versus control B$ \to$B comparison considered in Table \ref{table:ResultsTable}. We interpreted the estimated ATO as an estimate for the average net effect of rotating using the precrop A on a proxy for the yield of crop B, where the net effect incorporates the direct effect of rotation and the indirect effect of rotation mediated by changes in fertilizer and pesticide input use. While $V$ is merely a vegetation index-based proxy for yield, the analysis in recent works such as \cite{CREO_paper,Tillage_CausalForest} which apply causal forests using yield estimates from the Scalable Crop Yield Mapper \cite{originalSCYM,SCYM_CornJillVersion,SCYM_SoyDadoVersion} would, up to a normalization constant, have given the same results as using a vegetation index $V$ under a principled choice of the vegatation index $V$ and weather controls $X$ (see Appendix \ref{sec:VI_versus_SCYM} for a mathematical justification of this claim). Further, we convert estimates of the ATO to the crop yield scale in Section \ref{sec:ConvertingToYieldScale}, which in essence approximates the unknown normalization constant using subnational-level crop yield data.

\subsection{Analyzing heterogeneity of rotation benefits with weather}\label{sec:HeterogeneityAnalysisMethod}

For each country and rotation A$\to$B, we also assessed how the benefit of rotation varied with weather. Our weather variables of interest were growing season precipitation and growing season temperature. Growing season precipitation was defined as the total precipitation between May and September (or between January and June in cases where winter wheat was the outcome crop). Growing season temperature was defined as the mean of the daily maxima and minima temperature averaged across April-September (or across April-June in cases where winter wheat was the outcome crop).

To assess of how the benefit of the rotation A$\to$B versus the control B$\to$B varied with weather, we ran the following linear regression of $\omega(X)$ on the year $t$, the growing season precipitation $P$, and the growing season temperature $T$ \begin{equation}\label{eq:ModelToAssessHeterogeneity}
    \omega(X) = \beta_0 +\beta_{\text{Year}} t  + \beta_{\text{Precip}} P+\beta_{\text{Temp}} T + \varepsilon.
\end{equation} The above regression includes a linear time trend to control for gradually changing factors besides weather that may affect rotation benefits (e.g., soil quality, management practices, crop variety choices, etc.), and to reduce how much such factors can confound our estimates of how rotation benefits vary with weather. The regression was fit on the subsample of points where the two year sequence was either A$\to$B or B$\to$B and the estimated propensity score $\hat{\pi}(X)$ was between $0.05$ and $0.95$. Instead of implementing this regression directly, we implemented the regression using the \texttt{best\_linear\_projection} function in the \texttt{grf} package \cite{GrfPackage} which provided standard errors and $p$-values that appropriately accounted for the fact that the same sample was being used to estimate the response variable ($\omega(X)$) and to estimate the regression coefficients themselves. The \texttt{best\_linear\_projection} function also appropriately accounted for the large errors (associated with nonparametric estimation) in the estimates of $\omega(\cdot)$ when computing standard errors and $p$-values of the regression coefficients in \eqref{eq:ModelToAssessHeterogeneity}. In addition, we used the default setting where standard errors and $p$-values for the regression coefficients were based on heteroskedasticity-robust (HC3) estimation, although they did not account for potential spatial correlations (see Section \ref{sec:AccountForSpatialCorrelations}). We refer readers to the documentation of the \texttt{grf} package \cite{GrfPackage} and \cite{SemenovaChernozhukovTheoryForBestLinearProjection} for more details on how this regression was fit using an alternative regression that appropriately accounts for errors in estimation of $\omega(X)$ and to \cite{MACKINNONandWhiteHC3} for more details on how HC3 standard errors are calculated in regression settings.

We subsequently rescaled the estimated regression coefficients for $\beta_{\text{Precip}}$ and $\beta_{\text{Temp}}$ to be on a more interpretable scale. To do so, we again considered the subsample of points that had a two year crop sequence of either A$ \to$B or B$\to$B and had an estimated propensity score $\hat{\pi}(X)$ between $0.05$ and $0.95$. On this subsample, we computed the 25th percentile and 75th percentile for growing season precipitation denoted by $P_{0.25}$ and $P_{0.75}$ as well as these percentiles for growing season temperatures denoted by $T_{0.25}$ and $T_{0.75}$. Letting $\hat{\beta}_{\text{Precip}}$ and $\hat{\beta}_{\text{Temp}}$ denote the estimated regression coefficients from the regression in Equation \eqref{eq:ModelToAssessHeterogeneity}, our rescaled heterogeneity coefficients for temperature and weather were calculated with $$ \hat{\beta}^{\text{(resc.)}}_{\text{Precip}} = \hat{\beta}_{\text{Precip}} \times (P_{0.75}-P_{0.25})\quad \text{and} \quad \hat{\beta}^{\text{(resc.)}}_{\text{Temp}} = \hat{\beta}_{\text{Temp}} \times (T_{0.75}-T_{0.25}).$$

We repeated this process for each precrop A, outcome crop B, and country listed in Table \ref{table:ResultsTable}. The rescaled heterogeneity coefficients can be roughly interpreted as the amount by which the estimated benefit of rotation on our normalized yield proxy are expected to change as the growing season precipitation (or temperature) increases from the 25th percentile to the 75th percentile in the regions where both the rotation A$\to$B and the no rotation control B$\to$B are commonly observed.

We remark that our approaches do not depend upon linearity assumptions. In spite of our use of a linear model in Equation \eqref{eq:ModelToAssessHeterogeneity} to study heterogeneity with weather, we do not use linear approaches for the initial estimation of $\omega(X)$ or the treatment effects (e.g., \cite{DoublyRobustCILinearModel}) to ensure that our results on the precrop and diversification effects (Figures \ref{fig:TwoYearRotBenefit} and \ref{fig:DiversificationBenefit}; Table \ref{table:ResultsTable} (Columns 6 and 7)) do not lean on any linearity assumptions. Moreover, the population regression coefficients giving the best fit to model \eqref{eq:ModelToAssessHeterogeneity} are still well defined estimands of interest regardless of whether or not $\omega(\cdot)$ is linear in temperature and precipitation. Regardless of whether $\omega(\cdot)$ is linear, the estimands are the well-defined solutions to $$\argmin_{b_0,b_1,b_2,b_3 \in \mathbb{R}} \e \big[ ( \omega(X)- b_0 +b_1 t  + b_2 P+ b_3 T)^2 \big],$$ and moreover, such estimands still provide useful summary statistics about how $\omega(X)$ varies with precipitation and temperature.
We use linear approaches to study heterogeneity in $\omega(\cdot)$ in part because doing so allows valid $p$-value and confidence interval construction. In particular, the \texttt{best\_linear\_projection} function we use returns confidence intervals that account for the fact that $\omega(\cdot)$ is not known precisely and was estimated from the same sample, whereas existing software for nonlinear approaches (e.g., partial dependence plots, accumulated local effect plots, Shapley values) do not return such confidence intervals.

\subsection{Assessment of diverse crop rotations}\label{sec:RotDiversityMethodDescription}

For each two year crop sequence A$\to$B with A and B being distinct, we also assessed whether additional rotational diversity benefits crop yields. To do this, we considered samples with a three year cropping sequence of B$\to$A$\to$B as simple rotation control units and set $Z=0$ on such samples. All samples with a three year cropping sequence of D$\to$A$\to$B, where D is any croptype that is distinct from A and B were considered as treated units with $Z=1$ (we remark that samples of the form D$\to$A$\to$B where D was fallow, pasture or some other non-crop land classification were not included as treated units and were removed from this analysis (see Appendix \ref{sec:nonCropDefinitions})). With a binary treatment variable $Z$ distinguishing simple and diverse rotations and an outcome variable $V$ giving the GCVI-based proxy (defined in Section \ref{sec:SPH_GCVI}) for B's yield in the third year of the crop sequence, we ran a causal forest and estimated the ATO using the same software and weather controls described in Section \ref{sec:CausalForestAnalysis}. The ATO estimates provided estimates of the benefit of additional rotational diversity compared to a less diverse B$\to$A$\to$B cropping sequence. While there are many possible diverse rotations with different choices of crops to be grown prior to the A$\to$B sequence that can be studied, our ATO estimate gave a weighted average of the diversification benefit across all the crops distinct from A and B that are grown before the A$\to$B sequence (with weights proportional to the prevalence in our sample). Finally, as a supplementary analysis, we used the same approach described in Section \ref{sec:HeterogeneityAnalysisMethod} to study how the impact of switching from $Z=0$ (a simplified sequence of the form B$\to$A$\to$B) to $Z=1$ (a diversified sequence of the form D$\to$A$\to$B) varied with weather.

We remark that in China, we did not attempt to estimate the additional effect of rotational diversity for two reasons. First, our crop type classifications from \cite{YouChinaCropMap} only contained the categories of maize, soybean, rice, or other, with the other category not distinguishing between crops and noncrops (e.g., fallow or pastoral land). Second, there were only three years of crop sequence data in China, so had we used the above approach it would essentially have been a cross-sectional study, with all of the complete data coming from the same year.

\subsection{Converting results to the crop yield scale}\label{sec:ConvertingToYieldScale}

Our estimated effects of rotation and diversification described in Sections \ref{sec:CausalForestAnalysis} and \ref{sec:RotDiversityMethodDescription} and our rescaled heterogeneity coefficients described in Section \ref{sec:HeterogeneityAnalysisMethod} each used normalized peak GCVI as the outcome variable. To improve the interpretability of our results, we converted these estimated effects and heterogeneity coefficients from a unitless GCVI scale to the crop yield scale by using a post-hoc, linear calibration with subnational-level crop yield data.

In particular, to calibrate our results to the yield scale in the United States, for each crop type $j$, county $c$, and year $t$ we computed the average normalized peak GCVI, which we call $\bar{V}_{c,t}^{(j)}$ using the sample described in Section \ref{sec:DatasetAndSample}. We then extracted annual county-level crop yield data from the National Agricultural Statistics Service of the United States Department of Agriculture \cite{NASSQuickStats}, which we call $\bar{Y}_{c,t}^{(j)}$. Then for each crop type $j$, we fit the following weighted linear regression to estimate calibration coefficients \begin{equation}\label{eq:VItoYieldCalibration}
\bar{Y}_{c,t}^{(j)}= \alpha^{(j)} +\lambda^{(j)} \bar{V}_{c,t}^{(j)}+ \varepsilon_{c,t}^{(j)},    
\end{equation} where the weights were proportional to the number of samples in our dataset with crop type $j$ from county $c$ and year $t$. We let $\hat{\alpha}^{(j)}$ and $\hat{\lambda}^{(j)}$ be the estimated coefficients from the above regression and further let $\bar{Y}^{(j)}$ be the weighted average of the county-level yields of crop type $j$ across each county and year (using the same weights as used in the linear regression). For each outcome crop $j$, we then converted all of our estimates for rotation effects, heterogeneity coefficients, and diversification effects from the normalized peak GCVI scale to the crop yield scale by multiplying these estimates by $100 \times \hat{\lambda}^{(j)}/\bar{Y}^{(j)}$. We also converted the standard errors of these estimates to the crop yield scale by multiplying the standard errors by $100 \times \hat{\lambda}^{(j)}/\bar{Y}^{(j)}$.

Due to lack of availability of sufficiently high resolution subnational-level yield data in China, we were unable to deploy the above approach in each country. Instead, we used the scaling factors of $100 \times \hat{\lambda}^{(j)}/\bar{Y}^{(j)}$ learned in the US to convert our results to the yield scale (in units of percent of average yield) in each country for each outcome crop $j$. We used subnational-level yield data in Canada and France to validate this approach and found that subnational-level yield predictions in Canada and France that were based on coefficients from a calibration fit in the US  were only slightly less accurate than those based on a calibration fit with local (i.e., from Canada or France) subnational-level yield data (see Appendix \ref{sec:TransferabilityOfUSYieldCalib} for details).

We remark that we used weighted regression in our calibration to avoid giving too much weight to counties and years that were underrepresented in our US data sample. We also note that we rescale our estimates and standard errors by a factor of $100 \times \hat{\lambda}^{(j)}/\bar{Y}^{(j)}$ rather than by $\hat{\lambda}^{(j)}$ for two reasons. First, the factor that we use converts our estimated average effects (or estimated average changes in effects with weather) to be in units of percent of average yield rather than units of tons per hectare. The former units have been used in recent work studying the impacts of tillage on crop yield \cite{Tillage_CausalForest} and are more interpretable in comparisons across outcome crops. Second, dividing by $\bar{Y}^{(j)}$ leads to a more robust calibration in the sense that the calibration learned in the United States using $\hat{\alpha}^{(j)}/\bar{Y}^{(j)}$ and $\hat{\lambda}^{(j)}/\bar{Y}^{(j)}$ can more reliably be used in other countries than can a calibration using just $\hat{\alpha}^{(j)}$ and $\hat{\lambda}^{(j)}$ (Appendix \ref{sec:TransferabilityOfUSYieldCalib}). Finally, we remark that  interested readers can convert our results presented in units of percent of average yield to units of tons per hectare using the average yield estimates for our sample presented in Table \ref{Table:YieldToTonsPerHectare}.

\subsection{Net effects versus direct effects terminology}\label{sec:NetVSDirect}

Throughout the text, we present estimates of the net effects of cropping sequences on yield and how those net effects vary with weather. The net effects consist of the direct effect of rotation on crop yield as well as the presumably negative indirect effect of crop rotation on yield that is mediated by changes in chemical input use. Throughout the text, in cases where it is important to distinguish between the net effects on yields that we estimate and the direct effects on yields (i.e., the cropping sequence effects at fixed chemical input levels) that are often studied in field experiments, we refer to the latter as ``direct". The net precrop (or diversification) effects we estimate are essentially the precrop (or diversification) effects on yield when the treated units (i.e., the rotated (or diversified) units) are defined to also use their actual reduction in chemical inputs compared to the control units (i.e., the nonrotated (or non-diversified) units). The precrop (and diversification) effects we study are causal quantities of interest, but we note that they only capture the yield benefits of rotation and do not capture additional benefits associated with reduced chemical input costs. 

\section*{Supplementary Materials: Appendices}

\section{Discussion of Limitations}\label{sec:LimitationDiscussion}

The causal inference approach used relies upon the assumption that there are no unmeasured confounders. While we controlled for potential confounders such as weather and geolocation, we did not control for soil quality in our analysis. It is possible that on more fertile land farmers are less likely to rotate their crops, because in such land high crop yields can be obtained without rotation. In Appendix \ref{sec:AddingSoilCovariatesSensitivityCheck}, we find that adding soil covariates from the USDA Soil Survey Geographic Database \cite{SSURGO} as controls in the model barely changed our results. This sensitivity check was not conducted in Canada, China, and France due to lack of soil data, but our findings in the United States suggest that omission of soil covariates does not substantially bias our causal effect estimates and our estimates of how they vary with weather. 

A second potential concern is that fertilizer and pesticide use are unmeasured mediators that influence crop yields, and therefore we do not study the direct impacts of rotation or diversification on crop yield at fixed chemical input levels. Other unmeasured variables include tillage, cover cropping, and residue management, all of which can influence the effectiveness of crop rotation. However, it is less clear whether such practices should be considered unmeasured mediators or unmeasured confounders. Our estimates of the net benefits of rotation and diversification on crop yields very likely underestimate the direct benefits of rotation. In particular, fertilizer is well-known to improve crop yields and farmers in the US that rotate their crops are recommended to use less fertilizer for economic reasons \cite{sawyerNitrogenGuidelines}. Similarly, farms in France that have more crops rotated with winter wheat also tend to use less phytosanitary products \cite{FranceLongerRotationsLessPesticide}. Therefore, we expect the omission of fertilizer and pesticide as mediator variables to lead to conservative underestimates of the direct rotation benefits that are often studied in randomized field experiments. Indeed, a recent metanalysis of matanalyses on various farm management practices found that crop rotation led to a median increase of 16\% on the production level of the subsequent crop \cite{BeillouinMetaMetaAnalysis}, whereas the median estimated net rotation benefit on yield across all 36 precrop, outcome crop and country combinations in our study was 4.5\%. 

Nonetheless, our estimates for the benefits of rotation are still useful for understanding the effects of rotating crops on crop yield when using the typical reduction in fertilizer and pesticide use for a given rotation and region. For example, we still find a positive benefit of crop rotation even though less fertilizer and pesticide inputs are typically used in rotated fields. While our estimates do not capture the cost savings from using reduced chemical inputs, they can be coupled with work assessing the impact of rotation on chemical input use and how those impacts vary with weather to obtain a more complete picture of the comprehensive benefits of rotation and how they vary with weather.  Furthermore, as shown mathematically in Appendix \ref{sec:HeterogeneityAnalysisWithUnmeasuredConfounders}, even when there are unmeasured mediators such as fertilizer and pesticide use, our estimates of how the rotation benefits vary with weather could (under certain assumptions) still give unbiased estimates of how the direct rotation benefits vary with weather.

Third, the outcome and treatment variables used do not perfectly capture the actual outcome and treatment variables of interest. The outcome variable is a remotely sensed greenness measure that is merely a proxy for crop yield. In Appendix \ref{sec:VI_versus_SCYM}, we demonstrate that even though we do not use actual yield estimates as the outcome variable, our approach of using a vegetation index-based outcome variable coupled with an appropriate selection of weather covariates would give the same results, up to an unknown scaling factor, as the observational approach taken in recent works that assess the impacts of crop rotation \cite{CREO_paper} and tillage \cite{Tillage_CausalForest} on crop yield. Our rescaling of the estimated effects (Section \ref{sec:ConvertingToYieldScale}) mitigates the discrepancy between our approach and those taken in \cite{CREO_paper,Tillage_CausalForest} by approximating the unknown scaling factor using subnational-level crop yield data. In Appendix \ref{sec:QDANNcomparison}, we find that even when using modern, satellite-based yield estimates \cite{QDANNPaper} instead of a vegetation index-based proxy as the outcome variable, the results still roughly differ by an unknown scaling factor, and moreover, for most rotations, county-level regressions of average estimated yield on the average of the proxy gave good approximations of the unknown scaling factors.
In addition, in Appendix \ref{sec:UsingNIRvInsteadOfGCVI}, we find that our results were robust to the choice of vegetation index that was used as a yield proxy. 

A similar issue is that the treatment variable was determined based on crop type maps, which for some crop types and countries are somewhat inaccurate. In causal forest-based analyses, misclassification errors in the treatment variable can lead to attenuation bias \cite{CREO_paper,lewbel2007estimation} under the assumption that classification errors are independent of the outcome variable conditional on the actual treatment variable and the controls used in the causal forest. As with the issue of unmeasured fertilizer and pesticide use, measurement error would therefore likely lead to overly conservative estimates of rotation benefits. 

Fourth, our estimates for rotation benefits are not perfectly comparable across precrop choices. This is because even within the same country, we only estimated the effect of a rotation on the overlap region where both the treatment and control were commonly observed. For example, when comparing the Soybean$\to$Corn rotation to the Winter Wheat$\to$Corn rotation in the US, the estimates for the former are for parts of the US where both Soybean$\to$Corn and Corn$\to$Corn are commonly observed whereas the latter are for parts of the US where both Winter Wheat$\to$Corn and Corn$\to$Corn are commonly observed. A similar issue is that the rotation benefits cannot be directly compared across countries even for the same precrop and outcome crop choices. In particular, the analyses in each country corresponded to a different time period based on crop type map availability (Table \ref{Table:CropMapInfo}), so observed differences in the rotation benefits between two countries may be partially explained by climatic differences between their study periods. Moreover, the analyses in different countries used crop type maps with different accuracies. For example, we expect rotation benefit estimates from France to suffer from the least attenuation bias compared to those in other countries due to the higher accuracy of crop maps in France (unlike in the US, Canada, and China the crop type maps in France were not based on remote sensing predictions).

Fifth, unlike with metanalyses of long-term field experiments, our study is unable to assess the long-term versus short-term benefits of crop rotation or quantify how the benefit of crop rotation varies with fertilizer use. The benefits of some crop rotations may grow over time \cite{EuropeERL_WeatherInteractions7LTE,SmithEtAlFromLabMeeting}, likely due to soil health improvements. Recent metanalyses are also able to quantify how the benefit of rotation varies with fertilizer use \cite{LegumePrecropMetanalysis,SmithEtAlFromLabMeeting}. Without access to fertilizer data, and with a shorter study duration, we do not assess how the rotation benefits vary with fertilizer use or with the number of years that the rotation had been in place. 

Sixth, our analysis only considers the impact of rotation on one crop in a crop sequence rather than all crops in a crop sequence. Ultimately, farmers make decisions about which rotations to use based on many factors, including the expected yield and price of each of the crops in the rotation sequence. 
Our results concerning heterogeneity with weather cannot always tell us whether a cyclical rotation will be increasingly beneficial or decreasingly beneficial (when aggregating across all crops and years in the cycle) as the climate changes. For example, our results suggest that a 2-year rotation alternating between spring wheat and soybean in Canada would be more beneficial for soybean yields but less beneficial for spring wheat yields in a warming climate. We deem questions about the benefits of cyclical rotations on all crops in the rotation cycle and the heterogeneity of such benefits with weather beyond the scope of the current study. We leave such questions for future work and note that our results and our satellite-based causal machine learning approach can be used to answer such questions and can be used to build upon existing crop rotation decision support systems \cite{DecisionSupportROTAT,DecisionSupportROTOR,DecisionSupportCropRota,DecisionSupportFinland,DecisionSupportFruchtfolge}.

\section{Robustness Checks}

\subsection{Adding Soil Covariates in the United States}\label{sec:AddingSoilCovariatesSensitivityCheck}

In this appendix, we check whether our results from the United States are sensitive to the inclusion of soil covariates. Soil quality is a potential confounder that can influence both crop yield and rotation choices and was controlled for in previous studies that assessed the impact of crop rotation on crop yield using observational data \citep{CREO_paper,BealCohenObservationalRotUS}. In the United States, we considered 5 soil covariates taken from the USDA Soil Survey Geographic Database (SSURGO) \cite{SSURGO}: rootzone available water storage, available water storage in the top meter, and the National Commodity Crop Productivity Indices for corn, soybean and all crops. The first four of these soil covariates were used in \cite{CREO_paper}. These soil covariates were static variables measured once at each site and did not vary from year to year. In the other 3 countries in our study, we did not have access to a soil database, and therefore, our robustness check was limited to the United States.

\begin{figure}[!t]
    \centering
    \includegraphics[width=0.95 \hsize]{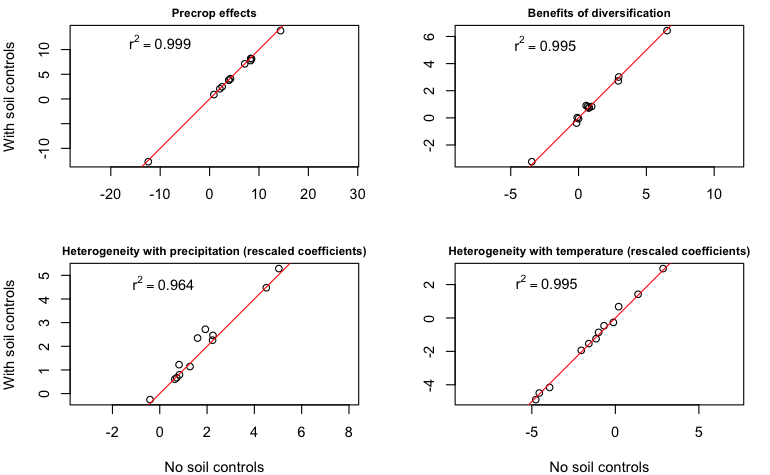}
    \caption{Sensitivity check of whether including soil covariates influences our results in the United States. In each scatter plot, each point corresponds to one of the 12 outcome crop and precrop pairs in the United States that are considered in our study. Each scatter plot gives the estimates for a quantity of interest when using soil covariates (y-axis) versus without using soil covariates (x-axis). The scatter plots depict estimates of the precrop effects (top left) calculated using the method in Section \ref{sec:CausalForestAnalysis}, the benefits of diversification (top right) calculated using the method in Section \ref{sec:RotDiversityMethodDescription}, and the rescaled heterogeneity coefficients for precipitation (bottom left) and temperature (bottom right) based on the method in Section \ref{sec:HeterogeneityAnalysisMethod}. All plotted quantities were converted from the unitless normalized peak GCVI scale to the crop yield scale (in units of percent of the mean crop yield for the corresponding outcome crop in the US) using the approach described in Section \ref{sec:ConvertingToYieldScale}. The red reference lines go through the origin and have slope equal to 1, and the text on the plot gives the coefficient of determination in the univariate regression between the plotted quantities.}
    \label{fig:WithVersusWithoutSoil}
\end{figure}

We found that including soil covariates as controls did not substantially influence the results from the causal forest-based analyses in the United States. Figure \ref{fig:WithVersusWithoutSoil} shows scatter plots of the results from our causal forest-based analyses when excluding soil covariates from the control vector $X$ (as is done in the main text of the paper) versus when including the 5 aforementioned soil covariates in the control vector $X$. As can be seen in the figure, the results do not change much when including soil covariates in the control vector. In addition, in nearly all estimates visualized in Figure \ref{fig:WithVersusWithoutSoil}, controlling for soil covariates did not change whether the 95\% confidence interval for that estimate lies above $0$, below $0$, or contains $0$. The only exception is that for the Spring Wheat$\to$Soybean rotation, the confidence interval for the temperature heterogeneity coefficient ($\beta_{\text{Temp}}$ from Equation \eqref{eq:ModelToAssessHeterogeneity}) lies above $0$ when controlling for soil covariates but contains $0$ when not controlling for soil covariates.

\subsection{Using a different vegetation index than GCVI}\label{sec:UsingNIRvInsteadOfGCVI}

In this appendix we check whether our results were sensitive to the choice of using peak GCVI as a proxy for yield. In particular, we repeated the analyses described in Sections \ref{sec:SPH_GCVI}--\ref{sec:ConvertingToYieldScale}, except we used the near infrared reflectance of vegetation (NIRv) as the vegetation index rather than GCVI. NIRv, a vegetation index developed in \cite{NIRvPaper}, is the product of the Normalized Difference Vegetation Index and the NIR band, and NIRv has shown strong performance for predicting crop yields \cite{ChinaNIRvGoodPerformance}. For this senstivity analysis, after removing observations with outlier values for peak harmonic NIRv, we set normalized peak harmonic NIRv to be our outcome variable and ran the causal forest analyses as described in Sections \ref{sec:CausalForestAnalysis}--\ref{sec:RotDiversityMethodDescription}. We then converted our results from the normalized NIRv scale to the crop yield scale using the approach described in Section \ref{sec:ConvertingToYieldScale} (by setting $\bar{V}_{c,t}^{(j)}$ to be the average normalized peak NIRv for crop type $j$ in each year $t$ and subnational unit $c$ when fitting the regression in Equation \eqref{eq:VItoYieldCalibration}).  We remark that when fitting the calibration model in Equation \eqref{eq:VItoYieldCalibration}, including with subnational-level yield data in France and Canada (Section \ref{sec:TransferabilityOfUSYieldCalib}), peak GCVI generally had a higher correlation with the subnational-level crop yield data  than did peak NIRv, and thus GCVI-based results are presented in the main text rather than NIRv-based results.

\begin{figure}[!t]
    \centering
    \includegraphics[width=0.95 \hsize]{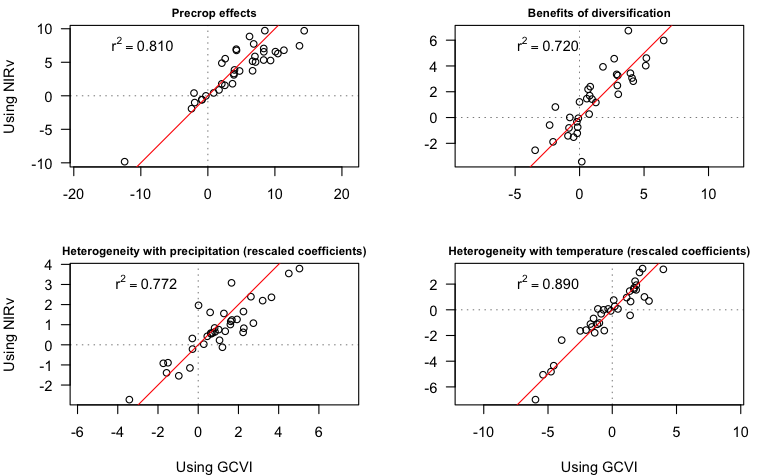}
    \caption{Sensitivity check of whether using NIRv instead of GCVI influences our results. In each scatter plot, each point corresponds to one of the 36 outcome crop, precrop, and country combinations that are considered in our study. Each scatter plot gives the estimates for a quantity of interest when using NIRv (y-axis) versus when using GCVI (x-axis). The scatter plots depict estimates of the precrop effects (top left) calculated using the method in Section \ref{sec:CausalForestAnalysis}, the benefits of diversification (top right) calculated using the method in Section \ref{sec:RotDiversityMethodDescription}, and the rescaled heterogeneity coefficients for precipitation (bottom left) and temperature (bottom right) based on the method in Section \ref{sec:HeterogeneityAnalysisMethod}. All plotted quantities were converted to the crop yield scale (in units of percent of the mean crop yield for the corresponding outcome crop in the corresponding region) using the approach described in Section \ref{sec:ConvertingToYieldScale}. The red reference lines go through the origin and have slope equal to 1, and the text on the plot gives the coefficient of determination in the univariate regression between the plotted quantities.}
    \label{fig:NIRvVersusGCVI}
\end{figure}

We found that using GCVI as a vegetation index rather than NIRv did not substantially influence the results from the causal forest-based analyses. Figure \ref{fig:NIRvVersusGCVI} shows scatter plots of the results from our causal forest-based analyses when using GCVI as the vegetation index (as is done in the main text of the paper) versus when using NIRv as the vegetation index. As can be seen in Figure \ref{fig:NIRvVersusGCVI}, there is generally good agreement between estimates based on GCVI and those based on NIRv. Although there are some disagreements in the estimates, we remark that these disagreements do not substantially change the main GCVI-based conclusions presented in the paper. In particular, none of our estimates change from being positive and statistically significant to being negative and statistically significant (or vice-versa) when switching from GCVI to NIRv. Additionally, the statistical significance and sign of the precrop effect estimates were the same for all 36 precrop, outcome crop, and country combinations with the exception of the two rotations in Canada with ``Pasture and Forages" as the precrop and the Winter Wheat$\to$Corn rotation in France. In these three cases, the two year sequence was estimated to have a statistically significant, negative effect on yield in the main text, but the estimates were rendered non-statistically significant when using NIRv instead of GCVI. 

\subsection{Converting outcome variable to the log scale}\label{sec:HeterogeneityWithWeatherLogScale}

We repeated our analyses when using the log of the estimated yield as an outcome variable rather than using a greeness proxy as the outcome variable and subsequently converting the results to the yield scale. This enabled us to check if our findings about heterogeneity with weather were merely driven by the impacts of weather on yield or greenness rather than by fundamental differences in the effectiveness of crop rotation under different weather conditions. For example, in the analysis in the main text we found that at higher precipitation the benefit of rotation tended to be greater. However, because benefit was measured in an absolute sense rather than in a relative sense, the findings in the main text did not alone rule out the possibility that the crop yield was higher at higher precipitation levels while the percent effect of rotation on crop yield did not vary with precipitation.

In particular we used the fitted coefficients in the model in Equation \eqref{eq:VItoYieldCalibration} relating normalized harmonic peak GCVI to subnational-level yield data in order to estimate normalized crop yields. We let $\tilde{Y}_{\text{normalized}}$ denote these normalized estimates of crop yield and reran the causal forest analyses when $\log(\tilde{Y}_{\text{normalized}})$ was the outcome variable. Note that we are more interested in the non-normalized yield estimates given by $\tilde{Y}=c \tilde{Y}_{\text{normalized}}$ for some (possibly unknown) constant $c$ but the constant $c$ is irrelevant when using $\log(\tilde{Y}_{\text{normalized}})$ as an outcome variable in a causal forest analysis.

Taking the log of the outcome variable allows us to study the relative rather than absolute rotation benefits. To see this, note that when $\log(\tilde{Y}_{\text{normalized}})$ is the outcome variable, the causal forest learns a function to estimate the following function $$\begin{aligned} \omega_{\log}(x) & \equiv \ \e[\log(\tilde{Y}_{\text{normalized}})|X=x,Z=1]-\e[\log(\tilde{Y}_{\text{normalized}})|X=x,Z=0] 
\\ & =\e[\log(\tilde{Y}) -\log(c)|X=x,Z=1]-\e[\log(\tilde{Y})-\log(c)|X=x,Z=0] 
\\ & = \e[\log(\tilde{Y})|X=x,Z=1]-\e[\log(\tilde{Y})|X=x,Z=0].\end{aligned}$$ Letting $\tilde{Y}(1)$ and $\tilde{Y}(0)$ denote the potential outcomes for the (estimated) yield under rotation versus no rotation, if we assume there are no unmeasured confounders, $$\begin{aligned}\omega_{\log}(x) & = \e \big[ \log \big(\tilde{Y}(1) \big) \big|X=x \big] -\e \big[ \log \big(\tilde{Y}(0) \big) \big|X=x \big]
\\ & = \e\Big[\log \Big(1+\frac{\tilde{Y}(1)-\tilde{Y}(0)}{\tilde{Y}(0)} \Big) \big| X=x \Big]. \end{aligned}$$ Because $\log(1+r) \approx r$ for $r$ near $0$, the above quantity on the right hand side is roughly the average percent effect of rotation on the yield estimates (divided by a factor of 100) written as a function of the covariates in $x$. Therefore under the assumption of no unmeasured confounders, $\omega_{\log}(x)$ can roughly be thought of as the average percent effect of rotation on yields, written as a function of the covariates in $x$. Similar to the analyses in Sections \ref{sec:CausalForestAnalysis} and Sections \ref{sec:RotDiversityMethodDescription} we use overlap weighted means of $\omega_{\log}(X)$ to estimate average rotation benefits and diversification benefits (with different choices of $Z$ for each estimate). Similar to the analysis in Section \ref{sec:HeterogeneityAnalysisMethod}, we assessed heterogeneity with weather on the sample where the propensity scores were between 0.05 and 0.95 by using the \texttt{best\_linear\_projection} function in the \texttt{grf} package \cite{GrfPackage} to fit the following modification of the model in equation \eqref{eq:ModelToAssessHeterogeneity} \begin{equation}\label{eq:ModelToAssessHeterogeneity_logScale}
        \omega_{\log}(X) = \beta_0 +\beta_{\text{Year}} t  + \beta_{\text{Precip}} P+\beta_{\text{Temp}} T + \varepsilon.
\end{equation}  We then rescaled the fitted estimates for $\beta_{\text{Precip}}$ and $\beta_{\text{Precip}}$ using the same approach as in Section \ref{sec:HeterogeneityAnalysisMethod}. 

\begin{figure}[!t]
    \centering
    \includegraphics[width=0.95 \hsize]{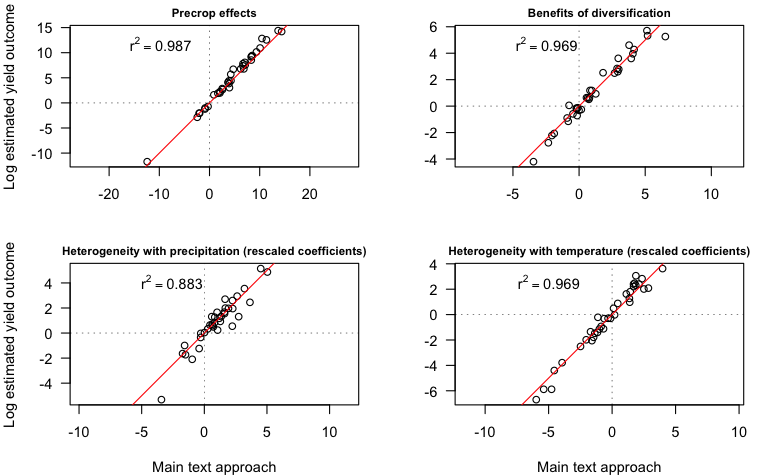}
   \caption{Sensitivity check of whether using the log of the estimated yield as the outcome variable influences our results. In each scatter plot, each point corresponds to one of the 36 outcome crop, precrop, and country combinations that are considered in our study. Each scatter plot gives the estimates for a quantity of interest when using the log of the estimated yield as the outcome variable  (y-axis) versus when using the approach in the main text which uses normalized peak GCVI as the outcome variable and then converts to the yield scale (x-axis). The y-axis is in units of percent while the x-axis is units of precent of the mean yield of the corresponding outcome crop in the corresponding study region. The scatter plots depict estimates of the precrop effects (top left) calculated using the method in Section \ref{sec:CausalForestAnalysis}, the benefits of diversification (top right) calculated using the method in Section \ref{sec:RotDiversityMethodDescription}, and the rescaled heterogeneity coefficients for precipitation (bottom left) and temperature (bottom right) based on the method in Section \ref{sec:HeterogeneityAnalysisMethod}. The red reference lines go through the origin and have slope equal to 1, and the text on the plot gives the coefficient of determination in the univariate regression between the plotted quantities.}
    \label{fig:logEstimatedYieldOutcome}
\end{figure}

Our estimates from these analyses using log estimated yield as the outcome variable were multiplied by a factor of 100 and are plotted in Figure \ref{fig:logEstimatedYieldOutcome} against our estimates from the main text. We find a high agreement between the two approaches meaning that our reported estimates can also be interpreted as relative effects rather than just absolute effects. Notably, this suggests that our finding of higher rotation benefits at higher precipitation levels was not merely driven by greater crop productivity at higher preciptation levels and was instead driven by fundamental increases in the effectiveness of crop rotation in rainier conditions.

\subsection{Controlling for geography}\label{sec:HeterogeneityWithWeatherGeographyControls}

To check whether our findings regarding the heterogeneity of rotation benefits with weather were driven by variations in weather across geographical locations rather than variations in weather within geographical locations, we considered modifications of the linear model in Equation \eqref{eq:ModelToAssessHeterogeneity}, with fixed effects that control for geography. In our first robustness check, we included a fixed effects term for the level-1 administrative unit. 
In our second robustness check, we partitioned the sample into 500 km $\times$ 500 km grid cells, and included fixed effects for the grid cells. This second robustness check had more fixed effect terms (except in France where the first robustness check had 2 more fixed effects terms). 
In our third robustness check, we included fixed effects terms for the level-2 administrative unit, which had even more fixed effect terms.

To conduct these robustness checks, we first fit the same causal forests with the approach described in Section \ref{sec:CausalForestAnalysis} and used the same seed for random number generation that was used for the analysis presented in the main text. We then used a modified version of the \texttt{best\_linear\_projection} function in the \texttt{grf} package \cite{GrfPackage} to fit each of the following three modifications of the model in Equation \eqref{eq:ModelToAssessHeterogeneity} \begin{equation}\label{eq:ModelToAssessHeterogeneity_StateFE}
        \omega(X) = \beta_0 +\beta_{\text{Year}} t  + \beta_{\text{Precip}} P+\beta_{\text{Temp}} T + c_{\text{State}} + \varepsilon, \quad \text{       }
\end{equation}  \begin{equation}\label{eq:ModelToAssessHeterogeneity_GridCellFE}
        \omega(X) = \beta_0 +\beta_{\text{Year}} t + \beta_{\text{Precip}} P+\beta_{\text{Temp}} T + c_{\text{Grid-cell}} + \varepsilon, \quad \text{and}
\end{equation} \begin{equation}\label{eq:ModelToAssessHeterogeneity_CountyFE}
        \omega(X) = \beta_0 +\beta_{\text{Year}} t  + \beta_{\text{Precip}} P+\beta_{\text{Temp}} T + c_{\text{County}} + \varepsilon, \quad \text{       }
\end{equation} where $c_{\text{state}}$, $c_{\text{grid-cell}}$, and $c_{\text{County}}$ are fixed effects for the level-1 administrative unit, the grid cell, and the level-2 administrative unit, respectively. As in the main text, these regressions were fit on the subset of samples where the estimated propensity score was between $0.05$ and $0.95$. Note that under the hood, the \texttt{best\_linear\_projection} function in the \texttt{grf} package fits a linear regression of ``doubly-robust scores" on the inputted covariates using the \texttt{lm} function, and in so doing constructs confidence intervals for the regression coefficients that appropriately account for the fact that the same sample was used to estimate the response variable $\omega(X)$ and appropriately account for the errors in the nonparametric estimates of the function $\omega(\cdot)$ \cite{GrfPackage,SemenovaChernozhukovTheoryForBestLinearProjection}. We modified the \texttt{best\_linear\_projection} to fit a fixed effects regression (using the \texttt{feols} function in the \texttt{fixest} package \cite{FixestPackage}) where the outcome variable was still the doubly-robust scores. This allowed us to use fixed effects that control for geography, while still constructing confidence intervals that appropriately accounted for the fact that the same sample was used to estimate the response variable $\omega(X)$ and appropriately accounted for the errors in the nonparametric estimates of the function $\omega(\cdot)$.

\begin{figure}[!t]
    \centering
    \includegraphics[width=0.95 \hsize]{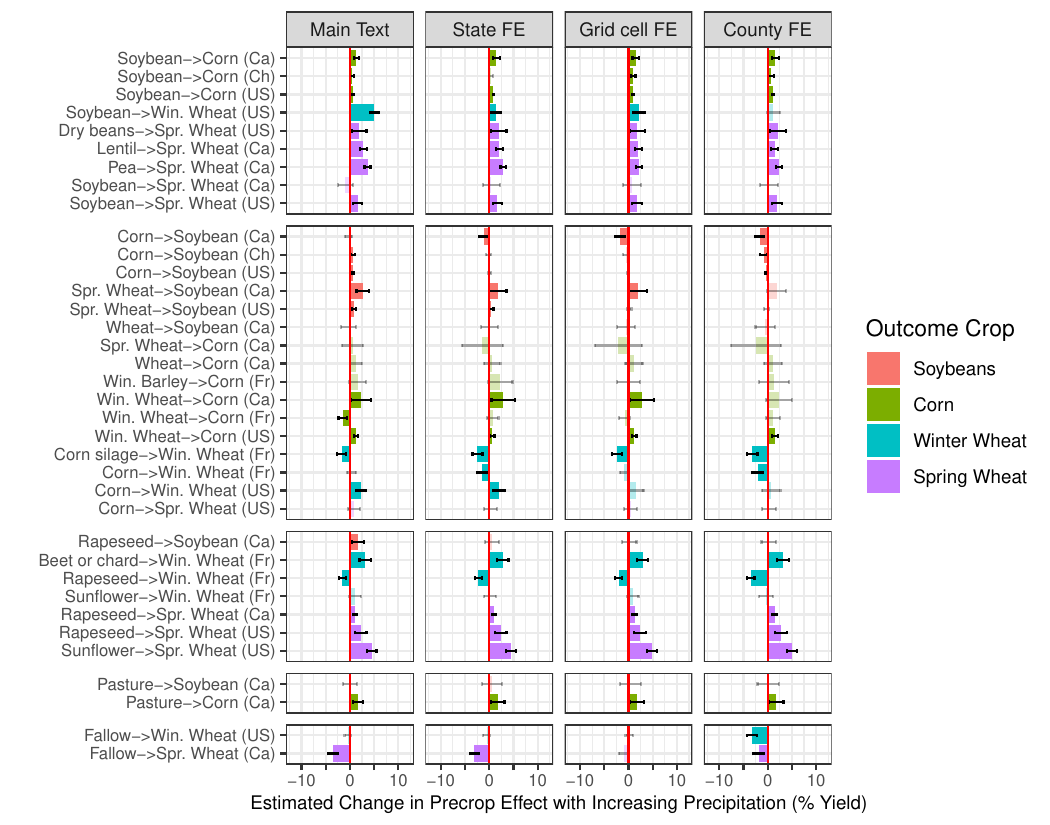}
    \caption{Sensitivity analyses for the estimates of the heterogeneity of precrop effects with precipitation. Each column gives estimates of the rescaled precipitation heterogeneity coefficients where the precipitation heterogeneity coefficient $\beta_{\text{Precip}}$ is calculated using different model specifications. In particular we plot rescaled coefficients from model \eqref{eq:ModelToAssessHeterogeneity} used in main text which does not include geographical fixed effects (first column), model \eqref{eq:ModelToAssessHeterogeneity_StateFE} which has fixed effects for level-1 administrative units (second column),  model \eqref{eq:ModelToAssessHeterogeneity_GridCellFE} which has fixed effects for each 500 km $\times$ 500 km grid cell (third column), and model \eqref{eq:ModelToAssessHeterogeneity_CountyFE} which has fixed effects for level-2 administrative units (fourth column). Each row also gives the 2-year crop sequence, the country from which the study sample was taken (Ca=Canada, Ch=China, Fr=France, US=United States), and for each of the four models, the rescaled regression coefficients which estimate how much the effect of crop rotation on crop yield changes when the precipitation increases from the 25th percentile to the 75th percentile of observed precipitation values. The x-axis values are in units of percent of the mean yield of the corresponding outcome crop in the corresponding region. The error bars give 95\% confidence intervals, and coefficient estimates that are not statistically significant (level $\alpha=0.05$ based on two-sided testing) are given a lighter, translucent color. The rows are grouped into blocks based on 5 precrop categories of interest: legumes, cereals, broadleaf crops, pasture/forages, and fallow.}
    \label{fig:SensitivityCheckHeterogeneityWithPrecip}
\end{figure}

\begin{figure}[!t]
    \centering
    \includegraphics[width=0.95 \hsize]{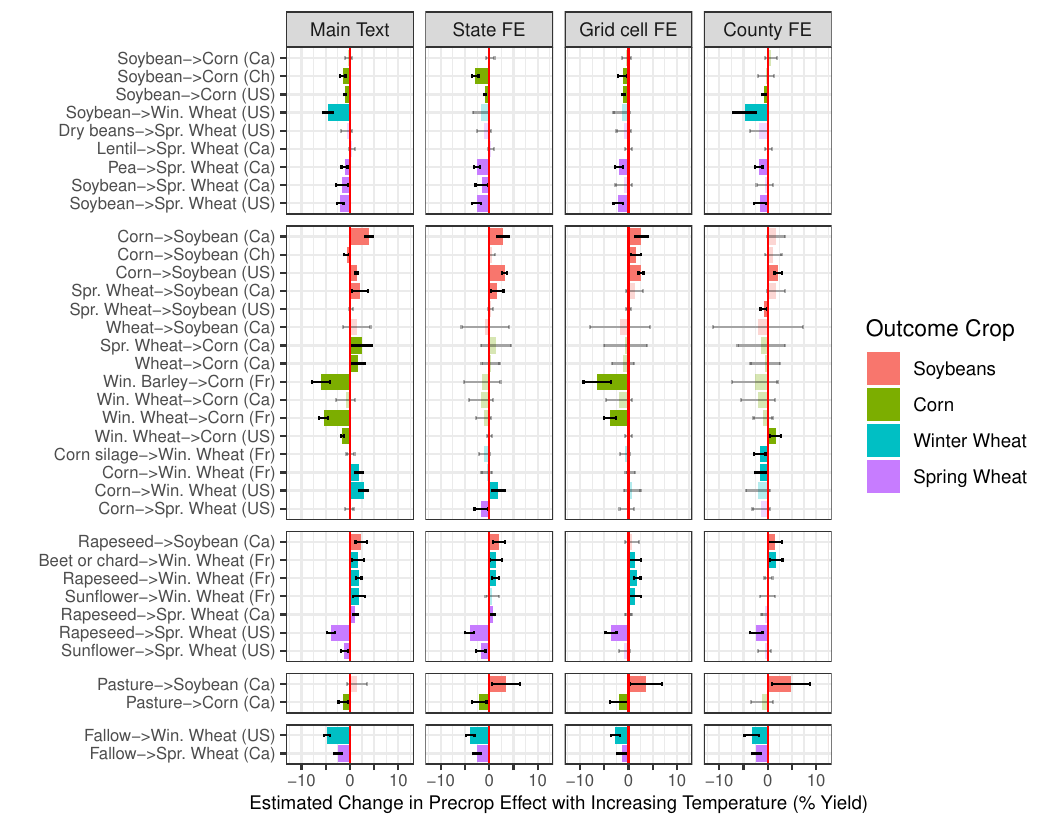}
    \caption{Sensitivity analyses for the estimates of the heterogeneity of precrop effects with temperature. This figure is the same as Figure \ref{fig:SensitivityCheckHeterogeneityWithPrecip}, except it plots estimates of how the precrop effects vary with temperature rather than precipitation. In particular, for each of the four models, we plot the rescaled regression coefficients which estimate how much the benefit of crop rotation on crop yield changes when the temperature increases from the 25th percentile to the 75th percentile of observed temperature values.}
    \label{fig:SensitivityCheckHeterogeneityWithTemp}
\end{figure}

We used the same approach as in Section \ref{sec:HeterogeneityAnalysisMethod} to rescale the fitted estimates of $\beta_{\text{Precip}}$ and $\beta_{\text{Temp}}$ from models \eqref{eq:ModelToAssessHeterogeneity_StateFE}, \eqref{eq:ModelToAssessHeterogeneity_GridCellFE}, and \eqref{eq:ModelToAssessHeterogeneity_CountyFE} and further rescaled these estimates to the yield scale using the approach in Section \ref{sec:ConvertingToYieldScale}. The rescaled coefficient estimates are plotted in the last three columns of Figures \ref{fig:SensitivityCheckHeterogeneityWithPrecip} and \ref{fig:SensitivityCheckHeterogeneityWithTemp}.

\subsection{Accounting for spatial correlations?}\label{sec:AccountForSpatialCorrelations}

In this appendix, we discuss and explore whether it makes sense to account for spatial correlations in the analysis. The samples used in each analysis in the paper were relatively well spread out (Table \ref{table:DistanceBetweenSamples}), with median distance to the nearest sample being at least 1km in every causal forest analysis conducted. This suggests that the choice of whether or not one accounts for spatial correlations may not have much influence on the ultimate results and conclusions.  As a sensitivity check, we test whether clustering the data based on 0.1° $\times$ 0.1° grid cells and calculating standard errors accounting for such clustering leads to much larger confidence intervals. In this sensitivity analysis we find that the confidence intervals increase in size by at most a factor of 1.7.

\begin{table}
 \footnotesize
 \caption{Summary statistics on the proximity of the samples used in the analyses. For each country and outcome crop in the study, Columns 3--5 give summary statistics describing the proximity of the pixels in the sample in the year 2019. The 3rd column gives the median distance to the nearest sample that had the same croptype classification in 2019. Columns 4 and 5, give the average number of nearby sampled pixels that had the same 2019 crop type classifications (based on drawing a circle of radius 1km and 4km, respectively, around each sample). We expect that these summary statistics would be similar in years other than 2019.}\label{table:DistanceBetweenSamples}
\begin{tabular}{llccc}
  \hline
Country & Croptype & Median distance & Average  \# of nearby  & Average \# of nearby  \\ 
 & & to nearest sample (km) & samples (within 1km) & samples (within 4km) \\
  \hline
Canada & Soybean & 1.41 & 0.46 & 4.53 \\ 
  Canada & Corn & 1.56 & 0.42 & 3.99 \\ 
  Canada & Spring Wheat & 1.07 & 0.68 & 6.14 \\ 
  China & Corn & 1.04 & 0.70 & 9.25 \\ 
  China & Soybean & 1.27 & 0.53 & 6.67 \\ 
  France & Winter Wheat & 1.49 & 0.39 & 4.81 \\ 
  France & Corn & 2.19 & 0.25 & 2.61 \\ 
  US & Corn & 1.37 & 0.45 & 5.42 \\ 
  US & Soybean & 1.48 & 0.40 & 4.65 \\ 
  US & Winter Wheat & 1.74 & 0.34 & 3.42 \\ 
  US & Spring Wheat & 1.25 & 0.54 & 5.50 \\ 
   \hline
\end{tabular}
\end{table}

It is well known that spatial correlations in environmental datasets can lead to unrealistically narrow confidence intervals; however, the question of whether to correct for spatial correlations is subtle. Many approaches to accounting for spatial correlations involve picking a type of bandwidth parameter that can be hard to reason about, but arbitrarily picking such a parameter can lead to confidence intervals that are invalid because they are either too small or too large. In addition, it is unclear whether accounting for spatial correlations is warranted when the data is a uniform random sample from the geography of interest (accounting for spatial correlations is certainly warranted when the sampled points are spatially clustered or when the geography of interest expands beyond the area from which the points were sampled). In this paper, we do not account for spatial correlations in the data given that we do not have prior knowledge to inform a bandwidth choice and that with the exception of China (where only data in the northeast was available), our samples are non-clustered random samples from the entire geography of interest. Nonetheless, in this appendix, we explore whether our results would be sensitive to accounting for spatial correlations.

To explore how accounting for spatial correlations impacts the final results, we considered each 0.1° $\times$ 0.1° grid cell to be a unique cluster. We then repeated the causal forest analyses assessing precrop effects (Section \ref{sec:CausalForestAnalysis}), diversification effects (Section \ref{sec:RotDiversityMethodDescription}), and heterogeneity fo precrop effects with weather (Section \ref{sec:HeterogeneityAnalysisMethod}), and set the ``cluster" parameter \texttt{causal\_forest} function based on the latitude and longitude of each sample and these 0.1° $\times$ 0.1° grid cells. The \texttt{average\_treatment\_effect} function used to estimate treatment effects and the \texttt{best\_linear\_projection} used to estimate heterogeneity coefficients then automatically accounted for the clusters when calculating standard errors (as recommended by the \texttt{grf} package documentation \cite{GrfPackage}, we used HC1 standard errors \cite{MACKINNONandWhiteHC3} for the latter for computational reasons and since they are similar to HC3 standard errors in large sample sizes). 

\begin{figure}
    \centering
    \includegraphics[width=0.95\linewidth]{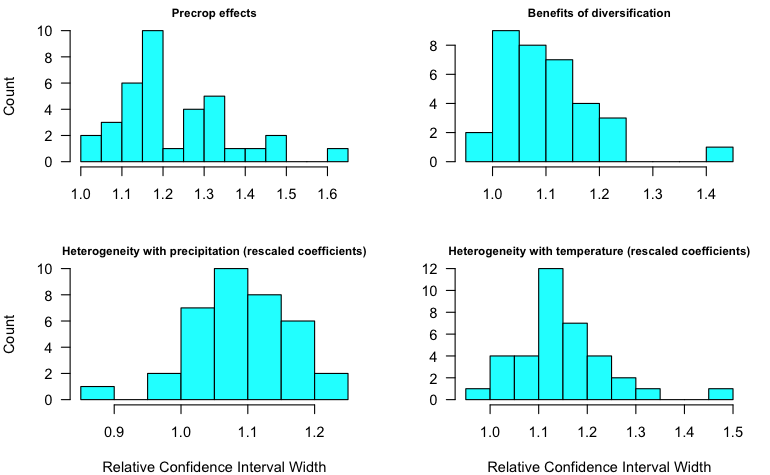}
    \caption{Histograms of how much the confidence intervals of quantities of interest grow when accounting for spatial correlations. The histograms give the ratio of the confidence interval width when clustering standard errors based on 0.1°$\times$0.1° grid cell assignments divided by the confidence interval widths from the main text. Each histogram is of 36 (or 34 (topright)) confidence interval width ratios accross all the precrop, outcome crop, and country combinations studied. The histograms depict the confidence interval ratios for the precrop effects (top left) calculated using the method in Section \ref{sec:CausalForestAnalysis}, the benefits of diversification (top right) calculated using the method in Section \ref{sec:RotDiversityMethodDescription}, and the rescaled heterogeneity coefficients for precipitation (bottom left) and temperature (bottom right) based on the method in Section \ref{sec:HeterogeneityAnalysisMethod}.}
    \label{fig:ClusterSE_CIchanges}
\end{figure}

Using clustered standard errors led to confidence intervals that were generally larger. In Figure \ref{fig:ClusterSE_CIchanges}, we show a histogram of the factors by which the confidence intervals grew when using clustered standard errors (relative to the confidence intervals in the main text). Overall, the confidence intervals did not grow substantially when accounting for spatial correlations using the aforementioned approach. Sometimes the confidence intervals shrunk by a small amount; however, we suspect this phenomenon was due to discrepancies with the random number generation in the causal forest analyses.

\subsection{Grid cell-level difference-in-means precrop effect estimates}\label{sec:DiffInMeanClustered}

In this subsection, we consider an intuitive, heuristic approach for estimating the precrop effects for comparison with the more formal causal forest approach. We restrict our attention to four precrop and outcome crop combinations in the US (Soybeans$\to$Corn, Corn$\to$Soybeans, Soybeans$\to$Spring Wheat, and Soybeans$\to$Winter Wheat).

For each precrop A and outcome crop B considered, we let $Z=1$ for samples (pixel-year pairs) that had the crop sequence A$\to$B and let $Z=0$ for samples (pixel-year pairs) that had the crop sequence B$\to$B, and removed all other samples from the analysis. We then partitioned the map of the US into 1°$\times$1° grid cells. Then for each 1°$\times$1° grid cell and each year that had at least 25 samples with $Z=1$ (corresponding to an A$\to$B rotation) and at least 25 samples with $Z=0$ (corresponding to a B$\to$B rotation), we took the average value of the normalized peak GCVI for the samples with $Z=1$ and for the samples with $Z=0$. In particular, we let $\bar{V}_{i,t}^{(Z=1)}$ and $\bar{V}_{i,t}^{(Z=0)}$ be the average of the normalized peak GCVI in the rotated group and non-rotated group, respectively, across the samples in grid cell $i$ and year $t$. A heuristic estimate of the precrop effect is $\bar{V}_{i,t}^{(Z=1)}-\bar{V}_{i,t}^{(Z=0)}$, and these estimates were then converted to the percent of average yield scale using the approach described in Section \ref{sec:ConvertingToYieldScale}. 
 
\begin{figure}[!t]
    \centering
    \includegraphics[width=0.95 \hsize]{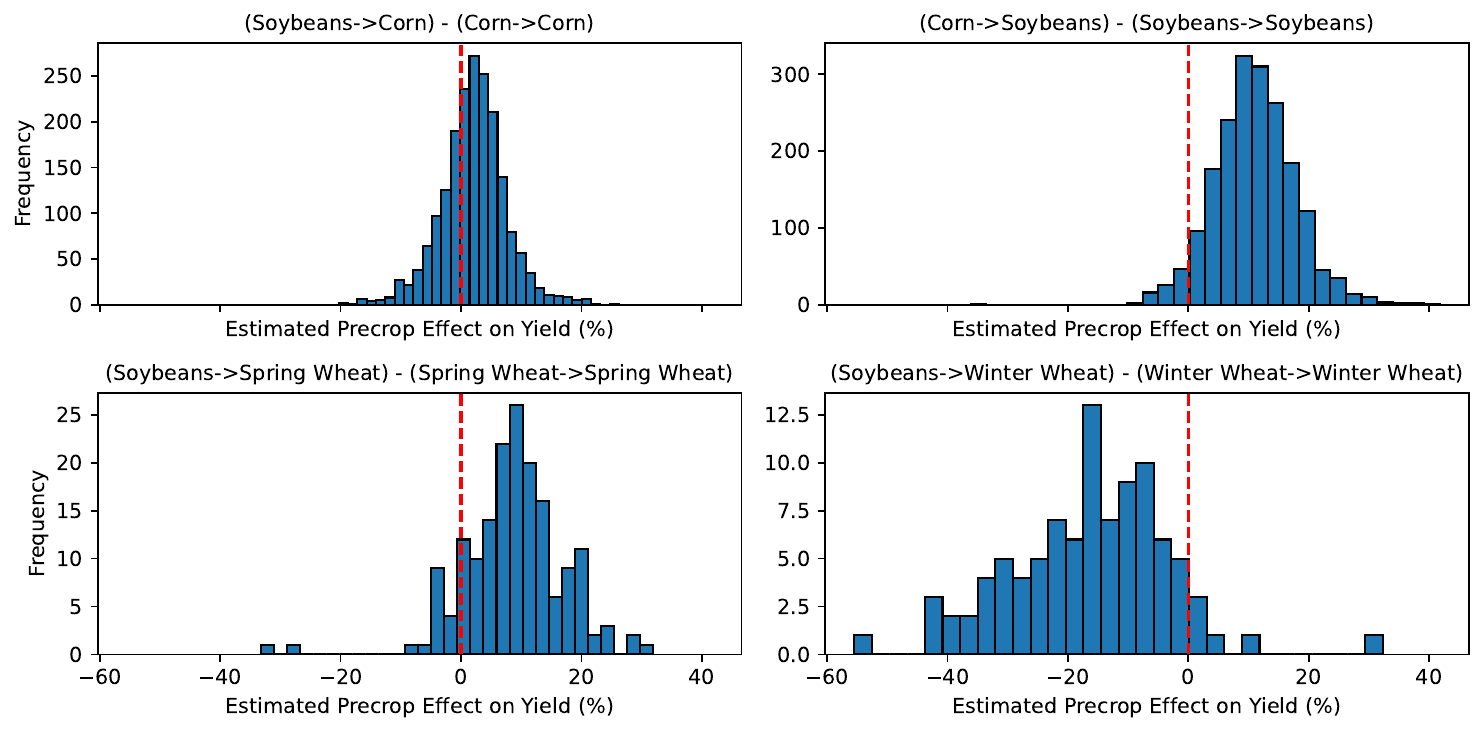}
   \caption{Heuristic estimates of precrop effects in the US using differences-of-means. Each panel gives a histogram of the differences in estimated yield between the rotated group and non-rotated group across all grid cell and year pairs (with a sufficient sample size). For more details see Appendix \ref{sec:DiffInMeanClustered}.}
    \label{fig:precrop_effect_heurestic_hist}
\end{figure}

For four different rotations in the US, Figure \ref{fig:precrop_effect_heurestic_hist} gives a histogram of these heuristic precrop effect estimates across all grid cells $i$ an years $t$ with at least 25 samples of both $Z=0$ and $Z=1$. The histograms suggest large positive average precrop effects for the Corn$\to$Soybeans and Soybeans$\to$Spring Wheat sequences in the US, a small positive average precrop effect for the Soybeans$\to$Corn sequence in the US, and substantial negative average precrop effect for the Soybeans$\to$Winter Wheat sequence in the US. These qualitative findings are consistent with the average estimated precrop effects estimated using the causal forest (see Table \ref{table:ResultsTable} (Column 6)).

\section{Sampling Scheme}\label{sec:samplingScheme}

For each country, we used a sampling scheme to ensure that our sampled pixels were a uniform random sample from croplands that grew a crop type of interest somewhat regularly (in at least 20$\%$ of sampled years). Due to differing crop types of interest in each country and due to differing number of years of crop map availability, the sampling scheme in each country was slightly different. In particular, after setting a (different) mask in each country for the croplands of interest, we gridded each country into 500 km $\times$ 500 km grid cells, drew a uniform random sample of $50{,}000$ pixels from each grid cell, and only retained the sampled pixels that were in this cropland mask. This resulted in a uniform random sample from the croplands of interest, and notably the distribution of the random sample did not depend on the grid cell size of 500 km $\times$ 500 km (the grid cell size was chosen to avoid high computational burden for each task in GEE).

In the United States, our crop types of interest were corn, soybean, and wheat, and we selected a random sample of pixels from the Eastern US in which one of those three crops was grown somewhat regularly according to the USDA's Cropland Data Layer \cite{CDL_Boryan}. In particular, we set a mask to only consider pixels for which during the 14 years between 2008 and 2021 (inclusive) the crop map classified either at least 3 years of corn, at least 3 years of soybean, or at least 3 years of wheat (any type of wheat). After setting these masks we randomly selected a sample of $324{,}777$ pixels from the croplands of interest by uniformly sampling 50,000 pixels in each 500 km $\times$ 500 km grid cell and removing pixels that did not meet our selection criteria. The region of interest used for sampling was the Eastern US (defined as east of the 100° W meridian), although because we included samples from grid cells that only partially overlapped with the region of interest, our westernmost sample had a longitude of 103.3° W.

In Canada, our crop types of interest where corn, soybean, wheat, canola, and barley, and we selected a random sample of pixels from Canada in which one of those five crops was grown somewhat regularly according to the Annual Crop Inventory \cite{CanadaACI}. In particular we set a mask to only consider pixels for which during the 10 years between 2011 and 2020 (inclusive) the crop map classified either at least 2 years of corn, at least 2 years of soybean, at least 2 years of wheat (any type of wheat), at least 2 years of canola, or at least 2 years of barley. After setting these masks we randomly selected a sample of $164{,}681$ pixels from the croplands of interest by uniformly sampling 50,000 pixels in each 500 km $\times$ 500 km grid cell and removing pixels that did not meet our selection criteria.

In northeastern China our crop types of interest where corn, soybean, and rice as these were the only three crops classified in the crop map from You et al. \cite{YouChinaCropMap} and the crop map only covers northeastern China (Jilin, Heilongjiang, Liaoning, and parts of Inner Mongolia). We selected a random sample of pixels in which the crop map classified one of these three crops at least once during the years 2017, 2018, and 2019. In particular we set a mask to exclude pixels that did not correspond to cropland and to exclude pixels for which the category ``other crop" was classified in each year between 2017 and 2019 (other crop denotes a crop other than corn, soybean, or rice). After setting these masks, we randomly selected a sample of $127{,}262$ pixels from the croplands of interest by uniformly sampling 50,000 pixels in each 500 km $\times$ 500 km grid cell and removing pixels that did not meet our selection criteria.

In France our crop types of interest were wheat, corn, barley, and rapeseed, and our sample came from all administrative regions in mainland France, with the exception of Provence-Alpes-Côte d'Azur (which is in the south east corner and had little cropland area for our outcome crops of interest). To set the sampling mask we used the French Parcel Dataset \cite{FrenchParcel} and data from 2010-2021. In particular, we set a mask to only consider pixels in France for which during the 12 years between 2010 and 2021 (inclusive) the parcel dataset had either at least 3 years of wheat (any type of wheat), at least 3 years of corn, at least 3 years of barley, or at least 3 years of rapeseed. After setting these masks, we randomly selected a sample of $57{,}473$ pixels from the croplands of interest by uniformly sampling 50,000 pixels in each 500 km $\times$ 500 km grid cell and removing pixels that did not meet our selection criteria. Note that because the data from 2010 to 2014 was at a coarser spatial resolution and the crop type labels had fewer, but broader categories, we used the crop type classifications from 2010 to 2014 to help set a mask for sampling, but did not use the classifications from these years in our analyses.

\section{Removing outlier peak GCVI values}\label{sec:outlierGCVI}

\begin{figure}[!t]
    \centering
    \includegraphics[width=0.95 \hsize]{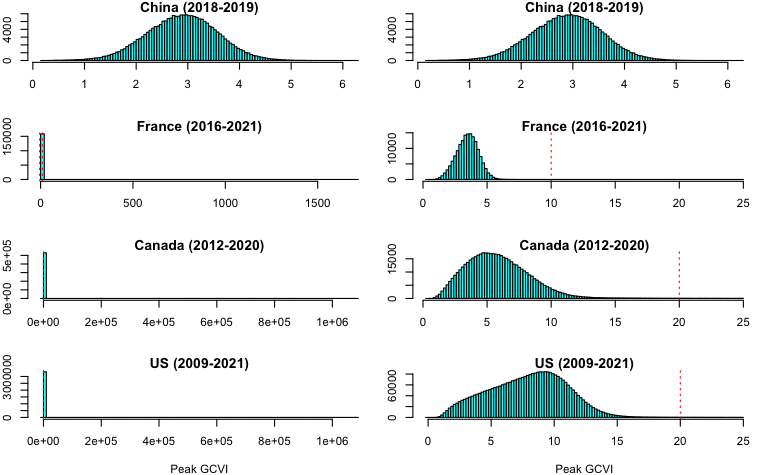}
    \caption{Histograms of estimated peak GCVI values in each country. The left panels depict the raw histograms of the peak GCVI values estimated by harmonic regressions without removing any outliers. The right panels depict zoomed in versions of these histograms that ignore estimated peak GCVI values outside of the -10 to 25 range. The vertical dashed lines depict the thresholds beyond which a point was deemed an outlier and removed from the analysis. In each country, these histograms only included peak GCVI estimates from pixel-year pairs in which an estimate of the previous year crop type was available and in which the current year crop type was classified as one of the outcome crop types of interest (see Table \ref{Table:ProportionOfSamplesRemoved} for outcome crops of interest in each country).}
    \label{fig:PeakGCVIHistAndThresh}
\end{figure}

 In China and France where Sentinel-2 was used, observations with peak GCVI less than $-5$ or greater than $10$ were deemed as outliers and removed. In the US and Canada where Landsat was used, observations with peak GCVI less than $-5$ or greater than $20$ were deemed as outliers and removed. These thresholds were determined by looking at histograms of the estimated peak GCVI values in each country. Figure \ref{fig:PeakGCVIHistAndThresh} depicts histograms for each country of the peak GCVI values estimated via harmonic regression. In cases where points were removed due to being outliers with peak GCVI estimates above a threshold, a red dotted line depicts the threshold used.

 For each country and outcome crop, the following table describes how many samples were removed due to having outlier peak GCVI values (Column 5).

\begin{table}[!h]\caption{Number of samples and proportion of samples removed for each outcome crop and country pair. For each outcome crop and country of interest, the third column gives the number of samples (unique pixel-year pairs) in our data sample, excluding those from the first year of data in each country (2008 in the US, 2011 in Canada, 2015 in France, and 2017 in China). The proportion of those samples which were removed due to missing values and removed due to being outliers are presented in the 4th and 5th columns, respectively. In our analysis, we do not use peak GCVI values from the first year of data within each country, and therefore do not consider those years in this table.}\label{Table:ProportionOfSamplesRemoved}
\centering
\begin{tabular}{llrrr}
  \hline
Country & Outcome Crop  & Count & NA Proportion & Outlier Proportion \\ 
  \hline
Canada & Corn & 61{,}028 & 0.015 & 0.025 \\ 
  Canada & Soybean & 85{,}469 & 0.011 & 0.035 \\ 
  Canada & Spring Wheat & 385{,}316 & 0.000 & 0.005 \\ 
  China & Corn & 141{,}698 & 0.000 & 0.000 \\ 
  China & Soybean & 58{,}772 & 0.000 & 0.000 \\ 
  France & Corn & 35{,}570 & 0.000 & 0.002 \\ 
  France & Winter Wheat & 123{,}176 & 0.000 & 0.002 \\ 
  US & Corn & 1{,}490{,}543 & 0.002 & 0.005 \\ 
  US & Soybean & 1{,}330{,}914 & 0.002 & 0.009 \\ 
  US & Spring Wheat & 182{,}672 & 0.002 & 0.011 \\ 
  US & Winter Wheat & 345{,}614 & 0.001 & 0.002 \\ 
   \hline
\end{tabular}
\end{table}

\section{Irrigation Imputation Scheme}\label{sec:IrrigationImputationScheme}

In the United States, we used annual irrigation classifications from the Landsat-based Irrigation Dataset \cite{IrrigationDat_ESSD} to construct a covariate for irrigation status $X_{\text{irr}}$. Let $W \in \{ \text{Irrigation}$, $ \text{No Irrigation} \}$ denote the classifications from this dataset. $W$ was available for each Landsat pixel and year until 2017.   Because classifications from this data product were available until 2017, for each year until 2017 we set $X^{(\text{irr})}$ to be $1$ for pixel-year pairs that where that were classified as irrigated and $X^{(\text{irr})}$ to be $0$ otherwise. Due to lack of estimates from the data product after the year 2017, we imputed $X^{(\text{irr})}$ using the number of years between 2011--2017 in which the pixel was classified as having irrigation. In particular, letting $N^{\text{(irr)}}_{\text{2011--2017}}$ be the number of years out of the years 2011--2017 that the pixel was classified as having irrigation and letting $\epsilon=0.01$, we defined our irrigation covariate to be

$$X^{(\text{irr})} \equiv \begin{cases}
    1 & \text { if Year } \leq 2017 \text{ and } W= \text{Irrigation}, \\
    (N^{\text{(irr)}}_{\text{2011--2017}}+\epsilon)/(7+2 \epsilon) & \text { if Year } > 2017, \text { and} \\
    0 & \text { if Year } \leq 2017 \text{ and } W= \text{No Irrigation}.
\end{cases}$$

For years after 2017, $X^{(\text{irr})}$ for each pixel was imputed to be roughly the proportion of years in the 2011--2017 range in which that pixel was classified as irrigated, with a slight modification so that the porportion would never be $0$ or $1$. The constructed covariate $X^{(\text{irr})}$ could be interpreted as a rough confidence score that the pixel would be classified as irrigated land, had the irrigation data product been extended beyond 2017. Notably, because we ultimately used $X^{(\text{irr})}$ as a feature in a causal forest, and did not study heterogeneity with respect to this feature, only the ordering of the $X^{(\text{irr})}$ covariate should have influenced the ultimate analyses and the specific values of $X^{(\text{irr})}$ should have been irrelevant (this is because the causal forest is a tree based algorithm where the exact spacing between the distinct $X^{(\text{irr})}$ values should not have influenced the node splits in any of the trees). Therefore, other choices of $\epsilon$ or approaches to estimate the probability of an irrigation classification after 2017 as a function of $N^{\text{(irr)}}_{\text{2011--2017}}$ were not considered.

\begin{figure}[!t]
    \centering
    \includegraphics[width=0.95 \hsize]{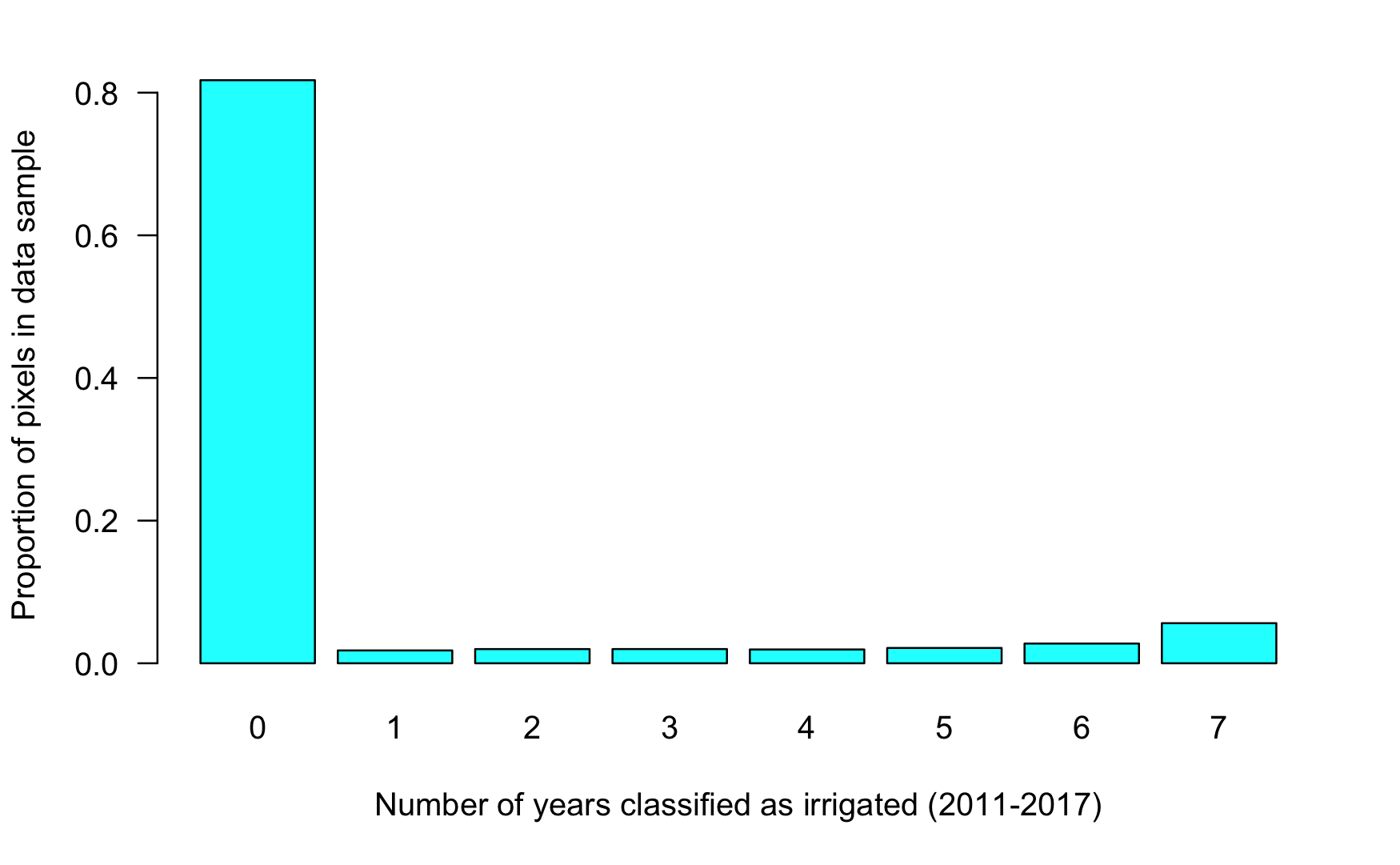}
    \caption{Histogram of the number of years between 2011 and 2017 in which each pixel was classified as irrigated. The classifications for each pixel and year in our dataset was taken from the Landsat-based Irrigation Dataset \cite{IrrigationDat_ESSD}.}
    \label{fig:IrrigationData}
\end{figure}

Our results are likely not sensitive to our chosen scheme for imputing post-2017 irrigation classifications using pre-2018 classifications. This is because farmers in the US do not frequently switch irrigation status from year to year. The distribution of $N^{\text{(irr)}}_{\text{2011--2017}}$ for samples in the US is plotted in Figure \ref{fig:IrrigationData}. It can be seen that $N^{\text{(irr)}}_{\text{2011--2017}}$ is often either $0$ or $7$, suggesting that switching irrigation status is indeed rare. Note that many of the cases where $N^{\text{(irr)}}_{\text{2011--2017}} \in \{1,..,6\}$ could be driven by measurement errors, as the irrigation data product had an overall accuracy of 90--97$\%$ depending on the region \citep{IrrigationDat_ESSD}.

\section{Heterogeneity analysis in the presence of unmeasured confounders or mediators}\label{sec:HeterogeneityAnalysisWithUnmeasuredConfounders}

In this appendix we show that even if the causal effect estimates provided in this paper are biased due to unmeasured variables, it is possible that our analysis of the heterogeneity of the rotation benefit with weather will not be biased. In particular, we present assumptions under which the causal effect estimates provided by a causal forest would be biased due to unmeasured confounders, but the estimates describing how those causal effects vary with weather would not be biased. An identical mathematical argument shows that if there are also unmeasured, manipulable mediators, then under similar assumptions, the causal forest gives biased estimates for the direct effects, but the estimates describing how those effects vary with weather would not be biased estimates of how the direct causal effects vary with weather (see Appendix \ref{sec:UnmeasuredMediatorSettingeWhereYouCanStillGetDirectEffects}). 

\subsection{Notation and assumptions}

Let $Z$ denote the binary treatment variable (in our case $Z=1$ if the precrop differs from the outcome crop and $Z=0$ otherwise), let $Y$ denote the outcome variable (in our case crop yield or a proxy for the yield of the outcome crop) and let $X \in \mathbb{R}^p$ be a vector of measured confounders. We let $U \in \mathbb{R}^k$ denote a vector of unmeasured confounders which can be associated with both $Z$ and $Y$. Finally we let $Y(1)$ and $Y(0)$ denote the potential outcomes under the treatment and under the control (assuming $X$ and $U$ remain fixed and $Z$ is forced to be $1$ or $0$ respectively). Since $(X,U)$ together comprise of all measured and unmeasured confounders, the potential outcomes and the treatment $Z$ are conditionally independent given $(X,U)$ (i.e., $\big( Y(1),Y(0) \big) \indep Z | (X,U)$).

We now introduce two assumptions, under which, the presense of unmeasured confounders would not bias our treatment effect heterogeneity analyses. 
\begin{enumerate}[(i)]
    \item $\e[Y|X,Z,U]=h(X)+\tau(X) Z +\eta^\tran U$, for some functions $h,\tau: \mathbb{R}^p \to \mathbb{R}$ and some vector $\eta \in \mathbb{R}^k$.
    \item $\e[U|X,Z]=g(X) + Z \gamma $, for some function $g: \mathbb{R}^p \to \mathbb{R}^k$ and $\gamma \in \mathbb{R}^k$.
\end{enumerate}

In words the first assumption states the model for the conditional mean of the outcome variable given all other variables in the model is linear in the unmeasured confounders $U$ and has no interaction between $U$ and either the treatment variable $Z$ or the observed confounders $X$. The second assumption states that the conditional means of each unmeasured confounder given the measured confounders $X$ and the treatment variable $Z$ are separable in $Z$ and $X$ (i.e., they have no interaction term between $Z$ and $X$).

\subsection{The desired causal estimands and the causal forest estimands}

Ultimately, we suppose the investigator is interested in estimating the conditional average treatment effect (CATE) and how the CATE varies with weather. Under Assumptions (i) and (ii), the CATE is given by $$\begin{aligned}
    \text{CATE}(x) & \equiv \e[Y(1)-Y(0)|X=x]
    \\ & = \e[Y(1)|X=x]-\e[Y(0)|X=x]
    \\ & = \e\Big[ \e[Y(1)|X,U] \Big| X=x \Big]-\e\Big[ \e[Y(0)|X,U] \Big| X=x \Big] 
    \\ & = \e\Big[ \e[Y(1)|X,U,Z=1] \Big| X=x \Big]-\e\Big[ \e[Y(0)|X,U,Z=0] \Big| X=x \Big]
    \\ & = \e\Big[ \e[Y|X,U,Z=1] \Big| X=x \Big]-\e\Big[ \e[Y|X,U,Z=0] \Big| X=x \Big]
    \\ & = \e[h(X)+\tau(X)+\eta^\tran U | X=x ]- \e[h(X)+\eta^\tran U | X=x ]
    \\ & =h(x) +\tau(x)+\e[\eta^\tran U|X=x] -h(x) -\e[\eta^\tran U|X=x] 
    \\ & = \tau(x).
\end{aligned}$$ Above the third equality follows from the tower property for conditional expectation, the fourth equaility holds because $\big( Y(1),Y(0) \big) \indep Z | (X,U)$, and the sixth equality follows from Assumption (i). The other steps follow from linearity of conditional expectation and combining terms.

On the other hand the causal forest estimates the following quantity $$\begin{aligned}
    \omega(x) & \equiv \e[Y|X=x,Z=1]-\e[Y|X=x,Z=0]
    \\ & = \e \Big[ \e[Y|X,Z,U] \Big| X=x,Z=1 \Big]-\e \Big[ \e[Y|X,Z,U] \Big| X=x,Z=0 \Big] 
    \\ & = \e[h(X)+\tau(X) Z +\eta^\tran U|X=x,Z=1]-\e[h(X)+\tau(X) Z +\eta^\tran U|X=x,Z=0]
    \\ & = h(x)+\tau(x)+ \eta^\tran \e[ U|X=x,Z=1]- h(x)- \eta^\tran \e[ U|X=x,Z=0]
    \\ & = \tau(x) + \eta^\tran \big( \e[U|X=x,Z=1]-\e[U|X=x,Z=0] \big)
        \\ & = \tau(x) + \eta^\tran \big( g(x)+\gamma -g(x) \big)
    \\ & = \tau(x)+\eta^\tran \gamma
    \\ & = \text{CATE}(x)+\eta^\tran \gamma.
\end{aligned}$$

Above the second equality follows from the tower property for conditional expectation, the third equality follows from Assumption (i), and the sixth equality follows from Assumption (ii), and the last equality follows from the previous result showing $\text{CATE}(x)=\tau(x)$ under Assumption (i) and (ii).  The other steps follow by linearity of conditional expectation and combining terms.

Under Assumptions (i) and (ii), we have shown that $\omega(x)=\text{CATE}(x) +\eta^\tran \gamma$, where $\omega(\cdot)$ is the function estimated by the causal forest and $\text{CATE}(\cdot)$ is the function that the investigator actually wants to estimate. These two quantities differ by an additive constant $\eta^\tran \gamma$. Therefore under Assumptions (i) and (ii), the average treatment effect (ATE) or the overlap-weighted average treatment effect (ATO) would suffer from a bias of $\eta^\tran \gamma$, but the estimates of how the treatment effect varies with weather obtained by regressing $\omega(X)$ on weather covariates (see model \eqref{eq:ModelToAssessHeterogeneity}) would not be biased. In particular, when $\omega(x)=\text{CATE}(x) +\eta^\tran \gamma$ the regression coefficients corresponding to weather covariates would be the same in model \eqref{eq:ModelToAssessHeterogeneity} had one used the ideal choice of the response variable $\text{CATE}(X)$ rather than the $\omega(X)$ response variable that we used in model \eqref{eq:ModelToAssessHeterogeneity}.

We remark that in practice Assumptions (i) and (ii) may not hold precisely. For example, if the model for $Y$ is nonlinear in the unmeasured confounders $U$, Assumption (i) would be violated, although one can always redefine the vector $U$ (e.g., by adding entries that are functions of $U$) such that the model for $Y$ will be linear in $U$. More problematically, if the model for $Y$ has an interaction between the unmeasured confounders $U$ and either the treatment variable $Z$ or the measured confounders $X$, Assumption (i) would be violated.  In addition, if the model for the unmeasured confounder vector $U$ has an interaction between the measured confounders $X$ and the treatment variable $Z$, Assumption (ii) would be violated. Nonetheless, even if such interactions exist rendering our assumptions violated, when such interactions are very small a sensitivity analysis can be done to show that the bias in our heterogeneity analysis would be very small. In addition, in the next subsection we show that if we modify Assumption (i) to allow for interactions between $Z$ and $U$ in the model for $Y$, our heterogeneity analysis confers unbiased estimates when considering meaningful causal estimands (although these estimands are different than the CATE). 

\subsection{Allowing for interactions between $Z$ and $U$ in model for $Y$}

Assumption (i) that $\e[Y|X,Z,U]=h(X)+\tau(X) Z +\eta^\tran U$ can be loosened, although it requires defining a slightly different causal estimand. In this subsection, we modify Assumption (i) to allow for an interaction between $U$ and $Z$ in the model for $Y$, but we keep Assumption (ii). Instead of Assumption (i) we assume \begin{enumerate}[(i')]
    \item $\e[Y|X,Z,U]=h(X)+\tau(X) Z +\eta^\tran U + \lambda^\tran U Z$, for some functions $h,\tau: \mathbb{R}^p \to \mathbb{R}$ and some vectors $\eta,\lambda \in \mathbb{R}^k$.
\end{enumerate}

While typically it is of interest to study how the CATE varies with weather, an investigator may also be interested in how the following two estimands vary with weather
$$\text{CATT}(x) \equiv \e[Y(1)-Y(0)|X=x,Z=1] \quad \text{and} \quad \text{CATC}(x) \equiv \e[Y(1)-Y(0)|X=x,Z=0].$$

The above estimands give the average treatment effect among the treated units and the average treatment effect among the control units, respectively, as a function of the covariates embedded in $x$. Observe that under Assumptions (i') and (ii),

$$\begin{aligned}
    \text{CATT}(x) & \equiv \e[Y(1)-Y(0)|X=x,Z=1]
    \\ & = \e[Y(1)|X=x,Z=1]-\e[Y(0)|X=x,Z=1]
    \\ & = \e \Big[ \e[Y(1)|X,Z,U] \Big| X=x,Z=1 \Big]-\e \Big[ \e[Y(0)|X,Z,U] \Big| X=x,Z=1 \Big]
     \\ & = \e \Big[ \e[Y(1)|X,U] \Big| X=x,Z=1 \Big]-\e \Big[ \e[Y(0)|X,U] \Big| X=x,Z=1 \Big]
    \\ & = \e \Big[ \e[Y(1)|X,Z=1,U] \Big| X=x,Z=1 \Big]-\e \Big[ \e[Y(0)|X,Z=0,U] \Big| X=x,Z=1 \Big]
     \\ & = \e \Big[ \e[Y|X,Z=1,U] \Big| X=x,Z=1 \Big]-\e \Big[ \e[Y|X,Z=0,U] \Big| X=x,Z=1 \Big]
    \\ & =   \e[h(X)+\tau(X) +\eta^\tran U + \lambda^\tran U |X=x,Z=1]-\e[h(X) +\eta^\tran U |X=x,Z=1]
    \\ & = \tau(x) + \lambda^\tran \e[U|X=x,Z=1]
    \\ & = \tau(x) + \lambda^\tran g(x) +\lambda^\tran \gamma.
\end{aligned}$$ Above the third equality holds by the tower property of conditional expectations, the fourth and fifth equalities hold because $\big( Y(1),Y(0) \big) \indep Z | (X,U)$ and, the seventh equality holds by Assumption (i') and the last equality holds by Assumption (ii). A similar argument shows that $$\text{CATC}(x) \equiv \e[Y(1)-Y(0)|X=x,Z=0] =\tau(x)+\lambda^\tran g(x).$$

On the other hand observe that under Assumptions (i') and (ii), the quantity that the causal forest estimates is $$\begin{aligned}
    \omega(x) & \equiv \e[Y|X=x,Z=1]-\e[Y|X=x,Z=0]
    \\ & = \e \Big[ \e[Y|X,Z,U] \Big| X=x,Z=1 \Big]-\e \Big[ \e[Y|X,Z,U] \Big| X=x,Z=0 \Big] 
    \\ & = \e[h(X)+\tau(X) Z +\eta^\tran U +\lambda^\tran U Z |X=x,Z=1]
    \\ & \ \ \ -\e[h(X)+\tau(X) Z +\eta^\tran U+\lambda^\tran U Z |X=x,Z=0]
    \\ & = \tau(x)+ (\eta+\lambda)^\tran \e[ U|X=x,Z=1] - \eta^\tran \e[ U|X=x,Z=0]
    \\ & = \tau(x) + (\eta+\lambda)^\tran \big( g(x)+\gamma) -\eta^\tran g(x)
    \\ & = \tau(x) +\lambda^\tran g(x) +\eta^\tran \gamma + \lambda^\tran \gamma 
    \\ & = \text{CATT}(x) + \eta^\tran \gamma = \text{CATC}(x)+ (\eta+\lambda)^\tran \gamma.
\end{aligned}$$

Above the second equality holds by the tower property of conditional expectations, the third equality holds by Assumption (i'), the fifth equality holds by Assumption (ii), and the last line hollds by the previous formulas for $\text{CATC}(x)$ and $\text{CATT}(x)$ under Assumptions (i') and (ii). 

Thus under Assumptions (i') and (ii), the causal forest estimand $\omega(x)$ only varies by additive constants from the estimands of interest $\text{CATC}(x)$ and $\text{CATT}(x)$. In particular, under these assumptions, $$ \omega(x)=\text{CATT}(x) + \eta^\tran \gamma = \text{CATC}(x)+ (\eta+\lambda)^\tran \gamma,$$ and further an analysis of the heterogeneity of $\omega(x)$ with weather (based on estimating the regression coefficients corresponding to weather in model \eqref{eq:ModelToAssessHeterogeneity}) should give an unbiased estimate for the heterogeneity of $\text{CATC}(x)$ or $\text{CATT}(x)$ with weather. 

\subsection{Setting where manipulable mediators are missing}\label{sec:UnmeasuredMediatorSettingeWhereYouCanStillGetDirectEffects}

We now consider a setting where the vector of unmeasured variables $U$ also contains unmeasured, manipulable mediators (and $U$ may or may not contain unmeasured confounders). Our motivating examples of unmeasured, manipulable mediators include fertilizer use and pesticide use. 

Similar to before, we let $Z$ denote the binary treatment variable (in our case $Z=1$ if the precrop differs from the outcome crop and $Z=0$ otherwise), let $Y$ denote the outcome variable (in our case crop yield or a proxy for the yield of the outcome crop), and let $X \in \mathbb{R}^p$ be a vector of measured confounders. We let $U \in \mathbb{R}^k$ denote a vector of unmeasured, manipulable mediators and (possibly) unmeasured confounders. The unmeasured, manipulable mediators impact $Y$ and are affected by $Z$, but have a value that can be chosen by the farmer. Finally we let $Y(1)$ and $Y(0)$ denote the potential outcomes under the treatment and under the control (assuming $X$ and $U$ remain fixed and $Z$ is forced to be $1$ or $0$ respectively). Since $(X,U)$ together comprise of all measured and unmeasured confounders, the potential outcomes and the treatment $Z$ are conditionally independent given $(X,U)$ (i.e., $\big( Y(1),Y(0) \big) \indep Z | (X,U)$). We now consider the same two assumptions considered in the previous subsection (except now $U$ contains manipulable mediators):

\begin{enumerate}[(i*)]
    \item $\e[Y|X,Z,U]=h(X)+\tau(X) Z +\eta^\tran U + \lambda^\tran U Z$, for some functions $h,\tau: \mathbb{R}^p \to \mathbb{R}$ and some vectors $\eta,\lambda \in \mathbb{R}^k$.
    \item $\e[U|X,Z]=g(X) + Z \gamma $, for some function $g: \mathbb{R}^p \to \mathbb{R}^k$ and $\gamma \in \mathbb{R}^k$.
\end{enumerate}

An identical mathematical argument to that seen in the previous subsection shows that under these two assumptions, $$ \omega(x)=\text{CATT}(x) + \eta^\tran \gamma = \text{CATC}(x)+ (\eta+\lambda)^\tran \gamma, \quad \text{where}$$

$$\text{CATT}(x) \equiv \e[Y(1)-Y(0)|X=x,Z=1] \quad \text{and} \quad \text{CATC}(x) \equiv \e[Y(1)-Y(0)|X=x,Z=0].$$

Note that since $Y(1)$ and $Y(0)$ denote the potential outcomes under the treatment and under the control (assuming $X$ and $U$ are fixed), $Y(1)-Y(0)$ is the random variable giving the direct effects. Hence $\text{CATT}(x)$ and $\text{CATC}(x)$ give the average direct effects (for treated and control units, respectively) as a function of the measured covaraites $x$. Meanwhile, our causal forest estimates the function $\omega(\cdot)$ and our heterogeneity analysis studies how $\omega(x)$ varies with $x$. In particular, our heterogeneity analysis estimates regression coefficients corresponding to the weather in a linear regression with $\omega(X)$ as the response.  Under Assumptions (i*) and (ii*), $\omega(x)$ differs from $\text{CATT}(x)$ and $\text{CATC}(x)$ by constants that do not vary with $x$. Thus, under Assumptions (i*) and (ii*), the estimates of regression coefficients from our heterogeneity analysis would still give unbiased estimates of how the average direct treatment effects (for either the treated or control units) vary with weather.

\section{A comparison of using a vegetation index rather than satellite-based crop yield estimates}\label{sec:VI_versus_SCYM}

In this appendix, we compare our approach to that of recent works which have used a similar causal forest methodology but instead used crop yield estimates from the Scalable Crop Yield Mapper (SCYM) \cite{originalSCYM,SCYM_CornJillVersion,SCYM_SoyDadoVersion} as the outcome variable. In particular, with SCYM yield estimates as the outcome variable, the causal forest method was used to study the effects of conservation tillage \cite{Tillage_CausalForest} and crop rotation \cite{CREO_paper} on corn and soybean yields. In the latter study, the causal effect effect estimates in various locations and years were validated against those from corresponding experimental studies, and were found to have a statistically significant correlation with causal effect estimates from randomized field experiments.

In this appendix, we will show that under certain conditions that the user can ensure by design, using the a vegetation index-based proxy as an outcome variable (like we use in this paper) is equivalent to using SCYM as an outcome variable up to a multiplicative constant. Therefore, if one trusts the causal forest-based approach in recent works that use SCYM as an outcome variable \cite{Tillage_CausalForest,CREO_paper}, they should also trust the approach in this paper that uses a vegetation index-based yield proxy. SCYM yield estimates at  30m $\times$ 30m  resolution are currently only available for corn and soybean in the midwestern United States, so using a vegetation index-based proxy is a highly attractive alternative for assessing the impact of farm management practices on crop yields in other regions or for crop types besides corn and soybean.

We conclude the appendix by noting that future analysis may use more modern yield maps based on neural networks and transfer learning approaches as opposed to SCYM. Therefore, we also provide an empirical comparison of our results to results for the same analysis when using recently produced pixel-level estimates of crop yield in the US \cite{QDANNPaper} as the outcome variable.

\subsection{Description of SCYM estimates}

To understand why using a vegetation index-based proxy as an outcome variable can give the same results as using SCYM yield estimates up to a multiplicative constant, we must first explain how SCYM yield estimates are produced. SCYM yields are estimated by running a process-based crop growth simulator called the Agricultural Production Systems sIMulator (APSIM) \cite{APSIM} several thousand times for various choices of input parameters that reflect distribution of sowing dates, soil types and qualities, and weather conditions in the target region of interest. These simulations produce both crop yield estimates as well as a time series of leaf area index, and the time series of leaf area index is subsequently transformed to a time series of GCVI using an equation from \cite{LeafAreaIndexToGCVI}. For each simulation, the estimated GCVI time series is then summarized by a vector $G_{\text{APSIM}}$ of a small number of features, the weather covariates used in the simulation are summarized by a vector $W_{\text{APSIM}}$, and the crop crop yield estimated in the simulation is defined to be $Y_{\text{APSIM}}$. Subsequently, the following linear regression model is fit to the thousands of simulated samples \begin{equation}\label{eq:APSIM_calib}
    Y_{\text{APSIM}}= \alpha_0 + \beta^\tran G_{\text{APSIM}} + \gamma^\tran W_{\text{APSIM}} +\varepsilon.
\end{equation} 

The estimated coefficients and coefficient vectors $\tilde{\alpha}_0$, $\tilde{\beta}$ and $\tilde{\gamma}$ are then stored and subsequently used to estimate crop yields using the actual observed weather covariates and satellite-based GCVI features. In particular, for a particular pixel and year in which one would like to make a yield estimate, let $W_{\text{obs}}$ denote the observed vector of the same weather covariates that were used in model \eqref{eq:APSIM_calib} and let $G_{\text{obs}}$ denote the corresponding vector of features that summarize the observed satellite-based GCVI time series. The SCYM yield estimates, which we call $\tilde{Y}_{\text{SCYM}}$ in that pixel and year are given by  \begin{equation}\label{eq:SCYM_ests}
    \tilde{Y}_{\text{SCYM}}= \tilde{\alpha}_0 + \tilde{\beta}^\tran G_{\text{obs}} + \tilde{\gamma}^\tran W_{\text{obs}}.
\end{equation} 

Above, we summarized the procedure for producing SCYM yield estimates used in the most recent version of SCYM, and more details can be found in \cite{SCYM_CornJillVersion}. Note that although in earlier versions of SCYM \cite{originalSCYM,SCYM_AzzarriVersion,SCYM_JinVersion} $G_{\text{obs}}$ was a vector of features of length greater than 1, \cite{SCYM_CornJillVersion} tested multiple choices of feature vectors $G_{\text{obs}}$ that summarized the GCVI time series and found that certain choices of scalars for $G_{\text{obs}}$ performed as well as (and sometimes better than) vectors of multiple features for the task of predicting corn yields. Therefore, in the preferred model in \cite{SCYM_CornJillVersion} as well as the most up to date versions of SCYM, $G_{\text{obs}}$ is a scalar feature rather than a vector of features which summarizes the GCVI time series.

\subsection{Imposable conditions that allow forgoing SCYM estimates}

Suppose an investigator wishes fit a causal forest on a dataset where the outcome variable is the SCYM yield estimate $\tilde{Y}_{\text{SCYM}}$, the treatment variable is some binary $Z \in \{0,1\}$, and the covariates are some vector $X \in \mathbb{R}^p$. If an investigator does not have access to $\tilde{Y}_{\text{SCYM}}$ or have the time or resources to generate $\tilde{Y}_{\text{SCYM}}$ by running a bunch of crop simulation models and implementing the procedure described in the previous subsection, they can save time and computational resources by simply fitting a causal forest with the same treatment variable $Z$ and covariates $X$, but with $G_{\text{obs}}$ 
rather than $\tilde{Y}_{\text{SCYM}}$ as the outcome variable. In particular, under the following two assumptions which can be imposed by design, the causal forest using $G_{\text{obs}}$ as an outcome variable will estimate a quantity that is simply a scalar multiple of the quantity that the desired causal forest using $\tilde{Y}_{\text{SCYM}}$ as an outcome variable would have estimated.

\begin{enumerate}[(i)]
\item $G_{\text{obs}}$ is a scalar rather than a vector. That is, the SCYM estimates are not based on more than one feature summarizing the GCVI time series.
\item The weather covariate $W_{\text{obs}}$ from \eqref{eq:SCYM_ests} is a deterministic (and measurable) function of covariate vector $X$ used in the causal forest.
\end{enumerate}

We remark that Assumption (i) holds for the most recent versions of SCYM \cite{SCYM_CornJillVersion}. This is because \cite{SCYM_CornJillVersion} found that a single well-chosen feature summarizing the GCVI timeseries was equally effective for producing yield predictions, if not better, than a number of choices of feature vectors summarizing the GCVI timeseries.

We also remark that Assumption (ii) is easy to impose by design. For example if the investigator fitting a causal forest includes all entries of $W_{\text{obs}}$ in their control vector $X$, Assumption (ii) is guaranteed to hold. In addition, any function that is continuous at all but finitely many points is measurable and also sums, products, quotients, limits, derivatives, and indefinite integrals of measurable functions are measurable. Therefore, it is highly implausible that anyone would design a future SCYM model where $W_{\text{obs}}$ is not a measurable function of some higher dimensional vector of all observed weather variables $W_{\text{full}}$. If an investigator includes $W_{\text{full}}$ (or at least the subset of the entries of $W_{\text{full}}$ used to compute $W_{\text{obs}}$) in their covariate vector $X$ for the causal forest, then Assumption (ii) will also hold.

\subsection{A theoretical argument that SCYM and GCVI-based outcomes give the same results up to a scalar multiple}

If an investigator runs a causal forest with SCYM yield $\YSCYM$ as the outcome variable, the causal forest would estimate the following quantity:

$$\omSCYM(x) \equiv \e [ \YSCYM | X=x, Z=1 ] - \e [ \YSCYM | X=x, Z=0 ].$$

If the investigator forgoes computing SCYM yield estimates and instead runs a causal forest with $G_{\text{obs}}$ (which can be extracted from time series of GCVI taken from satellite imagery), the causal forest instead estimates the quantity:

$$\omega_{G}(x) \equiv \e [ G_{\text{obs}} | X=x, Z=1 ] - \e [ G_{\text{obs}} | X=x, Z=0 ].$$

Hence, by plugging in formula \eqref{eq:SCYM_ests} observe that under Assumptions (i) and (ii)
$$\begin{aligned}
  \omSCYM(x) & \equiv \e [ \YSCYM | X=x, Z=1 ] - \e [ \YSCYM | X=x, Z=0 ] 
 \\ & = \e [ \tilde{\alpha}_0 + \tilde{\beta}^\tran G_{\text{obs}} + \tilde{\gamma}^\tran W_{\text{obs}} | X=x, Z=1 ] - \e [ \tilde{\alpha}_0 + \tilde{\beta}^\tran G_{\text{obs}} + \tilde{\gamma}^\tran W_{\text{obs}} | X=x, Z=0 ]  
 \\  & = \tilde{\alpha}_0+\e [ \tilde{\beta}^\tran G_{\text{obs}} | X=x, Z=1 ] +\tilde{\gamma}^\tran W_{\text{obs}}(x)
 \\ & \ \ \  -\tilde{\alpha}_0 - \e [  \tilde{\beta}^\tran G_{\text{obs}} | X=x, Z=0 ] -\tilde{\gamma}^\tran W_{\text{obs}}(x)
  \\  & = \e [ \tilde{\beta}^\tran G_{\text{obs}} | X=x, Z=1 ] - \e [  \tilde{\beta}^\tran G_{\text{obs}} | X=x, Z=0 ]
    \\ & = \tilde{\beta} \times \Big( \e [  G_{\text{obs}} | X=x, Z=1 ] - \e [  G_{\text{obs}} | X=x, Z=0 ] \Big)
    \\ & = \tilde{\beta} \times \omega_{G}(x).
\end{aligned}$$

Above the 3rd step follows by linearity of conditional expectation and Assumption (ii) which states that $W_{\text{obs}}$ is a measurable function of $X$. The penultimate step follows from Assumption (i) that $G_{\text{obs}}$ is a scalar (and therefore $\tilde{\beta}$ is also a scalar). Hence we have shown that under Assumptions (i) and (ii), $$\boxed{ \omSCYM(x) = \tilde{\beta} \times \omega_{G}(x).}$$ In the boxed result above, recall that $\omSCYM(\cdot)$ is the function that would be learned by the causal forest if SCYM were to be used as an outcome variable and $\omega_{G}(\cdot)$ is the function that is learned by the causal forest when the GCVI-based feature $G_{\text{obs}}$ is used as the outcome variable instead. 

Even though Assumption (ii) can be imposed by design, we also remark that the above formula will still hold in many cases where Assumption (ii) does not hold. In particular if we instead assume that conditional on $X$, $W_{\text{obs}}$ and $Z$ are statistically independent (i.e., $W_{\text{obs}} \indep Z |X$), the boxed result would still hold. However, we state Assumption (ii) rather than the looser condition about conditional independence because the former is more interpretable and can be imposed by design, while the assumption about conditional independence can be difficult to reason about or check empirically.

\subsection{Interpretation and implication of boxed result }

In words, the boxed result states that under Assumptions (i) and (ii), the quantity that a causal forest using SCYM as an outcome variable tries to estimate is just a scalar multiple of (i.e., $\tilde{\beta}$ times) the quantity that a causal forest tries to estimate when using the corresponding remotely-sensed feature that summarizes a vegetation index time series as an outcome variable. This further implies that under Assumptions (i) and (ii), the average treatment effect (ATE), the overlap-weighted average treatment effect (ATO), and the measures of treatment effect heterogeneity with weather that we considered in this paper (prior to rescaling) would be $\tilde{\beta}$ times as large had we produced SCYM yield estimates from crop growth simulators. This leads to nice conclusions which allow for savings in computational resources and enable researchers without access to SCYM yield predictions to conduct similar analyses. If an investigator is willing to trust the approach in recent papers that used causal forests and SCYM yield estimates \cite{Tillage_CausalForest,CREO_paper}, then that investigator should also be willing to trust the approach of fitting a causal forest when a vegetation index-based feature is used as the outcome variable, as is the approach in this paper. 

There are a few caveats about using a satellite-based vegetation index instead of SCYM yield estimates as an outcome variable in causal forest analyses worth mentioning. First, while this approach allows investigators to estimate the direction and statistical significance of the effects of a management practice on yield or the heterogeneity of that management practice with weather, it does not provide estimates in the interpretable scale of tons per hectare. If an investigator cares about estimating the magnitude of the rotation benefits and heterogeneity measures, and if data from randomized field experiments are available, they can instead use the calibration approach in \cite{CREO_paper}, which will simultaneously also remove some of the bias due to unmeasured confounders. Second, in cases where the investigator is interested in the output of a causal forest with the log of the SCYM yield as the outcome variable (as is used in \cite{CoverCrop_CausalForest} to study the impacts of cover cropping), the argument in the previous subsection does not hold. In particular, there is no simple multiplicative relationship between the outputs of causal forests ouptputs when the outcome variable is the log of SCYM yield versus that when the outcome variable is the log of a vegetation index. Third, some earlier versions of SCYM do not satisfy Assumption (i) and it is possible that in regions outside the US or croptypes other than corn and soybean the best SCYM-type yield estimates would require a vector of multiple features summarizing the GCVI time series. Fourth, $\tilde{\beta}$ is different for each outcome crop and for each region, so when studying multiple rotations in multiple regions, different scaling factors apply to different estimates. Similarly, it may be desirable to use a different vegetation index-based feature and a different weather covariate vector than those used for corn and soybean in the United States, whereas in this study we use the same weather covariates and vegetation index-based feature (peak harmonic GCVI) for each outcome crop type and region (with the exception of earlier season weather covariates for winter wheat). Fifth, in the analysis in this paper, we do not precisely meet Assumption (ii) even though we could have easily met it with little computational cost. For example, the most recent version of SCYM for estimating corn yields used the following four weather covariates to compose $W_{\text{obs}}$: June–August rainfall, June–August solar radiation, July VPD, and August maximum temperature \cite{SCYM_CornJillVersion}. These four weather covariates cannot be written as exact functions of the weather covariates we control for (July VPD can, but we control for slightly different temperature and precipitation variables, and we do not control for solar radiation). A similar caveat is that the most recent version of SCYM used a harmonic partial integral for the feature $G_{\text{obs}}$ summarizing the GCVI time series whereas we use the peak of the harmonic fit to the GCVI time series as our outcome variable for two reasons. First, using peak GCVI led to SCYM models with nearly as good performance in the midwestern United States \cite{SCYM_CornJillVersion}. Second, peak vegetation index values are commonly used proxies for yield in many regions throughout the world \cite{PeakVIMalawi,JinPeakGCVIKenya,PeakEVI2GoodProxyCanada}, and GCVI has sometimes been found to outperform other vegetation indices as a predictor of crop yield \cite{ChinaMaizeGCVIBetter}.

In spite of these caveats, the approach we present in this paper of using peak GCVI as an outcome variable rather than SCYM yield estimates can allow investigators to study the impacts of agronomic practices on yields without needing to run crop modeling simulations to calibrate the SCYM yield estimates. If an investigator has an idea of what GCVI time series-based summary statistic $G_{\text{obs}}$ and weather covariates $W_{\text{obs}}$ they would use to estimate crop yields for a particular crop type and region, they can ensure Assumptions (i) and (ii) are met and that forgoing fitting a SCYM model would only cost them an unknown scaling factor. This unknown scaling factor would be unnecessary if their goal is merely to assess directions and not magnitudes, or if they ultimately plan to calibrate their observational causal effect estimates with those from randomized experiments as in \cite{CREO_paper} or with subnational-level yield data as in Section \ref{sec:ConvertingToYieldScale}. 

\subsection{An empirical comparison when using nonlinear yield maps.}\label{sec:QDANNcomparison}

The previous theoretical results that the causal forest estimates when using SCYM versus a GCVI-based feature should only differ by an unknown scaling constant, rely crucially on the linear construction of the SCYM estimates (Equation \eqref{eq:SCYM_ests}). Given that we suspect that nonlinear crop yield predictions using satellite data will become increasingly used, in this subsection we also empirically check how our results using peak GCVI compare to those when using nonlinear yield estimates as an outcome variable. We use the corn, soybean, and winter wheat yield predictions in the United States that were produced in \cite{QDANNPaper}. \cite{QDANNPaper} uses a transfer learning approach called Quantile loss Domain Adversarial Neural Networks (QDANN) to estimate crop yields at 30m $\times$ 30m resolution, using satellite data, weather data, and county-level yield data as inputs. 

We compared the results of our causal forest analysis when using QDANN as the outcome variable versus when using normalized peak GCVI as the outcome variable (as is done in the main text of the paper). To create a head-to-head comparison, we ran the analyses using the same random seed for both outcome variable choices and when restricting our attention to the same exact samples (east of the 100° W line) such that both peak GCVI and QDANN estimates were available. 
We then rescaled the estimated effects that were based on peak GCVI to be in units of percent of the average QDANN yield using a county-level regression rescaling approach similar to that used in Section \ref{sec:ConvertingToYieldScale}. In particular, for each of 7 rotations in the US, we fit a linear regression of county-level averages of QDANN yield estimates on county-level averages of normalized peak GCVI with sample weights according to the number of the rotation-specific overlap samples in each county (i.e., the sample weights for each county were based on the number of samples $i$ in each county such that the propensity score $\hat{\pi}(x_i)$ satisfied $0.05 \leq \hat{\pi}(x_i) \leq 0.95$). We then used the estimated slope coefficients from these regressions and the overall average QDANN yield of each crop type to convert the peak GCVI-based results to be in units of the percent of average estimated QDANN yield (of the outcome crop of interest). Finally, we also used the average QDANN yield estimates to convert results based on those yield estimates to also be in units of percent of the average QDANN yield (of the outcome crop of interest). 

\begin{figure}
    \centering
    \includegraphics[width=0.95\linewidth]{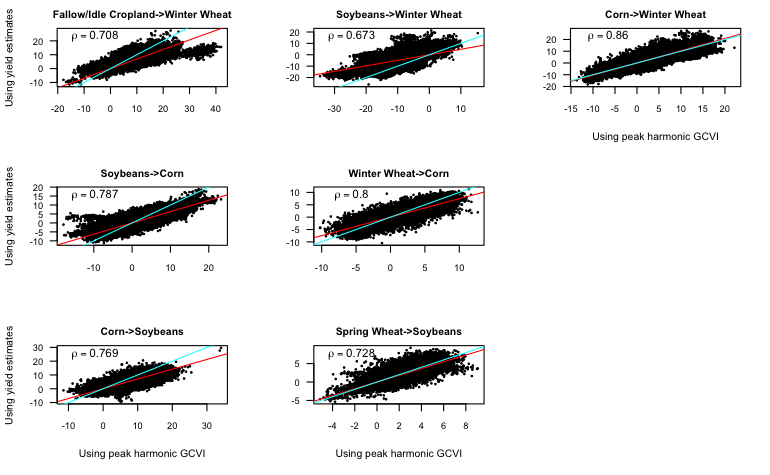}
    \caption{Comparison of CATE estimates when using estimated yield versus peak GCVI as the outcome variable. The x-axis plots CATE estimates when using peak GCVI as the outcome variable (after rescaling to be in units of \% of average estimated QDANN yield). The y-axis gives the CATE estimates when using QDANN yield estimates \cite{QDANNPaper} as the outcome variable (and is rescaled to be in units of \% of the average estimated QDANN yield). Only points with propensity scores between $0.05$ and $0.95$ are plotted. The text in the top left of each plot gives the Pearson correlation. The red line is the best fit line among lines that go through the origin. The cyan line has slope 1 and goes through the origin.}
    \label{fig:CATEscatterPlotsQDANN}
\end{figure}

In Figure \ref{fig:CATEscatterPlotsQDANN}, for each of 7 rotations in the US, we plot the estimated conditional average treatment effect (CATE) at each sample in the overlap region (i.e., we plot the values of $\hat{\omega}(x_i)$ for the samples $i$ such that $0.05 \leq \hat{\pi}(x_i) \leq 0.95$) when using peak GCVI as the outcome variable (x-axis) versus when using QDANN yield estimates as the outcome variable (y-axis). The correlations between the CATEs are quite high and moreover, lines through the origin fit the scatter plots quite well. This suggests that even though QDANN is a nonlinear approach to estimating yield, the theoretical conclusions from the previous subsections approximately hold in the sense that QDANN based results can be approximated by some unknown rescaling of peak GCVI-based results. In addition, the agreement between the best fit line through the origin (red) and the line through the origin with slope 1 (cyan) suggests that weighted regressions involving county-level averages (which were used in the main text and were used to rescale the x-axes in Figure \ref{fig:CATEscatterPlotsQDANN}) often give good approximations of this unknown rescaling factor.

\begin{figure}
    \centering
    \includegraphics[width=0.95\linewidth]{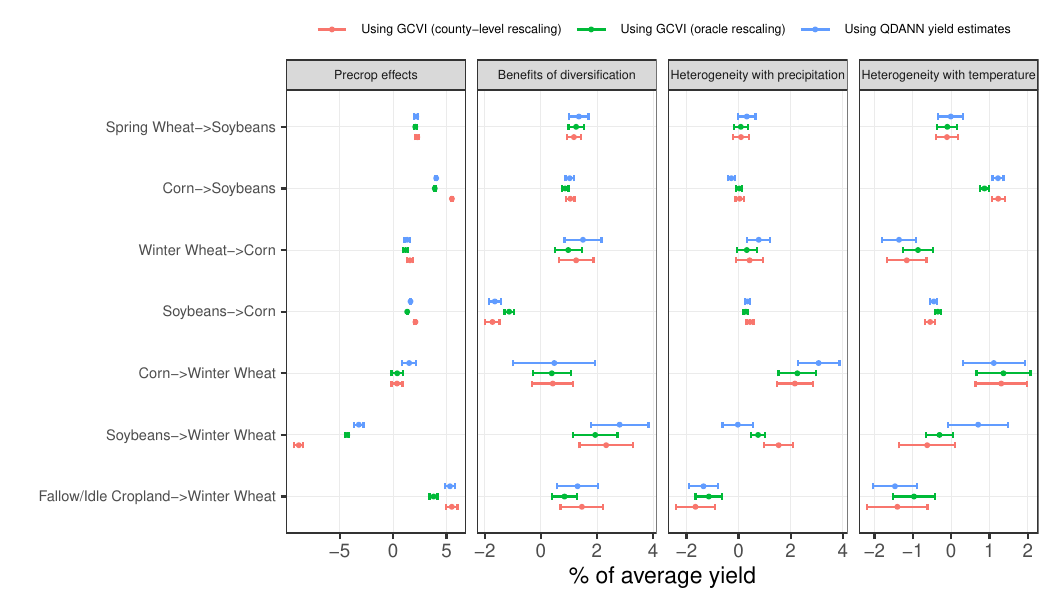}
    \caption{Comparison of results when using QDANN yield estimates as the outcome variable. The results presented are based on a subset of our sample in the US. The precrop and outcome crop studied are given on the y-axis. The first column gives the estimated precrop effects (Section \ref{sec:CausalForestAnalysis}), the second column gives estimated diversification effects (Section \ref{sec:RotDiversityMethodDescription}), and the final two columns give the rescaled heterogeneity coefficients (Section \ref{sec:HeterogeneityAnalysisMethod}). All results were rescaled to be in units of \% of average estimated QDANN yield. Results in blue were based on using QDANN yield estimates from \cite{QDANNPaper} as the outcome variable and normalizing by average QDANN estimates of yield for the outcome crop of interest. Results in green used peak GCVI as the outcome variable but assumed prior knowledge of the rescaling factor relating the CATE estimates based on QDANN with CATE estimates based on peak GCVI. Results in red also used peak GCVI as the outcome variable but assumed no prior knowledge of the rescaling factor. For the results in red, the rescaling factor for each cropping sequence was derived by regressing county-level averages of QDANN yield estimates on county-level averages using peak GCVI.}
    \label{fig:QDANNversusGCVIrescaledResults}
\end{figure}

In Figure \ref{fig:QDANNversusGCVIrescaledResults}, for the 7 rotations where QDANN yield estimates were available, we plot the main outcomes studied in the text (estimated precrop effects, benefits of diversification, and rescaled heterogeneity coefficients with weather). The results are based on the previously mentioned US subsample and either use QDANN yield estimates (blue) or peak GCVI as the outcome variable (red, green). We consider two approaches for rescaling the GCVI-based results: one is to calibrate using the coefficients from a county-level regression of QDANN yield averages on peak GCVI averages (red) and the other is to use the slope of the best fit line through the origin from the scatter plots in Figure \ref{fig:CATEscatterPlotsQDANN} (green). The results show that whether one uses QDANN yield estimates as an outcome variable or peak GCVI does not change the sign of the results and typically does not impact whether or not they are statistically significant. Results based on QDANN roughly differ from those based on peak GCVI by some scalar as evidenced by the agreement between the estimates plotted in green versus those plotted in blue. However, when county-level yield estimates are used for rescaling, the results based on QDANN can have lower magnitude.

These results suggest that if one uses satellite-based estimates of crop yields instead of peak GCVI as the outcome variable, the results can change by an unknown scaling factor. 
The approach taken in this paper of fitting county-level regressions of average estimated yield on average peak GCVI typically does a good job of recovering the unknown scaling factor.
Finally, we remark that estimated effects and heterogeneity coefficients that were based on QDANN yield estimates in this appendix should not be viewed as the ground truth values of the effects and coefficients for a validation. In particular, QDANN yield estimates are based on satellite imagery and the prediction errors could be consequential (the QDANN estimates have $R^2=0.48, 0.32$ and $0.39$ when compared to ground truth yield data for corn, soybeans and winter wheat, respectively \cite{QDANNPaper}). In addition, the estimates presented this appendix can differ from those in the main text because only a subset of the US study region was considered in this appendix.

\section{Labels considered non-crop in each country}\label{sec:nonCropDefinitions}

In the benefit of diversification analysis described in Section \ref{sec:RotDiversityMethodDescription}, we considered rotations of the form B$\to$A$\to$B to be control units and the diverse treated units to be all samples of the form D$\to$A$\to$B, where D was any croptype that was distinct from A and B. When D was not an actual crop, we did not include it in the sample of treated units. In particular, in cases where D had any of the following classifications given in Table \ref{table:nonCrop_labels}, the sample was not included in the analysis.

\begin{table}[!ht]
    \centering\renewcommand\cellalign{lc}
        \caption{Classifications in crop maps from each country that were deemed as non-crop. When the first year crop in a 3-year rotation had one of these classifications, the sample was not considered in our analysis of the benefit of rotational diversity. The classifications in the US were taken from the USDA Cropland Data Layer \cite{CDL_Boryan}. The classifications in Canada were taken from the Annual Crop Inventory \cite{CanadaACI}. The classifications in France were taken from the French Parcel dataset \cite{FrenchParcel} and subsequently translated to English using Google Translate.}
    \setcellgapes{3pt}\makegapedcells
    \begin{tabular}{| c | l |}
    \hline 
        Country & Classifications that were deemed as non-crop \\
        \hline
        US &  \makecell{Forest, Wetlands , Deciduous Forest, Developed/Open Space, \\ Developed/Low Intensity, Woody Wetlands, Shrubland, Open Water, \\
        Evergreen Forest, Developed/Med Intensity, Barren, Mixed Forest, \\
        Aquaculture, Developed/High Intensity, Christmas Trees, \\ Fallow/Idle Cropland, and Grassland/Pasture} \\ \hline 
        Canada & \makecell{Forest (undifferentiated), Peatland, Cloud, Coniferous, Water, \\ Exposed Land and Barren, Shrubland, Urban and Developed, Wetland,\\ Too Wet to be Seeded, Grassland, Fallow, Pasture and Forages} \\ \hline
        France & \makecell{Other temporary grassland 5 years or less, Fallow of 5 years or less, \\ Wooded area on former farmland, \\ Pastoral area - predominant ligneous fodder resources, \\ black fallow, Eligible strip along a forest without production, \\ 
        Cocksfoot 5 years or younger, Fallow of 6 years or more, \\
        Long rotation meadow (6 years or more), \\ Agricultural area temporarily not exploited, \\ Fallow land of 6 years or more declared as Ecological Focus Area, \\ buffer strip, and Permanent grassland - predominant grass (woody \\ fodder resources absent or scarce)} \\ \hline
    \end{tabular}
    \label{table:nonCrop_labels}
\end{table}

\section{Validating transferability of US-based GCVI to yield calibration}\label{sec:TransferabilityOfUSYieldCalib}

In this appendix, we compare the calibration of GCVI to yield that uses subnational-level yield data only from the US considered in the main text with a calibration of GCVI to yield that also uses subnational-level yield data from Canada and France. We also demonstrate that the US-based calibration is more robust for use in other regions when, as is done in the main text, the calibration coefficients in the US are divided by the average yield the US.

The subnational-level yield data for Canada and France were downloaded from government websites. For Canada, we downloaded the historical average yield data for each crop type, Census Agricultural Region (CAR), and year published by Agriculture and Agri-Food Canada \cite{CanadaSurveyData}. For France, we downloaded the historical average yield data for each crop type, Department, and year published by the French Ministry of Agriculture and Food, Agreste \cite{FranceSurveyData}.

To compare a US-based calibration with those from Canada and France, for each outcome crop $j$ and year $t$, we let $\bar{Y}_{c,t}^{(j,\text{Ca})}$ be the average yield in each CAR $c$ based on the Canadian survey data and $\bar{Y}_{c,t}^{(j,\text{Fr})}$ denote the average yield in each Department $c$ based on the French survey data. Similarly, for each outcome crop $j$ and year $t$, we let $\bar{V}_{c,t}^{(j,\text{Ca})}$ be the average normalized peak GCVI in each CAR $c$ in Canada and $\bar{V}_{c,t}^{(j,\text{Fr})}$ be the average normalized peak GCVI in each Department $c$ in France based on the sample described in Section \ref{sec:DatasetAndSample}. For each crop type $j$ we then fit the following weighted linear regressions

$$\bar{Y}_{c,t}^{(j,\text{Ca})} = \alpha^{(j,\text{Ca})} +\lambda^{(j,\text{Ca})} \bar{V}_{c,t}^{(j,\text{Ca})} + \varepsilon_{c,t}^{(j,\text{Ca})} \quad \text{and} \quad \bar{Y}_{c,t}^{(j,\text{Fr})} = \alpha^{(j,\text{Fr})} +\lambda^{(j,\text{Fr})} \bar{V}_{c,t}^{(j,\text{Fr})} + \varepsilon_{c,t}^{(j,\text{Fr})},$$ where the weights were proportional to the sample sizes in our satellite-based dataset. We let $\hat{\alpha}^{(j,\text{Ca})}$, $\hat{\lambda}^{(j,\text{Ca})}$, $\hat{\alpha}^{(j,\text{Fr})}$, and $\hat{\lambda}^{(j,\text{Fr})}$ denote the slope and intercept coefficients learned from the above regressions and further let $\bar{Y}^{(j,\text{Ca})}$ and $\bar{Y}^{(j,\text{Fr})}$ be the weighted average of yield of crop type $j$ based on subnational-level yield data in Canada and France (using the same weights as used in the linear regressions). Finally we let $\hat{\alpha}^{(j,\text{US})}$ and $\hat{\lambda}^{(j,\text{US})}$ denote the calibration coefficients learned from the regression in Equation \eqref{eq:VItoYieldCalibration} in the main text and let $\bar{Y}^{(j,\text{US})}$ be the weighted average of crop yield based on county-level yield data in the US described in Section \ref{sec:ConvertingToYieldScale}.

We then considered the following three approaches for estimating the average crop yield for each crop type $j$, year $t$, and subnational unit $c$.

\begin{enumerate}
    \item \textbf{US-based calibration (non-normalized):} Estimate $\bar{Y}_{c,t}^{(j,\text{Ca})}$, $\bar{Y}_{c,t}^{(j,\text{Fr})}$ and $\bar{Y}_{c,t}^{(j,\text{US})}$ with $$\hat{\alpha}^{(j,\text{US})}+\hat{\lambda}^{(j,\text{US})}\bar{V}_{c,t}^{(j,\text{Ca})}, \quad \hat{\alpha}^{(j,\text{US})}+\hat{\lambda}^{(j,\text{US})}\bar{V}_{c,t}^{(j,\text{Fr})}, \quad \text{and } \hat{\alpha}^{(j,\text{US})}+\hat{\lambda}^{(j,\text{US})}\bar{V}_{c,t}^{(j,\text{US})},$$ respectively. 
    
    \item \textbf{US-based calibration (normalized):} Estimate $\bar{Y}_{c,t}^{(j,\text{Ca})}$ and $\bar{Y}_{c,t}^{(j,\text{Fr})}$ with $$\big( \frac{\hat{\alpha}^{(j,\text{US})}}{\bar{Y}^{(j,\text{US})}}+\frac{\hat{\lambda}^{(j,\text{US})}}{\bar{Y}^{(j,\text{US})}} \times \bar{V}_{c,t}^{(j,\text{Ca})} \big) \times \bar{Y}^{(j,\text{Ca})} \quad \text{and} \quad \big( \frac{\hat{\alpha}^{(j,\text{US})}}{\bar{Y}^{(j,\text{US})}}+\frac{\hat{\lambda}^{(j,\text{US})}}{\bar{Y}^{(j,\text{US})}} \times \bar{V}_{c,t}^{(j,\text{Fr})} \big) \times \bar{Y}^{(j,\text{Fr})},$$ respectively. By cancelation, estimate $\bar{Y}_{c,t}^{(j,\text{US})}$ with $\hat{\alpha}^{(j,\text{US})}+\hat{\lambda}^{(j,\text{US})}\bar{V}_{c,t}^{(j,\text{US})}$.

    \item \textbf{Local calibration:} Estimate $\bar{Y}_{c,t}^{(j,\text{Ca})}$, $\bar{Y}_{c,t}^{(j,\text{Fr})}$ and $\bar{Y}_{c,t}^{(j,\text{US})}$ with $$\hat{\alpha}^{(j,\text{Ca})}+\hat{\lambda}^{(j,\text{Ca})}\bar{V}_{c,t}^{(j,\text{Ca})}, \quad \hat{\alpha}^{(j,\text{Fr})}+\hat{\lambda}^{(j,\text{Fr})}\bar{V}_{c,t}^{(j,\text{Fr})}, \quad \text{and } \hat{\alpha}^{(j,\text{US})}+\hat{\lambda}^{(j,\text{US})}\bar{V}_{c,t}^{(j,\text{US})},$$ respectively. 
\end{enumerate}

We compared the accuracy of the three approaches for estimating $\bar{Y}_{c,t}^{(j,\text{Ca})}$, $\bar{Y}_{c,t}^{(j,\text{Fr})}$ and $\bar{Y}_{c,t}^{(j,\text{US})}$ by using a weighted RMSE, where the weights used for computing the RMSE were proportional to the number of samples in our satellite-based dataset. The weighted RMSEs for the 3 approaches are presented in Table \ref{table:GCVItoYieldRMSEtable}.

\begin{table}[ht] \caption{RMSE for various calibration approaches. For each country and crop type presented in the table, the weighted RMSE for 3 different calibration approaches to estimating subnational-level crop yields are reported. The 3rd column gives the weighted RMSE for the US-based calibration approach that does not normalize by the weighted average of crop yield (Approach 1), the 4th column gives the weighted RMSE for the US-based calibration approach that does normalize by the weighted average of crop yield (Approach 2), and the 5th column uses local subnational-level yield data to fit the calibration model (Approach 3). All weighted RMSE values printed in the table are in units of tons per hectare. }\label{table:GCVItoYieldRMSEtable}
\footnotesize
\centering
\begin{tabular}{llrrr}
  \hline
Country & Crop type & US-based (non-normalized) & US-based (normalized)  & Local calibration \\ \hline \hline
Canada & Corn & 1.326 & 0.963 & 0.958 \\ 
  Canada & Soybean & 0.492 & 0.324 & 0.324 \\ 
  Canada & Spring Wheat & 0.511 & 0.510 & 0.506 \\ \hline
  France & Corn & 1.631 & 1.249 & 1.227 \\ 
  France & Winter Wheat & 4.583 & 1.357 & 1.349 \\ \hline
  US & Corn & 1.249 & 1.249 & 1.249 \\ 
  US & Soybean & 0.472 & 0.472 & 0.472 \\ 
  US & Spring Wheat & 0.532 & 0.532 & 0.532 \\ 
  US & Winter Wheat & 0.799 & 0.799 & 0.799 \\ 
   \hline
\end{tabular}
\end{table}

The results in Table \ref{table:GCVItoYieldRMSEtable} show that the first approach, which does not normalize by average US crop yields, leads to far worse performance than the second approach, which normalizes by average US crop yields. Meanwhile, we see that the third approach, which uses a locally fit calibration model, performs only slightly better than the second approach, which uses a normalized US-based calibration. Therefore, in the main text we use the second approach because of its superior performance to the non-normalized approach and because, unlike the local calibration approach which performs only slightly better, it can readily be used in countries such as China where we do not have access to subnational-level yield data at sufficiently high resolution. 

We also remark that the weighted RMSEs in Table \ref{table:GCVItoYieldRMSEtable} for winter wheat yields in France are relatively high compared to those in the US. One explanation for this is that the French subnational-level yield data downloaded from \cite{FranceSurveyData} included spelt and soft winter wheat in the same category whereas our data sample from which we calculated the average normalized peak GCVI only included soft winter wheat. Thus, the weighted RMSE reported for winter wheat in France using the US-based calibration approach may be artificially high. Further, given that the subnational-level yield data in France does not distinguish between spelt and soft winter wheat, a local calibration with such data in France could be unreliable and less accurate than a US-based calibration. 

\section{Checking fit of causal forests}\label{sec:CausalForestGoodnessOfFits}

We considered a few metrics to assess the goodness of fit of the fitted causal forest and propensity score models that were used in this study. The goodness of fit metrics are described below and presented in Table \ref{table:ModelGoodnessOfFitMetrics} for each causal forest that was used to study 2-year cropping sequences.

First, we to assess how well calibrated the fitted propensity score function was, we used the Expected Calibration Error (ECE) evaluated on the training data. As noted in \cite{ReviewPaperWithECEmetric}, ECE is a widely used binning-based measure of how well calibrated are the estimated probabilities that binary variables take on certain values. In our case, a random forest is used to fit the propensity model $\hat{\pi}(\cdot)$, where $\hat{\pi}(\cdot)$ estimates $\pi(x)=\mathbb{P}(Z=1 \giv X=x)$, which is the probability that a rotation occurred given a specific realization of the covariates. We computed the ECE of $\hat{\pi}(\cdot)$ by splitting the overlap samples into 9 equally spaced bins. For each $k=1,\dots,9$ we let $B_k$ denote all training samples $i$ for which $\hat{\pi}(x_i) \in [\frac{2k-1}{20},\frac{2k+1}{20})$. Note that if $\hat{\pi}(\cdot)$ were well calibrated, we would expect the average of the $Z_i$ in bin $B_k$ (given by $\frac{1}{\vert  B_k \vert } \sum_{i \in B_k} Z_i$) to be approximately equal to the $\hat{\pi}(\cdot)$-based estimate of its expectation (given by $\frac{1}{\vert B_k \vert} \sum_{i \in B_k} \hat{\pi}(x_i)$). The ECE is a weighted mean absolute error of such approximations where the weights are proportional to the bin sizes. In our case we use $$\text{ECE}\big(\hat{\pi}(\cdot)\big)=\sum_{k=1}^9 w_k \big| \frac{1}{\vert B_k \vert} \sum_{i \in B_k} \hat{\pi}(x_i)-\frac{1}{\vert  B_k \vert } \sum_{i \in B_k} Z_i \big| \quad  \text{where } w_k \equiv \frac{\vert B_k \vert}{\sum_{k'=1}^9 \vert B_{k'} \vert}.$$ The ECE values are reported in the final column of Table \ref{table:ModelGoodnessOfFitMetrics} and ranged from 0.005--0.027, suggesting the estimated propensity score function was well calibrated in all examples in the study.

Next we assessed how accurate predictions of the outcome variable $V$ were based on the causal forest. Note that the causal forest does not involve fitting a model to estimate the function $\mu_V(x,z)=\e[V \giv X=x,Z=z]$, but this function can be estimated using random forest models that the causal forest did fit. In particular, recall that when training the causal forest estimates of the function $\omega(x)=\e[V \giv X=x,Z=1]-\giv X=x,Z=0]$ and the function  $\pi(x)=\mathbb{P}(Z=1 \giv X=x)$. Further the \texttt{causal\_forest} function in R automatically returned a random forest-based estimate of the function $\bar{\mu}_V(x)=\e[V \giv X=x]$. Since $Z$ is a binary variable, note that by the law of total expectation and reordering terms $$\mu_V(x,1)=\bar{\mu}_V(x) + (1-\pi(x)) \omega(x) \quad \text{and} \quad \mu_V(x,0)=\bar{\mu}_V(x) -\pi(x) \omega(x).$$ Using the estimates of the functions $\bar{\mu}_V(\cdot)$, $\pi(\cdot)$, and $\omega(\cdot)$, returned by the \texttt{causal\_forest} which we call $\hat{\bar{\mu}}_V(\cdot)$, $\hat{\pi}(\cdot)$, and $\hat{\omega}(\cdot)$, we constructed estimates of $V$ as a function of $x$ and $z$ using $$\hat{V}(x,z)= \hat{\bar{\mu}}_V(x) + (z-\hat{\pi}(x)) \hat{\omega}(x) = \begin{cases} \hat{\bar{\mu}}_V(x) + (1-\hat{\pi}(x)) \hat{\omega}(x) & \text{ if } z=1, \\
\hat{\bar{\mu}}_V(x) -\hat{\pi}(x) \hat{\omega}(x) & \text{ if } z=0.
\end{cases}$$ We computed these causal forest based estimates of $\hat{V}$ on all the training samples and calculated the RMSE relative to the actual $V$. The RMSE was then rescaled to be in units of \% of the average yield for that particular outcome crop and country (using the approach in Section \ref{sec:ConvertingToYieldScale}). The rescaled RMSEs are reported in the penultimate column of Table \ref{table:ModelGoodnessOfFitMetrics}, and ranged from $13.2\%$--$31.8\%$.

Finally we used the \texttt{test\_calibration} function in the \texttt{grf} package to assess how well calibrated the fitted causal forests were. The \texttt{test\_calibration} function computes the best fit in a linear regression (on held-out data) with a rescaled estimated ATO and a rescaled gap between $\omega(X)$ and the estimated ATO as the sole two regressors. The rescaling factors and outcome variable were chosen such that if the causal forest were perfectly calibrated and if $\omega(X)$ were nonconstant (i.e. there were real treatment effect heterogeneity), the two regression coefficients would be $1$. In particular, the \texttt{test\_calibration} automatically fit the following weighted linear regression on held out data with weights proportional to the overlap weights $\hat{\pi}(X)(1-\hat{\pi}(X))$ used in the study $$V-\hat{\bar{\mu}}_V(X)= \beta_{\bar{\omega}} \big( Z-\hat{\pi}(X) \big) \overline{\hat{\omega}(X)} + \beta_{\Delta} \big( Z-\hat{\pi}(X) \big) \big(
\hat{\omega}(X) - \overline{\hat{\omega}(X)} \big) + \varepsilon.$$ In the above linear regression, $\hat{\bar{\mu}}_V(x)$ is the previously mentioned random forest-based estimate of $\e[V|X=x]$, $\hat{\pi}(x)$ is the propensity score estimating $\mathbb{P}(Z=1 \giv X=x)$, $\hat{\omega}(x)$ is the causal forest based estimate of $\omega(x)$ defined in Equation \eqref{eq:omegaDefinition}, $\overline{\hat{\omega}(X)}$ is the overlap weighted sample mean of $\hat{\omega}(X)$ (which gives the estimated ATO), and $\varepsilon$ is mean $0$ error. In the above model, we would expect $\beta_{\bar{\omega}}$ and $\beta_{\Delta}$ to be near 1 if the causal forests are well calibrated.
Indeed, in Table \ref{table:ModelGoodnessOfFitMetrics}, we find that the estimates of $\beta_{\bar{\omega}}$ and $\beta_{\Delta}$ are near 1 (with the exception of the Winter Wheat$\to$Corn rotation in the US and some of the rotations in Canada involving corn).

Overall, the results in Table \ref{table:ModelGoodnessOfFitMetrics} suggest that the fitted causal forests and propensity score models were well calibrated. The estimated values of $\beta_{\bar{\omega}}$ and $\beta_{\Delta}$ being near $1$ suggested that the causal forests were generally well calibrated and that the estimated rotation benefits were heterogeneous. The ECE metric being generally close to $0$ suggested that the propensity score models were well calibrated. The RMSE for the predicted yields ranging from $13.2\%$--$31.8\%$ of the average yield do not suggest the causal forest model was poorly calibrated and merely suggested that the features and rotation status did not explain all of the observed variation in yield (or peak GCVI). 

For the casual forests fit in Section \ref{sec:RotDiversityMethodDescription}, the RMSEs were a bit higher (ranging from $17.5\%$--$39.9\%$) and the ECE scores still suggested a well calibrated propensity score models (ranging from $0.006$--$0.028$). In some instances the causal forests used to assess diversification effects had estimates of $\beta_{\bar{\omega}}$ and $\beta_{\Delta}$ that were not near $1$, likely due to small sample sizes, although because we did not study heterogeneity of diversification effects with weather, we suspect such cases of poor calibration were not problematic for our analysis as we merely considered mean effects of diversification.

\section{How our heterogeneity with weather findings relate to existing claims in the literature}\label{sec:HeterogeneityWithWeatherUSversusLiterature}

We found higher rotation benefits at higher precipitation levels. Consistent with our findings, a recent metanalysis of 45 experiments in China \cite{ChinaExpirementalMetanalysis} found that experiments with mean annual precipitation above 550mm had higher rotation benefits averaging 25\%, while experiments with mean annual precipitation between 400mm and 550mm had much lower benefits averaging 9\%. However, \cite{ChinaExpirementalMetanalysis} was unable to conclude whether the finding was driven by actual changes in rotation benefit with weather rather than by variation in the type of rotations studied across different climatic regions. In contrast with our findings, an analysis of seven long term experiments in Europe \cite{EuropeERL_WeatherInteractions7LTE} found that there was a statistically significant negative interaction between precipitation and crop rotation in a model for winter cereal yields; however, they did not find such an interaction for spring cereal yields. Moreover, their result for winter cereal yields was based on only three experimental sites as well as a comparison of monoculture to highly diverse rotations, rather than the comparison of monoculture to simple rotations upon which we focus our heterogeneity analysis. Our findings are consistent with a recent satellite-based study in Finland that found higher precipitation during the outcome crop's growing season often increased the rotation benefit, while drought during either the precrop's or outcome crop's growing season tended to reduce the rotation benefits \cite{FinlandSatellite_based2024}. However, that study did not leverage a causal machine learning approach and averaged the yearly precipitation and drought metrics across large geographical regions within Finland. Our findings are also consistent with a satellite-based study in Belgium that found a strong negative Spearman correlation between the rotation benefit and climate water deficit \cite{BelgiumRotation_NPP_Giannarkis}, although that study did not look at different precrop and outcome crop combinations separately. 

We found that at higher temperatures rotation benefits were higher when the outcome crop was soybean but lower when the precrop was a legume. Consistent with our results, an analyses using a combination of experimental data and observational data \cite{CREO_paper,JunxiongPaper} found that under increasing temperatures the Corn-Soybean rotation in the US is more beneficial to soybean yields but less beneficial to corn yields. An analysis of seven long term experiments in Europe \cite{EuropeERL_WeatherInteractions7LTE} did not find a statistically significant interaction between temperature and crop rotation in a model for cereal yields, although their results average across experiments with different precrops and outcome crops. A recent metanalysis of 45 experiments in China \cite{ChinaExpirementalMetanalysis} found that experiments with low mean annual temperature had much higher rotation benefits on average than did experiments with high mean annual temperature; however, the experiments in colder regions did not have the same precrop and outcome crops as those in warmer regions. The authors explain that the higher benefits from rotation in cooler regions can be explained by the more frequent selection of legumes as precrops in cooler regions. 

Other recent works suggest rotation benefits vary with weather, but do not provide precise insight into how the benefits vary with weather. A recent global metanalysis of 462 field experiments involving legume precrops \cite{LegumePrecropMetanalysis} found that mean annual temperature and mean annual precipitation were relatively important predictors of the legume precrop effect on yield (nearly as import as the outcome crop type); however, the study did not assess the direction of the heterogeneity of the rotation benefit with these climatic variables. Another study using 11 long-term experiments in North America \cite{BowlesLTEpaper} analyzes the effect of crop rotational diversity on corn yields, and after defining a weather favorability index they find a positive interaction between the favorability index and rotational diversity in 2 out of the 11 experiments (the other experiments did not have a significant interaction). This gives some evidence that crop rotation is slightly more effective in more favorable weather conditions, but because their weather favorability index is simply given by the yearly average of (detrended) corn yields at each site, their results do not directly say anything about how rotation benefit changes with precipitation at fixed temperatures or how rotation benefit changes with temperature at fixed precipitation levels. 

\section*{Supplemental Tables and Figures}

\begin{table}[hbt!]
    \caption{Limitations of existing satellite-based studies on the impacts of crop rotation. Each number corresponds to a different study. The first row indicates whether a study focused on a single country or multiple countries. The second row indicates whether more than 2 types of rotation were studied (where each rotation is defined by a precrop and outcome crop combination). The third row indicates whether a study separately assessed the rotation benefit for each outcome crop and precrop combination. The fourth row indicates whether a study assessed the benefits of switching from a simple rotation to a more diverse rotation by separately considering the impacts when 3 distinct crops are grown in 3 consecutive years. The fifth row indicates whether a study assessed how the benefit of rotation varies with weather and the final row indicates whether such an assessment included uncertainty quantifications. The sixth row indicates whether a study used a formal causal inference approach. In particular, many of these studies simply compare average vegetation indices (such as NDVI, transformed NDVI, RVI, etc.) in the rotated samples versus those in non-rotated samples \cite{ChinaRiceToBeanRotation_SatelliteBased,ChinaRiceToCottonRotation_SatelliteBased,FinlandSatellite_based,FinlandSatellite_based2024} or fit a machine learning or regression model that does not use a formal causal inference framework \cite{UkraineRotation_satelliteBased,ObservationalApproachAustria,LawesEtAl_AustraliaMLToStudyRotBenefitsNotCausal}. The seventh row indicates whether a confidence interval is provided for the estimated rotation effects.}\label{table:ObsSatLitReview}
    \centering \scriptsize
    \begin{tabular}{l|l|l}
         & No & Yes   \\ \hline \hline
        
        Multiple countries? & \cite{CREO_paper,ChinaRiceToBeanRotation_SatelliteBased,ChinaRiceToCottonRotation_SatelliteBased,UkraineRotation_satelliteBased,FinlandSatellite_based,FinlandSatellite_based2024,ObservationalApproachAustria,BelgiumRotation_NPP_Giannarkis,JunxiongPaper,LawesEtAl_AustraliaMLToStudyRotBenefitsNotCausal} & \\ \hline
        
        More than 2 types of rotation studied? & \cite{CREO_paper,ChinaRiceToBeanRotation_SatelliteBased,ChinaRiceToCottonRotation_SatelliteBased,JunxiongPaper} & \cite{UkraineRotation_satelliteBased,FinlandSatellite_based,FinlandSatellite_based2024,ObservationalApproachAustria,BelgiumRotation_NPP_Giannarkis,LawesEtAl_AustraliaMLToStudyRotBenefitsNotCausal} \\ \hline

        Stratifies based on precrop and outcome crop? & \cite{BelgiumRotation_NPP_Giannarkis}  & \cite{CREO_paper,ChinaRiceToBeanRotation_SatelliteBased,ChinaRiceToCottonRotation_SatelliteBased,UkraineRotation_satelliteBased,FinlandSatellite_based,FinlandSatellite_based2024,ObservationalApproachAustria,JunxiongPaper,LawesEtAl_AustraliaMLToStudyRotBenefitsNotCausal} \\ \hline

        Assesses benefit of diversification? & \cite{CREO_paper,ChinaRiceToBeanRotation_SatelliteBased,ChinaRiceToCottonRotation_SatelliteBased,UkraineRotation_satelliteBased,FinlandSatellite_based,FinlandSatellite_based2024,ObservationalApproachAustria,BelgiumRotation_NPP_Giannarkis,JunxiongPaper,LawesEtAl_AustraliaMLToStudyRotBenefitsNotCausal} & \\ \hline

        Assesses heterogeneity with weather? & \cite{ChinaRiceToBeanRotation_SatelliteBased,ChinaRiceToCottonRotation_SatelliteBased,UkraineRotation_satelliteBased,FinlandSatellite_based,ObservationalApproachAustria,LawesEtAl_AustraliaMLToStudyRotBenefitsNotCausal}\textsuperscript{\dag} & \cite{CREO_paper,FinlandSatellite_based2024,BelgiumRotation_NPP_Giannarkis,JunxiongPaper} \\ \hline

        Uses formal causal inference framework? & \cite{ChinaRiceToBeanRotation_SatelliteBased,ChinaRiceToCottonRotation_SatelliteBased,UkraineRotation_satelliteBased,FinlandSatellite_based,FinlandSatellite_based2024,ObservationalApproachAustria,LawesEtAl_AustraliaMLToStudyRotBenefitsNotCausal}  & \cite{CREO_paper,BelgiumRotation_NPP_Giannarkis,JunxiongPaper} \\ \hline

        Confidence intervals for estimated effects? & \cite{CREO_paper,FinlandSatellite_based,FinlandSatellite_based2024,ObservationalApproachAustria,JunxiongPaper,LawesEtAl_AustraliaMLToStudyRotBenefitsNotCausal} & \cite{ChinaRiceToBeanRotation_SatelliteBased,ChinaRiceToCottonRotation_SatelliteBased,UkraineRotation_satelliteBased,BelgiumRotation_NPP_Giannarkis} \\ \hline

        Confidence intervals for weather interactions? & \cite{CREO_paper,ChinaRiceToBeanRotation_SatelliteBased,ChinaRiceToCottonRotation_SatelliteBased,UkraineRotation_satelliteBased,FinlandSatellite_based,FinlandSatellite_based2024,ObservationalApproachAustria,BelgiumRotation_NPP_Giannarkis,LawesEtAl_AustraliaMLToStudyRotBenefitsNotCausal} & \cite{JunxiongPaper} 
    \end{tabular}
\vspace{-0.2cm}
\smallskip
\dag \cite{LawesEtAl_AustraliaMLToStudyRotBenefitsNotCausal} studied heterogeneity of rotation benefit with geographical location, discussing how it relates to weather conditions
\end{table}

\begin{table}[hbt!]
\footnotesize
\caption{Proportion of samples with low and high propensity scores. The 4th and 5th columns give the number of treated and control units used in each of our causal forest analyses, respectively. The proportion of samples (among the treated and control units combined) which were estimated to have low propensity score (less than $0.05$) or high propensity score (greater than $0.95$) are rounded to the nearest hundredth and displayed in the 6th and 7th columns, respectively. Here C=Corn, S=Soybean, WW=Winter Wheat, SW=Spring Wheat, P= Pasture and Forages, UW=Wheat (undistinguished between WW and SW), WB=Winter Barley, R=Rapeseed, Su=Sunflower, DB=Dry Beans, Le=Lentil, Pe=Pea, F=Fallow, Si=Corn Silage, Ch=Chard/non-fodder beet.}\label{table:OverlapTable}
\centering
\begin{tabular}{lllllcc}
  \hline
Country & Treatment & Control & $n_{\text{Treat}}$ &  $n_{\text{Control}}$ & $\hat{\pi}(x) < 0.05$ & $\hat{\pi}(x) > 0.95$  \\ 
  \hline
  \hline US & S$\to$C & C$\to$C & 811,541 & 432,622 & 0.04 & 0.05 \\ 
  US & WW$\to$C & C$\to$C & 61,653 & 432,622 & 0.62 & 0.00 \\ 
  Canada & S$\to$C & C$\to$C & 21,312 & 16,497 & 0.03 & 0.01 \\ 
  Canada & P$\to$C & C$\to$C & 4,437 & 16,497 & 0.03 & 0.00 \\ 
  Canada & WW$\to$C & C$\to$C & 3,016 & 16,497 & 0.65 & 0.00 \\ 
  Canada & UW$\to$C & C$\to$C & 2,380 & 16,497 & 0.58 & 0.00 \\ 
  Canada & SW$\to$C & C$\to$C & 2,015 & 16,497 & 0.64 & 0.00 \\ 
  China & S$\to$C & C$\to$C & 28,133 & 111,379 & 0.36 & 0.02 \\ 
  France & WW$\to$C & C$\to$C & 11,415 & 15,327 & 0.16 & 0.01 \\ 
  France & WB$\to$C & C$\to$C & 2,123 & 15,327 & 0.48 & 0.00 \\ 
   \hline US & C$\to$S & S$\to$S & 843,666 & 269,506 & 0.00 & 0.25 \\ 
  US & SW$\to$S & S$\to$S & 57,524 & 269,506 & 0.65 & 0.00 \\ 
  Canada & C$\to$S & S$\to$S & 26,706 & 19,068 & 0.00 & 0.01 \\ 
  Canada & SW$\to$S & S$\to$S & 9,107 & 19,068 & 0.48 & 0.01 \\ 
  Canada & R$\to$S & S$\to$S & 8,665 & 19,068 & 0.51 & 0.01 \\ 
  Canada & P$\to$S & S$\to$S & 3,412 & 19,068 & 0.20 & 0.00 \\ 
  Canada & UW$\to$S & S$\to$S & 2,829 & 19,068 & 0.60 & 0.00 \\ 
  China & C$\to$S & S$\to$S & 19,884 & 34,969 & 0.01 & 0.00 \\ 
   \hline US & S$\to$SW & SW$\to$SW & 68,494 & 35,515 & 0.16 & 0.23 \\ 
  US & R$\to$SW & SW$\to$SW & 18,061 & 35,515 & 0.21 & 0.02 \\ 
  US & Su$\to$SW & SW$\to$SW & 14,730 & 35,515 & 0.05 & 0.05 \\ 
  US & DB$\to$SW & SW$\to$SW & 7,624 & 35,515 & 0.62 & 0.00 \\ 
  US & C$\to$SW & SW$\to$SW & 7,693 & 35,515 & 0.26 & 0.00 \\ 
  Canada & R$\to$SW & SW$\to$SW & 199,115 & 44,809 & 0.00 & 0.28 \\ 
  Canada & Le$\to$SW & SW$\to$SW & 33,892 & 44,809 & 0.28 & 0.05 \\ 
  Canada & Pe$\to$SW & SW$\to$SW & 30,637 & 44,809 & 0.01 & 0.05 \\ 
  Canada & S$\to$SW & SW$\to$SW & 11,332 & 44,809 & 0.68 & 0.05 \\ 
  Canada & F$\to$SW & SW$\to$SW & 10,139 & 44,809 & 0.37 & 0.03 \\ 
   \hline US & F$\to$WW & WW$\to$WW & 84,845 & 151,613 & 0.38 & 0.05 \\ 
  US & S$\to$WW & WW$\to$WW & 35,241 & 151,613 & 0.63 & 0.09 \\ 
  US & C$\to$WW & WW$\to$WW & 22,417 & 151,613 & 0.59 & 0.00 \\ 
  France & R$\to$WW & WW$\to$WW & 32,045 & 18,119 & 0.00 & 0.01 \\ 
  France & Si$\to$WW & WW$\to$WW & 14,917 & 18,119 & 0.11 & 0.06 \\ 
  France & C$\to$WW & WW$\to$WW & 13,378 & 18,119 & 0.01 & 0.02 \\ 
  France & Ch$\to$WW & WW$\to$WW & 9,720 & 18,119 & 0.29 & 0.00 \\ 
  France & Su$\to$WW & WW$\to$WW & 8,401 & 18,119 & 0.34 & 0.02 \\ 
   \hline
\end{tabular}
\end{table} 

\begin{table}[hbt!]
\footnotesize
\caption{Metrics assessing how well calibrated the causal forest models were. The 6th and 7th columns give the estimated coefficients in a regression on held out data which assessed how well calibrated the fitted causal forest was (coefficients near $1$ imply well calibrated models and the existence of heterogeneity). The penultimate column gives the RMSE of the causal forest-based prediction of yields (in units of percent of average yield). The final column gives the ECE measure of how well calibrated the propensity score model is, with ECE values nearer to $0$ implying better calibrated propensity scores. See Section \ref{sec:CausalForestGoodnessOfFits} for more details. Here C=Corn, S=Soybean, WW=Winter Wheat, SW=Spring Wheat, P= Pasture and Forages, UW=Wheat (undistinguished between WW and SW), WB=Winter Barley, R=Rapeseed, Su=Sunflower, DB=Dry Beans, Le=Lentil, Pe=Pea, F=Fallow, Si=Corn Silage, Ch=Chard/non-fodder beet.}\label{table:ModelGoodnessOfFitMetrics}
\centering
\begin{tabular}{lllllrrll}
  \hline
Country & Treatment & Control & $n_{\text{Treat}}$ &  $n_{\text{Control}}$  &  $\hat{\beta}_{\bar{\omega}}$ & $\hat{\beta}_{\Delta}$ & RMSE & $\text{ECE}\big(\hat{\pi}(\cdot)\big)$ \\ 
  \hline
\hline US & S$\to$C & C$\to$C & 811,541 & 432,622 & 1.00 & 0.84 & \ 18.8\% & 0.008 \\ 
  US & WW$\to$C & C$\to$C & 61,653 & 432,622 & 0.60 & 0.94 & \ 19.4\% & 0.011 \\ 
  Canada & S$\to$C & C$\to$C & 21,312 & 16,497 & 1.04 & 0.58 & \ 22.8\% & 0.007 \\ 
  Canada & P$\to$C & C$\to$C & 4,437 & 16,497 & 0.93 & 0.47 & \ 24.0\% & 0.006 \\ 
  Canada & WW$\to$C & C$\to$C & 3,016 & 16,497 & 1.02 & 0.54 & \ 23.7\% & 0.014 \\ 
  Canada & UW$\to$C & C$\to$C & 2,380 & 16,497 & 1.09 & 0.71 & \ 23.6\% & 0.015 \\ 
  Canada & SW$\to$C & C$\to$C & 2,015 & 16,497 & 0.56 & 0.73 & \ 23.6\% & 0.016 \\ 
  China & S$\to$C & C$\to$C & 28,133 & 111,379 & 1.00 & 0.90 & \ 13.2\% & 0.010 \\ 
  France & WW$\to$C & C$\to$C & 11,415 & 15,327 & 0.96 & 0.90 & \ 19.8\% & 0.009 \\ 
  France & WB$\to$C & C$\to$C & 2,123 & 15,327 & 1.06 & 1.00 & \ 19.6\% & 0.009 \\ 
   \hline US & C$\to$S & S$\to$S & 843,666 & 269,506 & 1.00 & 0.88 & \ 20.6\% & 0.010 \\ 
  US & SW$\to$S & S$\to$S & 57,524 & 269,506 & 1.01 & 0.85 & \ 21.5\% & 0.007 \\ 
  Canada & C$\to$S & S$\to$S & 26,706 & 19,068 & 0.99 & 0.75 & \ 27.2\% & 0.014 \\ 
  Canada & SW$\to$S & S$\to$S & 9,107 & 19,068 & 0.92 & 0.81 & \ 24.9\% & 0.018 \\ 
  Canada & R$\to$S & S$\to$S & 8,665 & 19,068 & 0.96 & 0.97 & \ 25.0\% & 0.017 \\ 
  Canada & P$\to$S & S$\to$S & 3,412 & 19,068 & 0.95 & 0.53 & \ 27.0\% & 0.005 \\ 
  Canada & UW$\to$S & S$\to$S & 2,829 & 19,068 & 1.05 & 0.55 & \ 26.4\% & 0.013 \\ 
  China & C$\to$S & S$\to$S & 19,884 & 34,969 & 1.01 & 0.92 & \ 18.3\% & 0.013 \\ 
   \hline US & S$\to$SW & SW$\to$SW & 68,494 & 35,515 & 1.02 & 0.85 & \ 23.4\% & 0.011 \\ 
  US & R$\to$SW & SW$\to$SW & 18,061 & 35,515 & 1.01 & 0.97 & \ 22.3\% & 0.011 \\ 
  US & Su$\to$SW & SW$\to$SW & 14,730 & 35,515 & 0.98 & 0.98 & \ 22.4\% & 0.019 \\ 
  US & DB$\to$SW & SW$\to$SW & 7,624 & 35,515 & 1.01 & 0.95 & \ 23.2\% & 0.024 \\ 
  US & C$\to$SW & SW$\to$SW & 7,693 & 35,515 & 1.04 & 0.91 & \ 22.7\% & 0.011 \\ 
  Canada & R$\to$SW & SW$\to$SW & 199,115 & 44,809 & 1.01 & 1.06 & \ 22.7\% & 0.008 \\ 
  Canada & Le$\to$SW & SW$\to$SW & 33,892 & 44,809 & 1.02 & 1.11 & \ 21.1\% & 0.018 \\ 
  Canada & Pe$\to$SW & SW$\to$SW & 30,637 & 44,809 & 1.00 & 1.08 & \ 22.4\% & 0.027 \\ 
  Canada & S$\to$SW & SW$\to$SW & 11,332 & 44,809 & 1.00 & 0.82 & \ 21.9\% & 0.013 \\ 
  Canada & F$\to$SW & SW$\to$SW & 10,139 & 44,809 & 0.93 & 1.05 & \ 21.9\% & 0.013 \\ 
   \hline US & F$\to$WW & WW$\to$WW & 84,845 & 151,613 & 0.89 & 1.08 & \ 30.1\% & 0.011 \\ 
  US & S$\to$WW & WW$\to$WW & 35,241 & 151,613 & 0.94 & 1.06 & \ 31.8\% & 0.006 \\ 
  US & C$\to$WW & WW$\to$WW & 22,417 & 151,613 & 1.01 & 1.13 & \ 31.2\% & 0.012 \\ 
  France & R$\to$WW & WW$\to$WW & 32,045 & 18,119 & 1.00 & 0.85 & \ 19.1\% & 0.009 \\ 
  France & Si$\to$WW & WW$\to$WW & 14,917 & 18,119 & 0.99 & 0.82 & \ 19.3\% & 0.022 \\ 
  France & C$\to$WW & WW$\to$WW & 13,378 & 18,119 & 1.01 & 0.95 & \ 20.1\% & 0.023 \\ 
  France & Ch$\to$WW & WW$\to$WW & 9,720 & 18,119 & 0.99 & 0.81 & \ 19.3\% & 0.012 \\ 
  France & Su$\to$WW & WW$\to$WW & 8,401 & 18,119 & 1.01 & 0.87 & \ 19.4\% & 0.012 \\ 
   \hline
\end{tabular}
\end{table}
\newpage

\begin{table}[hbt!]
\scriptsize
\caption{Estimates of precrop effects, impacts of diversification, and heterogeneity of the precrop effects with weather. Column 6 gives the estimated precrop effect based on the method described in Section \ref{sec:CausalForestAnalysis} where the treated and control units are defined in Columns 2 and 3. The last two columns give the estimated rescaled regression coefficients (Section \ref{sec:HeterogeneityAnalysisMethod}) that describe the heterogeneity of this precrop effect with weather, which can be thought of as a rough estimate of how much the precrop effect given in Column 6 would change as precipitation (or temperature) increases by the gap between the 25th percentile and 75th percentile of precipitation (or temperature). Column 7 gives estimated effects of diversification. In particular for a row in which Column 2 says A$\to$B, Column 7 gives an estimate of the effect of diversification on B's yield where the control units have the form B$\to$A$\to$B and the treated units have the form Other Crop$\to$A$\to$B. Columns 4 and 5 give the number of samples of treated units and control units that were available to compute the entries in Column 6 whereas Column 7 was computed on a subset of the samples counted in Column 4. All quantities in Columns 6--9 were converted from the unitless normalized peak GCVI scale to the crop yield scale using the approach described in Section \ref{sec:ConvertingToYieldScale}. In particular all of the quantities in Columns 6--9 are in units of percent of the mean yield of the corresponding outcome crop type in the corresponding study region and can be converted to units of tons per hectare using Table \ref{Table:YieldToTonsPerHectare}. Stars indicate statistical significance (level $\alpha=0.05$, two-sided). Here C=Corn, S=Soybean, WW=Winter Wheat, SW=Spring Wheat, P= Pasture and Forages, UW=Wheat (undistinguished wheat (WW or SW)), WB=Winter Barley R=Rapeseed, Su=Sunflower, DB=Dry Beans, Le=Lentil, Pe=Pea, F=Fallow, Si=Corn Silage, Ch=Chard/non-fodder beet.}\label{table:ResultsTable}
\centering
\begin{tabular}{lllrrllll}
  \hline
Country  & Treatment & Control & $n_{\text{Treat}}$ &  $n_{\text{Control}}$ &  $\text{Eff}_{\text{2-year}}$ & $\text{Eff}_{\text{diverse}}$ & $\hat{\beta}^{\text{(resc.)}}_{\text{Precip}}$ & $\hat{\beta}^{\text{(resc.)}}_{\text{Temp}}$ \\ \hline
   \hline US & S$\to$C & C$\to$C & 811,541 & 432,622 & \ \ \thinspace 2.5\%* & \ \thinspace -3.4\%* & \ \ \thinspace 0.6\%* & \ \thinspace -1.0\%* \\ 
  US & WW$\to$C & C$\to$C & 61,653 & 432,622 & \ \ \thinspace 0.9\%* & \ \ \thinspace 0.7\%* & \ \ \thinspace 1.3\%* & \ \thinspace -1.6\%* \\ 
  Canada & S$\to$C & C$\to$C & 21,312 & 16,497 & \ \ \thinspace 1.7\%* & \ \thinspace -2.3\%* & \ \ \thinspace 1.3\%* & \ \thinspace -0.3\% \\ 
  Canada & P$\to$C & C$\to$C & 4,437 & 16,497 & \ \thinspace -2.0\%* & \ \ \thinspace 0.2\% & \ \ \thinspace 1.7\%* & \ \thinspace -1.4\%* \\ 
  Canada & WW$\to$C & C$\to$C & 3,016 & 16,497 & \ \ \thinspace 3.7\%* & \ \ \thinspace 5.2\%* & \ \ \thinspace 2.3\%* & \ \thinspace -0.9\% \\ 
  Canada & UW$\to$C & C$\to$C & 2,380 & 16,497 & \ \ \thinspace 3.9\%* & \ \ \thinspace 2.9\%* & \ \ \thinspace 1.2\% & \ \ \thinspace 1.7\%* \\ 
  Canada & SW$\to$C & C$\to$C & 2,015 & 16,497 & \ \thinspace -0.3\% & \ \ \thinspace 1.3\% & \ \ \thinspace 0.6\% & \ \ \thinspace 2.5\%* \\ 
  China & S$\to$C & C$\to$C & 28,133 & 111,379 & \ \thinspace -2.4\%* & \ \ \thinspace NA & \ \ \thinspace 0.5\%* & \ \thinspace -1.5\%* \\ 
  France & WW$\to$C & C$\to$C & 11,415 & 15,327 & \ \thinspace -0.9\%* & \ \thinspace -2.1\%* & \ \thinspace -1.5\%* & \ \thinspace -5.4\%* \\ 
  France & WB$\to$C & C$\to$C & 2,123 & 15,327 & \ \thinspace -0.9\% & \ \thinspace -1.9\% & \ \ \thinspace 1.6\% & \ \thinspace -6.0\%* \\ 
   \hline US & C$\to$S & S$\to$S & 843,666 & 269,506 & \ \ \thinspace 8.5\%* & \ \ \thinspace 0.7\%* & \ \ \thinspace 0.7\%* & \ \ \thinspace 1.4\%* \\ 
  US & SW$\to$S & S$\to$S & 57,524 & 269,506 & \ \ \thinspace 4.2\%* & \ \ \thinspace -0.0\% & \ \ \thinspace 0.8\%* & \ \ \thinspace 0.2\% \\ 
  Canada & C$\to$S & S$\to$S & 26,706 & 19,068 & \ \ \thinspace 6.8\%* & \ \ \thinspace 2.7\%* & \ \thinspace -0.3\% & \ \ \thinspace 4.0\%* \\ 
  Canada & SW$\to$S & S$\to$S & 9,107 & 19,068 & \ \ \thinspace 2.6\%* & \ \ \thinspace 3.8\%* & \ \ \thinspace 2.6\%* & \ \ \thinspace 2.1\%* \\ 
  Canada & R$\to$S & S$\to$S & 8,665 & 19,068 & \ \ \thinspace 2.1\%* & \ \ \thinspace 1.8\%* & \ \ \thinspace 1.7\%* & \ \ \thinspace 2.4\%* \\ 
  Canada & P$\to$S & S$\to$S & 3,412 & 19,068 & \ \thinspace -2.1\%* & \ \thinspace -0.8\% & \ \ \thinspace 0.0\% & \ \ \thinspace 1.4\% \\ 
  Canada & UW$\to$S & S$\to$S & 2,829 & 19,068 & \ \ \thinspace 4.3\%* & \ \ \thinspace 0.8\% & \ \thinspace -0.3\% & \ \ \thinspace 1.4\% \\ 
  China & C$\to$S & S$\to$S & 19,884 & 34,969 & \ \ \thinspace 6.2\%* & \ \ \thinspace NA & \ \ \thinspace 0.6\%* & \ \thinspace -0.6\%* \\ 
   \hline US & S$\to$SW & SW$\to$SW & 68,494 & 35,515 & \ \ \thinspace 8.3\%* & \ \ \thinspace 2.9\%* & \ \ \thinspace 1.6\%* & \ \thinspace -2.0\%* \\ 
  US & R$\to$SW & SW$\to$SW & 18,061 & 35,515 & \ \ \thinspace 7.1\%* & \ \thinspace -0.1\% & \ \ \thinspace 2.2\%* & \ \thinspace -3.9\%* \\ 
  US & Su$\to$SW & SW$\to$SW & 14,730 & 35,515 & \ \ \thinspace 3.9\%* & \ \ \thinspace 1.0\% & \ \ \thinspace 4.5\%* & \ \thinspace -1.2\%* \\ 
  US & DB$\to$SW & SW$\to$SW & 7,624 & 35,515 & \ 14.4\%* & \ \ \thinspace 0.6\% & \ \ \thinspace 1.9\%* & \ \thinspace -0.7\% \\ 
  US & C$\to$SW & SW$\to$SW & 7,693 & 35,515 & \ \ \thinspace 2.1\%* & \ \thinspace -0.1\% & \ \ \thinspace 0.8\% & \ \thinspace -0.1\% \\ 
  Canada & R$\to$SW & SW$\to$SW & 199,115 & 44,809 & \ \ \thinspace 9.4\%* & \ \ \thinspace 3.0\%* & \ \ \thinspace 1.1\%* & \ \ \thinspace 1.1\%* \\ 
  Canada & Le$\to$SW & SW$\to$SW & 33,892 & 44,809 & \ 10.4\%* & \ \ \thinspace 4.2\%* & \ \ \thinspace 2.7\%* & \ \ \thinspace 0.4\% \\ 
  Canada & Pe$\to$SW & SW$\to$SW & 30,637 & 44,809 & \ 11.3\%* & \ \ \thinspace 4.1\%* & \ \ \thinspace 3.6\%* & \ \thinspace -1.1\%* \\ 
  Canada & S$\to$SW & SW$\to$SW & 11,332 & 44,809 & \ 13.7\%* & \ \ \thinspace 3.9\%* & \ \thinspace -1.0\% & \ \thinspace -1.7\%* \\ 
  Canada & F$\to$SW & SW$\to$SW & 10,139 & 44,809 & \ \ \thinspace 4.7\%* & \ \ \thinspace 5.1\%* & \ \thinspace -3.4\%* & \ \thinspace -2.5\%* \\ 
   \hline US & F$\to$WW & WW$\to$WW & 84,845 & 151,613 & \ \ \thinspace 8.3\%* & \ \ \thinspace 3.0\%* & \ \thinspace -0.4\% & \ \thinspace -4.8\%* \\ 
  US & S$\to$WW & WW$\to$WW & 35,241 & 151,613 & -12.4\%* & \ \ \thinspace 6.5\%* & \ \ \thinspace 5.0\%* & \ \thinspace -4.6\%* \\ 
  US & C$\to$WW & WW$\to$WW & 22,417 & 151,613 & \ \ \thinspace 4.0\%* & \ \ \thinspace 0.8\% & \ \ \thinspace 2.2\%* & \ \ \thinspace 2.9\%* \\ 
  France & R$\to$WW & WW$\to$WW & 32,045 & 18,119 & \ 10.1\%* & \ \thinspace -0.9\%* & \ \thinspace -1.6\%* & \ \ \thinspace 1.8\%* \\ 
  France & Si$\to$WW & WW$\to$WW & 14,917 & 18,119 & \ \ \thinspace 8.3\%* & \ \thinspace -0.2\% & \ \thinspace -1.7\%* & \ \ \thinspace 0.1\% \\ 
  France & C$\to$WW & WW$\to$WW & 13,378 & 18,119 & \ \ \thinspace 6.7\%* & \ \thinspace -0.5\% & \ \ \thinspace 0.3\% & \ \ \thinspace 1.9\%* \\ 
  France & Ch$\to$WW & WW$\to$WW & 9,720 & 18,119 & \ \ \thinspace 7.0\%* & \ \thinspace -0.2\% & \ \ \thinspace 3.2\%* & \ \ \thinspace 1.7\%* \\ 
  France & Su$\to$WW & WW$\to$WW & 8,401 & 18,119 & \ \ \thinspace 6.7\%* & \ \thinspace -0.8\% & \ \ \thinspace 1.0\% & \ \ \thinspace 1.9\%* \\ 
   \hline
\end{tabular}
\end{table}

\newpage

\begin{table}[ht]
\footnotesize
\caption{Estimated average yields in our data sample. The average yields in our data sample are estimated by using subnational-level yield data (CAR-level in Canada, province-level in China, department-level in France, and county-level in the US) drawn from the sources in the 4th column. In particular, for each crop type and country we took a weighted average of subnational-level yield data with the weights determined by the number of samples we had in our satellite-based dataset for each crop type in each subnational administrative unit. Readers interested in converting our reported results in Table \ref{table:ResultsTable} from units of percent of average yield to units of tons per hectare, can multiply those reported estimates by 0.01 times the corresponding average yield estimate reported in this table to obtain estimates in units of tons per hectare.}\label{Table:YieldToTonsPerHectare}
\centering
\begin{tabular}{llrr}
  \hline
Crop type & Country & Average yield estimate (t/ha) & Subnational-level yield data source \\ 
  \hline \hline
Corn & Canada & 9.35 & \cite{CanadaSurveyData}\\ 
  Corn & China & 6.78 & \cite{ChinaSurveyData}\\ 
  Corn & France & 9.01 & \cite{FranceSurveyData}\\ 
  Corn & US & 10.35 & \cite{NASSQuickStats}\\  \hline
  Soybean & Canada & 2.78 & \cite{CanadaSurveyData}\\ 
  Soybean & China & 1.87 & \cite{ChinaSurveyData}\\ 
  Soybean & US & 3.17 & \cite{NASSQuickStats}\\  \hline
  Spring Wheat & Canada & 3.08 & \cite{CanadaSurveyData}\\ 
  Spring Wheat & US & 3.16 & \cite{NASSQuickStats}\\  \hline
  Winter Wheat & France & 7.01 & \cite{FranceSurveyData}\\ 
  Winter Wheat & US & 2.69 & \cite{NASSQuickStats}\\ 
   \hline
\end{tabular}
\end{table} \newpage

\begin{figure}[!h]
    \centering
    \includegraphics[width=0.95 \hsize]{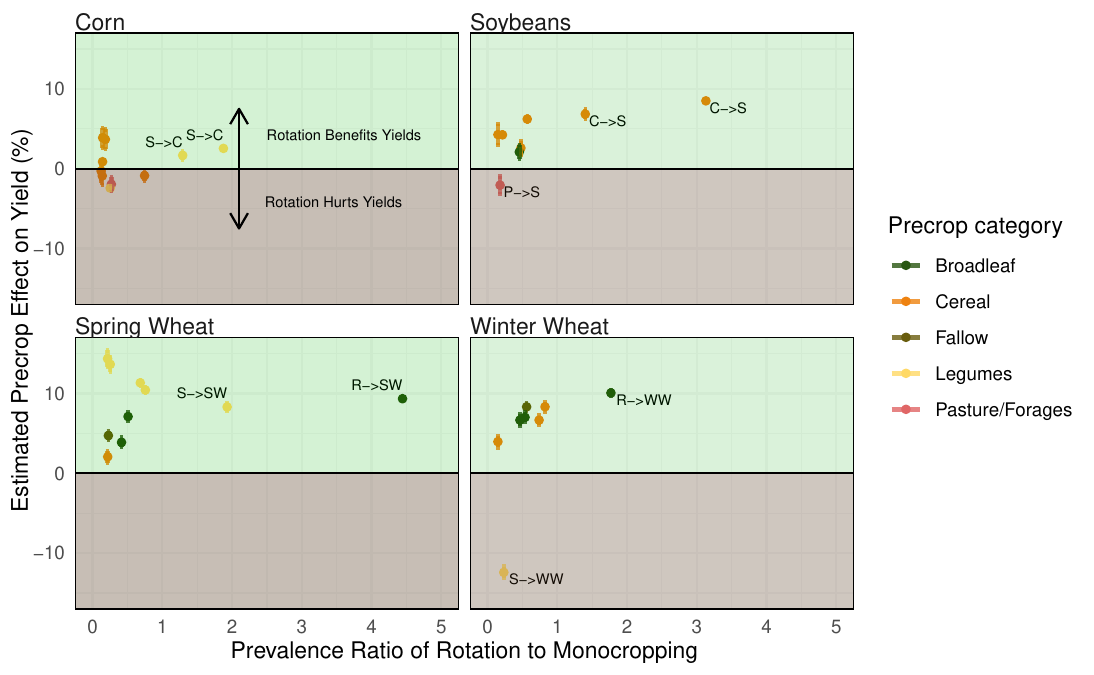}
    \caption{Estimated precrop effects colored by precrop category. This figure is the same as Figure \ref{fig:TwoYearRotBenefit}, except each point is colored by the precrop category rather than by country.}
    \label{fig:TwoYearRotBenefitPrecropCategory}
\end{figure}

\begin{figure}[hbt!]
    \centering
    \includegraphics[width=0.95 \hsize]{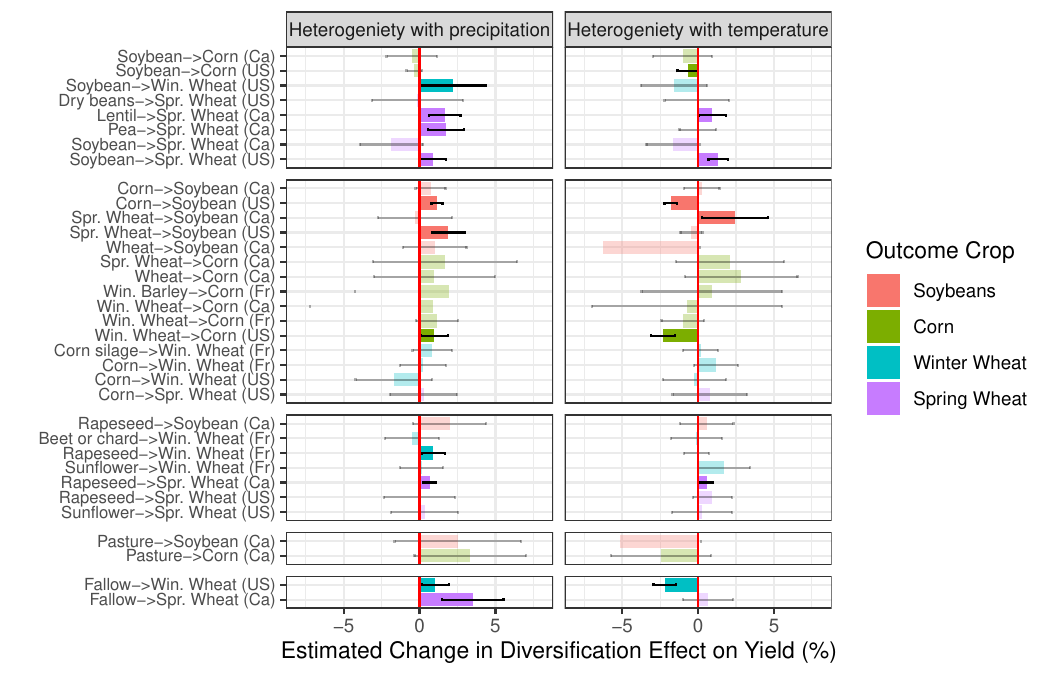}
    \caption{Estimates of the heterogeneity of the diversification effects with weather. Each row specifies the precrop, outcome crop, and the country from which the study sample was taken (Ca=Canada, Ch=China, Fr=France, US=United States). For each precrop A and outcome crop B, the figure depicts how the impact of switching from a simple rotation B$\to$A$\to$B to a diverse rotation of the form Other Crop$\to$A$\to$B varies with precipitation and temperature using the best fit regression coefficients to the linear model in Equation \eqref{eq:ModelToAssessHeterogeneity}. The figure depicts rescaled regression coefficients for growing season precipitation (left column) and temperature (right column) which estimate how much the diversification effect changes when the precipitation or temperature increases from the 25th percentile to the 75th percentile of observed values. The x-axis values are in units of percent of the mean crop yield of the corresponding outcome crop in the corresponding study region and can be converted to units of tons per hectare using the yield estimates in Table \ref{Table:YieldToTonsPerHectare}. The error bars give 95\% confidence intervals (based on heteroskedasticity-robust (HC3) estimation), and coefficient estimates that are not statistically significant (level $\alpha=0.05$ based on two-sided testing) are given a lighter, translucent color. The rows are grouped into blocks based on 5 precrop categories of interest: legumes, cereals, broadleaf crops, pasture/forages, and fallow.}
    \label{fig:HeterogeneityDiversificationFig}
\end{figure}


\begin{figure}[!t]
    \centering
    \includegraphics[width=0.95 \hsize]{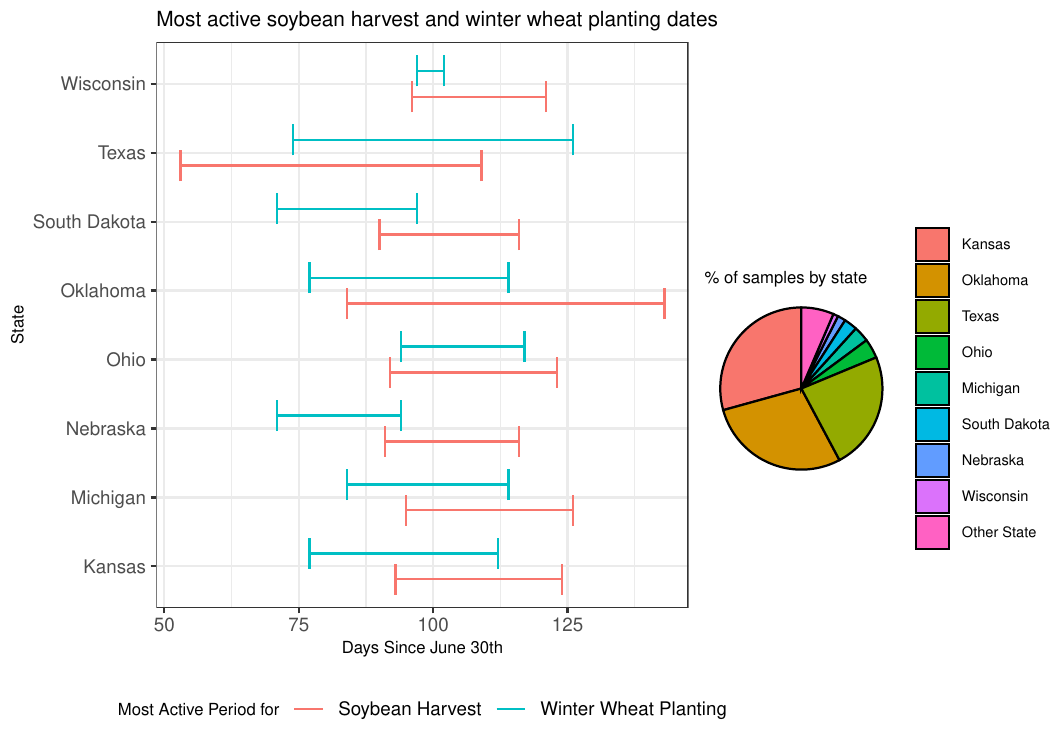}
    \caption{Planting dates for winter wheat versus harvest dates for soybean in the US. The date range for the most active period of soybean harvest and winter wheat planting from each state were extracted from a USDA National Agricultural Statistics Service 2010 report \cite{NASS_TypicalPlantingAndHarvestDateUS} and are plotted in the left panel above. Note that we only plotted planting and harvest periods for the 8 states that met the following criteria: (i) the soybean harvest and winter wheat planting dates were both included in the report (ii) the state contained at least 100 Soybean$\to$Winter Wheat and 100 Winter Wheat$\to$Winter Wheat samples that were used in our study (iii) the state contained at least
    1\% of the samples that were used in our causal forest analysis of the effect of the Soybean$\to$Winter Wheat sequence. The right panel gives a pie chart of the percentage of samples from each state that was used in our analysis of the Soybean$\to$Winter Wheat sequence. In all states except Texas the active period for the soybean harvest ended after the active planting period for winter wheat, although we remark that only a small number (0.18\%) of the Soybean$\to$Winter Wheat samples were from Texas.}
    \label{fig:HarvestPlantingDatesSoyWW}
\end{figure}
 
\end{document}